\documentclass[12pt]{iopart}
\usepackage{iopams}
\usepackage[square,numbers,sort&compress]{natbib}
\usepackage{graphicx,amssymb,bm}
\usepackage{csquotes}
\usepackage{color}
\usepackage[normalem]{ulem}

\newcommand{\refeq}[1]{(\ref{#1})}
\newcommand{\refsec}[1]{\ref{#1}}
\newcommand{\escaleS}{ES }
\newcommand{\iscaleS}{IS }
\newcommand{\iscaleNS}{IS}
\newcommand{\escaleNS}{ES}

\newcommand{\beqn}{\begin{eqnarray}}
\newcommand{\eeqn}{\end{eqnarray}}
\newcommand{\xp}{\times}
\newcommand{\kron}[1]{\delta_{#1}}
\newcommand{\vzero}{0}
\newcommand{\mb}[1]{\bi{#1}}
\newcommand{\fract}[2]{#1 / #2 }

\newcommand{\eye}{\mb{I}}
\newcommand{\jacob}{\mathcal{J}}
\newcommand{\expo}[1]{\exp{[#1]}}
\newcommand{\imag}{\rmi}
\newcommand{\order}[1]{\Or\left( #1 \right) }

\newcommand{\ltsp}{c}
\newcommand{\bu}{\mb{b}}
\newcommand{\bvec}{\mb{B}}
\newcommand{\evec}{\mb{E}}

\newcommand{\bmag}{B}
\newcommand{\cycf}{\Omega}
\newcommand{\wfreq}{\omega}

\newcommand{\fldl}{\alpha}
\newcommand{\fldls}{\alpha_{\rm{s}}}
\newcommand{\fldlf}{\alpha_{\rm{f}}}
\newcommand{\evfldl}{\mb{e}_\fldl}
\newcommand{\flxl}{\psi}
\newcommand{\flxls}{\psi_{\rm{s}}}
\newcommand{\flxlf}{\psi_{\rm{f}}}
\newcommand{\evflxl}{\mb{e}_\flxl}
\newcommand{\bmagform}{\nbl \fldl \xp \nbl \flxl}
\newcommand{\eone}{\mb{e}_1}
\newcommand{\etwo}{\mb{e}_2}

\newcommand{\kk}{\mb{k}}
\newcommand{\ks}{{\mb{k}_{\rm{s}}}}
\newcommand{\ksp}{{\mb{k}^\prime_{\rm{s}}}}
\newcommand{\kf}{{\mb{k}_{\rm{f}}}}
\newcommand{\kfp}{{\mb{k}^\prime_{\rm{f}}}}
\newcommand{\kkflxl}{k_\flxl}
\newcommand{\kkfldl}{k_\fldl}
\newcommand{\ekkflxl}{K_\flxl}
\newcommand{\ekkfldl}{K_\fldl}
\newcommand{\kperp}{k_\perp}
\newcommand{\kpara}{k_\|}

\newcommand{\rr}{{\mb{r}}}

\newcommand{\rs}{{\mb{r}_{\rm{s}}}}
\newcommand{\rf}{{\mb{r}_{\rm{f}}}}

\newcommand{\gc}{{\mb{R}}}
\newcommand{\gcp}{{\mb{R}_\perp}}
\newcommand{\gcs}{{\mb{R}_{\rm{s}}}}
\newcommand{\gcf}{{\mb{R}_{\rm{f}}}}

\newcommand{\ttime}{t}
\newcommand{\ts}{{t_{\rm{s}}}}
\newcommand{\tf}{{t_{\rm{f}}}}
\newcommand{\tcut}{\tau_{\rm{c}}}
\newcommand{\teql}{\tau_{\rm{P}}}

\newcommand{\xcut}{\Delta_{\flxl}}
\newcommand{\ycut}{\Delta_{\fldl}}
\newcommand{\esarea}{A_{\rm{e}}}
\newcommand{\isarea}{A_{\rm{i}}}

\newcommand{\nbl}{\nabla}
\newcommand{\nblper}{\nabla_\perp}
\newcommand{\nblpar}{\nabla_\|}
\newcommand{\nblf}{\nbl_{\rm{f}}}
\newcommand{\nbls}{\nbl_{\rm{s}}}

\newcommand{\pvel}{\mb{v}}
\newcommand{\gyrophase}{\gamma}
\newcommand{\phase}{\Psi}
\newcommand{\energy}{\varepsilon}
\newcommand{\pitch}{\lambda}
\newcommand{\sign}{\sigma}
\newcommand{\vpar}{v_\|}
\newcommand{\vmag}{v}
\newcommand{\vperp}{v_\perp}
\newcommand{\lpar}{\theta}
\newcommand{\kpar}{\bu \cdot \nbl \lpar}
\newcommand{\lscal}{a}
\newcommand{\toragl}{\zeta}
\newcommand{\saffac}{q}
\newcommand{\saffacprim}{\saffac^\prime_0}
\newcommand{\bcur}{I}
\newcommand{\bmagformaxi}{\bcur(\flxl) \nbl \toragl+\nbl \toragl \xp \nbl \flxl}

\newcommand{\func}{H}
\newcommand{\orbav}[1]{\left\langle{#1}\right\rangle^{\rm{o}}}
\newcommand{\turbav}[1]{\left\langle{#1}\right\rangle^{\mbox{turb}}}
\newcommand{\esav}[1]{\left\langle{#1}\right\rangle^{\mbox{ES}}}
\newcommand{\gav}[2]{\left\langle{#1}\right\rangle^{\gyrophase}_{#2}}
\newcommand{\intv}[1]{\int d^3 \mb{v} #1 }

\newcommand{\ovl}[1]{\overline{#1}}
\newcommand{\tld}[1]{\tilde{#1}}

\newcommand{\ptl}[1]{\phi_{#1}} 
\newcommand{\gptl}[1]{\varphi_{#1}}

\newcommand{\ciptl}{\ovl{\phi}_{\rm{c}}}
\newcommand{\iptl}[1]{\ovl{\phi}_{#1}} 
\newcommand{\igptl}[1]{\ovl{\varphi}_{#1}}
\newcommand{\igptli}[1]{\ovl{\varphi}_{\rm{i} #1}}
\newcommand{\igptle}[1]{\ovl{\varphi}_{\rm{e} #1}} 

\newcommand{\eptl}[1]{\tld{\phi}_{#1}} 
\newcommand{\egptl}[1]{\tld{\varphi}_{#1}} 
\newcommand{\egptli}[1]{\tld{\varphi}_{\rm{i} #1}} 
\newcommand{\egptle}[1]{\tld{\varphi}_{\rm{e} #1}} 

\newcommand{\fluc}[1]{\phi_{#1}}
\newcommand{\hh}[1]{h_{#1}}
\newcommand{\hhe}[1]{h_{\rm{e} #1}}
\newcommand{\hhi}[1]{h_{\rm{i} #1}}
\newcommand{\ihh}[1]{\ovl{h}_{#1}} 
\newcommand{\ihhi}[1]{\ovl{h}_{\rm{i} #1}} 
\newcommand{\ihhe}[1]{\ovl{h}_{\rm{e} #1}} 
\newcommand{\ehh}[1]{\tld{h}_{#1}}  
\newcommand{\ehhi}[1]{\tld{h}_{\rm{i} #1}} 
\newcommand{\ehhe}[1]{\tld{h}_{\rm{e} #1}}

\newcommand{\igge}[1]{\ovl{g}_{\rm{e} #1}} 
\newcommand{\egg}[1]{\tld{g}_{#1}}  
 
\newcommand{\egge}[1]{\tld{g}_{\rm{e} #1}} 

\newcommand{\fulldist}{f}
\newcommand{\fullptl}[1]{\Phi_{#1}}
\newcommand{\dlf}{\delta f}
\newcommand{\idlf}{\ovl{\delta f}}
\newcommand{\edlf}{\widetilde{\delta f}}

\newcommand{\idlfe}{\ovl{\delta f}_{\rm{e}}}
\newcommand{\edlfe}{\widetilde{\delta f}_{\rm{e}}}

\newcommand{\spe}{\nu}
\newcommand{\eqlba}{F_{0\spe}}
\newcommand{\eqlb}{F_{0}}
\newcommand{\eqlbi}{F_{0i}}
\newcommand{\eqlbe}{F_{0e}}
\newcommand{\eqlbf}{F}

\newcommand{\bes}[1]{ J_0(#1)}
\newcommand{\jies}{\tld{J}}

\newcommand{\gyrd}{\rho_{\rm{th}}}
\newcommand{\gyrdvec}{\bm{\rho}}
\newcommand{\gyrdvecs}{\bm{\rho}_\spe}
\newcommand{\gyrdvece}{\bm{\rho}_e}
\newcommand{\gyrdveci}{\bm{\rho}_i}
\newcommand{\gyrds}{\rho_{\rm{th},\spe}}
\newcommand{\gyrdi}{\rho_{\rm{th,i}}}
\newcommand{\gyrde}{\rho_{\rm{th,e}}}

\newcommand{\dens}{n}
\newcommand{\denss}{n_\spe}
\newcommand{\densi}{n_{\rm{i}}}
\newcommand{\dense}{n_{\rm{e}}}

\newcommand{\temp}{T}
\newcommand{\temps}{T_\spe}
\newcommand{\tempi}{T_{\rm{i}}}
\newcommand{\tempe}{T_{\rm{e}}}

\newcommand{\vther}{v_{\rm{th}}}
\newcommand{\vtheri}{v_{\rm{th,i}}}
\newcommand{\vthere}{v_{\rm{th,e}}}
\newcommand{\vthers}{v_{\rm{th},\spe}}

\newcommand{\zed}{Z}
\newcommand{\zeds}{Z_\spe}
\newcommand{\zedi}{Z_{\rm{i}}}
\newcommand{\charge}{e}

\newcommand{\drv}[2]{\frac{\partial #1}{\partial #2}}
\newcommand{\drvt}[2]{{\partial #1}/{\partial #2}}

\newcommand{\ma}{m_\spe}
\newcommand{\me}{m_{\rm{e}}}
\newcommand{\mi}{m_{\rm{i}}}
\newcommand{\massrt}{\left(m_{\rm{e}}/m_{\rm{i}}\right)^{1/2}}
\newcommand{\massrst}{\left(m_{\rm{e}}/m_{\rm{i}}\right)^{1/4}}
\newcommand{\meomi}{\left(\frac{m_{\rm{e}}}{m_{\rm{i}}}\right)}
\newcommand{\meominb}{\frac{m_{\rm{e}}}{m_{\rm{i}}}}
\newcommand{\massr}{\meomi^{1/2}}
\newcommand{\massru}{\left(\frac{m_{\rm{i}}}{m_{\rm{e}}}\right)^{1/2}}
\newcommand{\massruct}{\left({m_{\rm{i}}}/{m_{\rm{e}}}\right)^{3/2}}
\newcommand{\massrut}{\left({m_{\rm{i}}}/{m_{\rm{e}}}\right)^{1/2}}

\newcommand{\massrs}{\meomi^{1/4}}
\newcommand{\massrl}{\meominb}
\newcommand{\massrlt}{\me/\mi}

\newcommand{\vm}{\mb{v}^M}
\newcommand{\vme}{\mb{v}^M_{\rm{e}}}
\newcommand{\vmi}{\mb{v}^M_{\rm{i}}}

\newcommand{\ve}{\mb{v}^E}
\newcommand{\eve}{\tld{\mb{v}}^E}
\newcommand{\ive}{\ovl{\mb{v}}^E}
\newcommand{\cive}{\ovl{\mb{u}}}

\newcommand{\evee}{\tld{\mb{v}}^E_{\rm{e}}}
\newcommand{\ivee}{\ovl{\mb{v}}^E_{\rm{e}}}
\newcommand{\evei}{\tld{\mb{v}}^E_{\rm{i}}}
\newcommand{\ivei}{\ovl{\mb{v}}^E_{\rm{i}}}

\newcommand{\cycfs}{\Omega_\spe}
\newcommand{\cycfe}{\Omega_{\rm{e}}}
\newcommand{\cycfi}{\Omega_{\rm{i}}}

\newcommand{\ecorrper}{\tld{l}_{\perp}}
\newcommand{\icorrper}{\ovl{l}_{\perp}}
\newcommand{\ecorrpar}{\tld{l}_{\|}}
\newcommand{\icorrpar}{\ovl{l}_{\|}}

\newcommand{\source}{S}
\newcommand{\fsav}[1]{\left\langle #1 \right\rangle^{\rm{FT}}}
\newcommand{\hflux}{Q}
\newcommand{\ihflux}{\ovl{Q}}
\newcommand{\ihfluxi}{\ovl{Q}_{\rm{i}}}
\newcommand{\ihfluxe}{\ovl{Q}_{\rm{e}}}
\newcommand{\ehflux}{\tld{Q}}
\newcommand{\ehfluxi}{\tld{Q}_{\rm{i}}}
\newcommand{\ehfluxe}{\tld{Q}_{\rm{e}}}

\newcommand{\intrf}{\int d^2 \rf}
\newcommand{\intrs}{\int d^2 \rs}
\newcommand{\intrpar}{\int d \mb{r}_\| }
\newcommand{\isclvol}{V_{\rm{i}}}
\newcommand{\esclvol}{V_{\rm{e}}}
\newcommand{\nescl}{N_{\rs}}

\newcommand{\lparp}{\lpar^{+}}
\newcommand{\lparm}{\lpar^{-}}
\newcommand{\lparpm}{\lpar^{\pm}}

\newcommand{\copNL}{C_{}}
\newcommand{\cop}{C_{\rm{l}}}

\newcommand{\cope}{C_{\rm{le}}}
\newcommand{\copi}{C_{\rm{li}}}
\newcommand{\icope}{\ovl{C_{\rm{le}}}}
\newcommand{\ecopi}{\widetilde{C_{\rm{li}}}}
\newcommand{\icopi}{\ovl{C_{\rm{li}}}}
\newcommand{\ecope}{\widetilde{C_{\rm{le}}}}
\newcommand{\icopek}{\ovl{C_{\rm{le}}}_{\ks}}

\newcommand{\icopik}{\ovl{C_{\rm{li}}}_{\ks}}
\newcommand{\ecopek}{\widetilde{C_{\rm{le}}}_{\kf}}

\newcommand{\copnaii}{C_{\rm{ii}}}

\newcommand{\copnaee}{C_{\rm{ee}}}
\newcommand{\copnaei}{L_{\rm{ei}}}

\newcommand{\copii}[1]{C_{\rm{ii}}\left[#1\right]}

\newcommand{\copee}[1]{C_{\rm{ee}}\left[#1\right]}
\newcommand{\copei}[1]{L_{\rm{ei}}\left[#1\right]}
\newcommand{\cfreq}{\nu_{\rm{C}}}
\newcommand{\cfreqii}{\nu_{\rm{ii}}}
\newcommand{\cfreqie}{\nu_{\rm{ie}}}
\newcommand{\cfreqei}{\nu_{\rm{ei}}}
\newcommand{\cfreqee}{\nu_{\rm{ee}}}
\newcommand{\dvel}{\delta \vmag}
\newcommand{\dvels}{\delta \vmag_{\spe}}
\newcommand{\dveli}{\delta \vmag_{\rm{i}}}
\newcommand{\dvele}{\delta \vmag_{\rm{e}}}
\newcommand{\uveli}{\mb{u}_{\rm{i}}}
\newcommand{\iuveli}{\ovl{\mb{u}}_{\rm{i}}}

\newcommand{\euveli}{\tld{\mb{u}}_{\rm{i}}}

\newcommand{\inonlintime}{\ovl{\tau}_{\rm{nl}}}
\newcommand{\enonlintime}{\tld{\tau}_{\rm{nl}}}

\newcommand{\myoverbrace}[2]{\overbrace{#1}^{\displaystyle #2}} 
\begin{document}

\title[Scale-Separated Turbulence]{   A Scale-Separated Approach for Studying 
   Coupled Ion and Electron Scale Turbulence}
\author{M. R. Hardman$^{1,2}$, M. Barnes$^{1,2}$, C. M. Roach$^2$, \\ and F. I. Parra$^{1,2}$}
\address{ $^1$ Rudolf Peierls Centre for Theoretical Physics, University of Oxford, Oxford OX1 3PU, UK}
\address{ $^2$ Culham Centre for Fusion Energy, Abingdon OX14 3DB, UK}
\ead{\texttt{michael.hardman@physics.ox.ac.uk}} 

\begin{abstract}

Multiple space and time scales arise in plasma turbulence in
magnetic confinement fusion devices because of the
smallness of the square root of the electron-to-ion mass ratio $\massrt$ and the
consequent disparity of the ion and electron thermal 
gyroradii and thermal speeds. Direct simulations of 
this turbulence that include both ion and electron 
space-time scales indicate that there can be significant
 interactions between the two scales. The extreme 
 computational expense and complexity of these direct
 simulations motivates the desire for reduced treatment.
 By exploiting the scale separation between ion and electron scales,
 and expanding the gyrokinetic equations for the turbulence in $\massrt$,
 we derive such a reduced system of gyrokinetic equations
 that describes cross-scale interactions.
 The coupled gyrokinetic equations contain novel terms
 which provide candidate mechanisms for the observed
 cross-scale interaction. The electron scale turbulence
 experiences a modified drive due to gradients in the ion scale
 distribution function, and is advected by the ion scale
 $E \xp B$ drift, which varies in the direction parallel to the magnetic field line.
The largest possible cross-scale term in the ion scale equations is
 sub-dominant in our $\massrt$ expansion.
 Hence, in our model the ion scale turbulence evolves independently
 of the electron scale turbulence. To complete the scale-separated
 approach, we provide and justify a parallel
boundary condition for the coupled gyrokinetic equations
 in axisymmetric equilibria based on the standard \enquote{twist-and-shift}
 boundary condition. This approach allows one to simulate multi-scale turbulence
 using electron scale flux tubes nested within an ion scale flux tube.
 
\end{abstract}
\newline

\noindent Prepared for submission to: \PPCF
\maketitle

\section{Introduction}

 Anomalous transport of heat and particles is a major limiting
 factor in the performance of tokamaks. 
The dominant transport mechanism is
turbulence arising from micro-instabilities 
that are driven by macroscopic gradients 
of the plasma profiles.
 Whilst the plasma profiles have length scales of order
 the size of the device $a$ in all directions, 
 the characteristic turbulent wavenumbers perpendicular
 and parallel to the magnetic field
 $\kperp$ and $\kpara$ typically satisfy 
 $\kperp \gyrd \sim 1$ and $\kpara \lscal \sim 1$ respectively, where $\gyrd$ 
is the thermal gyroradius of a particular particle species. In many existing experimental
 devices, and in projected reactor conditions,
 $\gyrd / a \ll 1 $ due to the strong confining magnetic field.
Consequently, one can assume a separation
of spatial scales between the plasma equilibrium and the fluctuations
in the plane 
 perpendicular to the magnetic field line. In addition, 
 the equilibrium profiles typically evolve much more
slowly than the turbulence, which fluctuates at a characteristic
 frequency $\wfreq \sim \vther/a$,
 where $\vther$ the thermal speed. Therefore,
 one can assume scale-separation in time.
 Using the assumptions of scale-separation in space and in time
 it is possible to derive separate
 but coupled evolution equations for
 the plasma profiles and the turbulent 
fluctuations 
\cite{sugamaPoP97,SugamaPoP98,trantimecatto08PPCF,abelRPP13,candyPoP09,barnesPoP10b,longwavecalvo12PPCF}. \newline

 Whilst the turbulence is often scale-separated from the profiles,
 the turbulence itself contains subsidiary scales due to the
 presence of multiple species, which introduces multiple
 thermal speeds $\vthers = \sqrt{2\temps/\ma}$ and
 multiple thermal gyroradii $\gyrds = \vthers / \cycfs$, where
 $\temps$ is the species temperature, $\ma$ is the species mass,
 and $\cycfs = \zeds \charge \bmag / \ma \ltsp $
 is the species cyclotron frequency,
 where $\zeds$ is the species charge number, 
 $\charge$ is the proton charge, $\bmag$ 
 is the magnetic field strength and $\ltsp$ is the speed of light.
 Ions and electrons have vastly different masses, $\mi \gg \me$. 
 In the core of a magnetic confinement fusion device
 ions and electrons typically have temperatures of
 the same order $\tempi \sim \tempe \sim \temp$.
 As a consequence, distinct micro-instabilities exist at
 the ion scale, where $\wfreq \sim  \vtheri/a$ and
 $\kperp \gyrdi \sim 1$, and at the electron scale,
 where $\wfreq \sim  \vthere/a$ and $\kperp \gyrde \sim 1$,
 due to the differing dynamics at each scale.
The turbulence arising from the micro-instabilities 
can thus be scale-separated and have a multi-scale character.  \newline

Until recently, the main paradigm for understanding
turbulent transport was that the transport was due to the larger
wavelength modes in the turbulence (cf.~\cite{surkoS83,cowleyPoFB91,hortonRMP99}),
 $\kperp \gyrdi \lesssim 1$; i.e., where ion physics plays an important role. This paradigm
 and the computational cost of multi-scale simulations,
where one must resolve a wide range of space-time scales,
 has meant that investigations into turbulent transport have
 mostly studied ion scale turbulence in isolation through
single-scale simulation and theory with the implicit
 assumption that there are no interactions between 
 fluctuations at the disparate ion and electron scales in the turbulence. Electron
 scale turbulence has, until recently,
been studied independently of the ion scale turbulence under
 the same implicit assumption that there is no significant interaction
 between the ion and electron scales. This assumption is likely to be
 valid when the ion scale turbulence is suppressed but is otherwise questionable. 
 Examples may be found in 
 \cite{dorland2000electron,jenko2000electron,jenko2002prediction,STroach2009PPCF,ResolvingWGuttenfelder2011POP,ProgressWGuttenfelder2013NF,2016colyer}. 
 Nonetheless, it is known that electron scale turbulence can
  drive experimentally relevant levels of transport in some cases \cite{jenko2002prediction}.
Electron scale transport has been observed on NSTX \cite{ren2017recent},
 and is a candidate for anomalous transport on MAST \cite{STroach2009PPCF,2016colyer}. \newline
 
 Without directly simulating or observing the full multi-scale turbulence,
 it is difficult to assess to what extent
there are cross scale interactions in the turbulence,
and whether or not all scales will contribute significantly to the transport.
Unfortunately, studying multi-scale turbulence through direct
simulation is made very challenging by the size of $\massrt$ for a realistic deuterium
 plasma, $\massrt \sim 1/60$, which determines the separation
 of $\gyrde/\gyrdi \sim \massrt$ and $\vtheri/\vthere \sim \massrt$. 
 For example, if one wanted to extend the resolution
 of a well-resolved ion scale
 simulation to capture both the $\lscal / \vtheri$ and $\lscal / \vthere$
 time scales then one must increase the resolution in
 time by approximately $\vthere/\vtheri \sim \massrut$.
 To resolve length scales perpendicular to the magnetic field line
 comparable to both $\gyrdi$ and $\gyrde$
 one must increase the resolution in both the perpendicular
 directions by $\gyrdi/\gyrde \sim \massrut$.
 Overall the increased cost scales like $\massruct$.
 For the deuterium mass a well resolved multi-scale
 simulation could be expected to cost $60^3$ more than
 the well resolved  ion scale simulation.
 This cost is currently prohibitive for routine investigation. \newline

The earliest attempts to study multi-scale turbulence
via direct simulation were made in \cite{ waltz2007coupled, candy2007effect, gorler2008scale}
 using unphysically small ion to electron mass ratio.
 Recently, with improvements in computing power,
 it has been possible to perform small numbers of direct 
multi-scale simulations with the deuterium 
\cite{howard2014synergistic,howard2015fidelity,howard2016enhanced,howard2016comparison}
 and hydrogen \cite{maeyama2015cross,maeyama2017supression} mass ratio.
 The multi-scale 
 simulations allow us to observe 
 features of multi-scale turbulence: there can be a scale-separation; 
 the electron scale heat flux can be comparable to the ion scale heat flux, and even necessary 
 to match experimental results \cite{howard2016enhanced};
 and there are nontrivial interactions between the ion and electron scale.
 In \cite{ waltz2007coupled, candy2007effect, maeyama2015cross, howard2016comparison, howard2016enhanced}
 it is shown that the ion scale fluctuations can affect the electron scale fluctuations.
 This is demonstrated in Figure 3 of \cite{howard2016enhanced},
 where we see that varying the ion temperature gradient
drastically changes the electron scale fluctuations.
As the  ion response at electron scales is negligible,
 the only mechanism through which the ion temperature gradient can affect the electron scale
 fluctuations is by cross-scale interactions.
  Further evidence for this cross-scale interaction
 appears in Figure 2 of \cite{maeyama2015cross}, where high-$\kperp$ modes
 in the multi-scale simulation are suppressed
 compared to the modes in the high-$\kperp$ single scale simulation. 
 We also note that the electron scale turbulence can affect the ion scale
 turbulence. The presence of the electron
 scale turbulence can increase ion scale fluctuation amplitudes compared to an ion scale
 only simulation \cite{ howard2016comparison,maeyama2015cross,howard2014synergistic};
see, e.g., Figure 2b of \cite{maeyama2015cross} and
 Figure 3 of \cite{howard2014synergistic}. 
 In \cite{maeyama2017supression} the electron scale turbulence 
 is able to effectively suppress the microtearing mode,
 which exists in the low-$\kperp$ range.  \newline

 We note that using unphysically large values of $\massrt$  
 can lead to qualitatively unrealistic results in numerical experiments
 \cite{howard2014multi,howard2015fidelity}. For example, in Figure 3 of \cite{howard2015fidelity}
 we see that there are clearly defined, separated, ion and electron scale peaks
 in the electron heat flux for the physical mass ratio $\massrt = 1/60 $. 
 However, in the case with $\massrt = 1/20 $ 
 we see only a single peak in the electron heat flux spectrum,
 indicating that for the unphysical value of $\massrt$, for the parameters considered in \cite{howard2015fidelity},
 the ion and electron scales can no longer be distinguished or separated. \newline
 
\begin{figure}[htb]
\begin{center}

\includegraphics[width=0.7\textwidth]{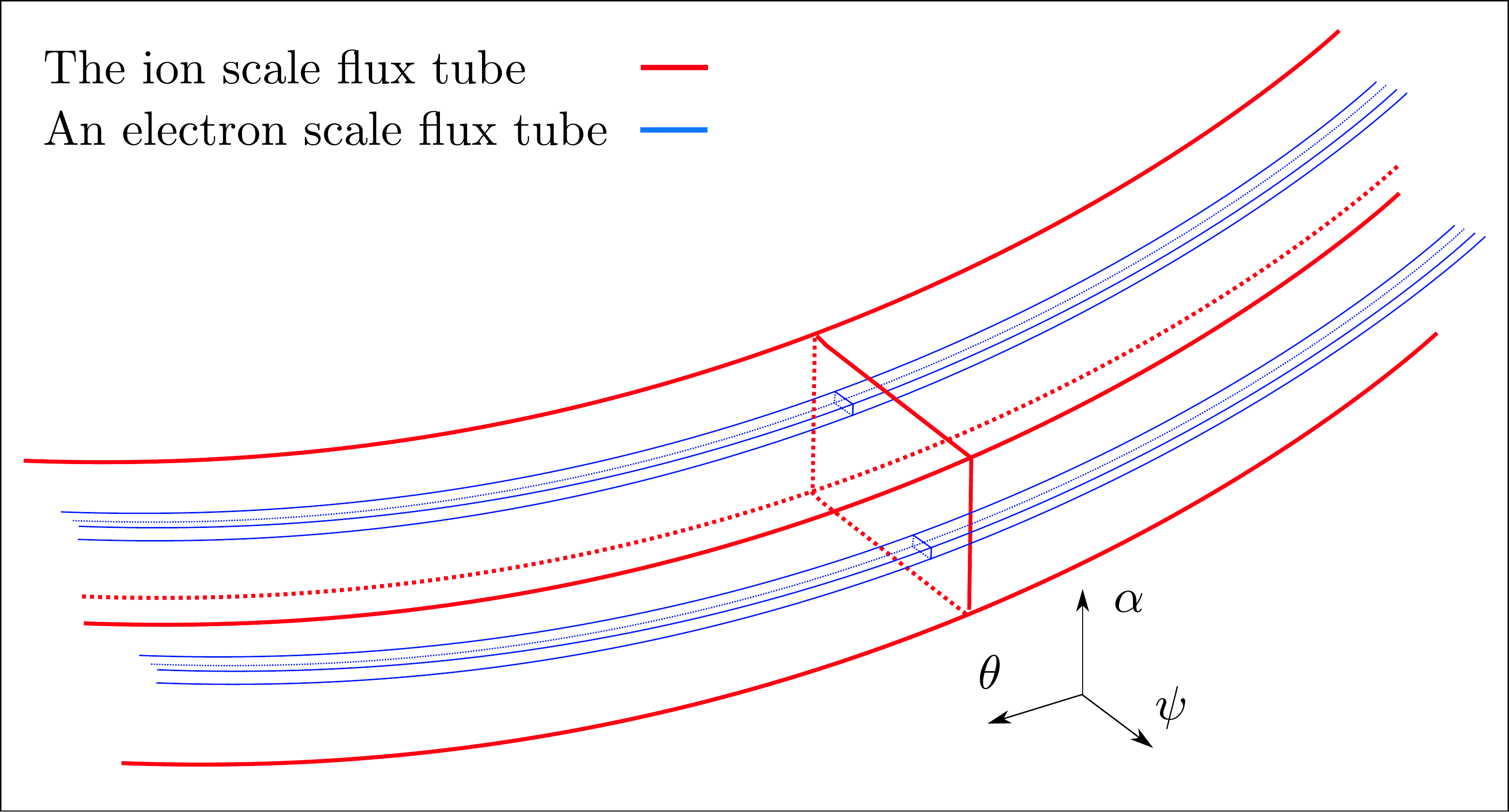}
\caption { Diagram illustrating how electron scale flux tubes (blue)
 may be nested within a larger ion scale flux tube (red).
 The coordinates $(\flxl,\fldl,\lpar)$ denote the usual magnetic-field-following
 radial, binormal, and parallel-to-the-field coordinates.}
\label{figure:cartoonfluxtubes}
\end{center}
\end{figure}

Direct multi-scale simulations demonstrate that there is a
 rich variety of physics to investigate.
 However, the high computational cost and the difficulty of 
diagnosing direct multi-scale simulations
 means that there is a need for analytic theory to help 
provide a theoretical understanding
 of the mechanisms of the cross-scale interactions. 
In this paper we will assume scale-separation between the ion and electron
 scales in the turbulence. 
By treating $\massrt $ as an asymptotically-small parameter 
we expand the gyrokinetic equation for
 the turbulence to find separate but coupled evolution
 equations for the ion and electron scale turbulence. These equations
 may be solved using a system of coupled flux tubes,
 visualised in Figure \ref{figure:cartoonfluxtubes}.
 This approach is reminiscent of the approach taken in 
 \cite{sugamaPoP97,SugamaPoP98,trantimecatto08PPCF,abelRPP13,candyPoP09,barnesPoP10b,longwavecalvo12PPCF}
 to study the evolution of the turbulence
 and the profiles in a scale-separated way. \newline
 
The remainder of the paper is organised as follows.
 In Section \refsec{section:derivation} we review
 the concepts used to derive the gyrokinetic equation that
 will be necessary for our separation of the ion and electron 
scales.
In Section \refsec{sec:separation} we state our orderings
for length and time scales and present the formalism which
 we will use to separate the ion and electron scales.
We also introduce the method of multiple scales as a technique
for deriving the coupled gyrokinetic equations and we define
 an electron scale average, which, along with the assumption
of statistical periodicity of the electron scale turbulence,
allows us to uniquely decompose the turbulence into ion
and electron scale pieces.
 In Section \refsec{sec:coupledgke} we apply the electron scale average
 to find the coupled ion scale and electron scale gyrokinetic
 equations, and retain the largest possible cross-scale interaction terms
 in our expansion in $\massrt$.
 We do this without explicitly assuming the size of the fluctuations,
to allow for the possibility of exotic orderings 
for the sizes of the fluctuations in the presence of cross-scale coupling. 
 In Section \refsec{sec:qn} we use the electron scale average to
find the quasineutrality relations which close the coupled equations.
By using dominant balance arguments, we find in Section 
\refsec{sec:size} the only self-consistently-allowed 
 ordering for the size of the ion and electron scale
 fluctuations: the usual gyro-Bohm ordering.
We show that it is possible to neglect the non-adiabatic response of the ion species at
 electron scales, which is necessary
 for a local description of the electron scale.
In Section \refsec{sec:sclsepcoupledeq} we use our results
 to obtain the maximally-ordered, scale-separated, coupled gyrokinetic equations. 
 At the ion scale we retain the usual ion physics.
 However, the equation for electrons at ion scales is averaged over the particle
 orbits in the direction parallel to the magnetic field. 
The ion scale fluctuations evolve independently of the electron scale turbulence,
 because there are no cross-scale terms appearing in the ion scale equations.
At the electron scale the response of the ion species can be modelled as Maxwell-Boltzmann.
 The electron equation at electron scales contains
 two new terms which depend on gradients of ion scale fluctuations.
 There is an $E \xp B$ advection due to the ion scale potential,
 which varies in the direction parallel to the magnetic field line.
 The effect of the $E \xp B$ advection is to
 shear electron scale eddies in the parallel-to-the-field direction
 by differential flows.
 The gradient of the equilibrium distribution function,
the usual drive of instability, is modified by the gradient
of the ion scale, electron distribution function.
In Section \refsec{sec:pbc} we provide and justify a 
parallel boundary condition for the electron scale
 gyrokinetic equation, consistent with the \enquote{twist-and-shift} 
 boundary condition \cite{beerPoP95} for the ion scale turbulence in a local flux tube domain.
Finally, in Section \refsec{sec:physics} we discuss the insights drawn from our scale-separated approach
 and the physics of the new cross-scale terms that explicitly appear in the coupled equations.

\section {The Gyrokinetic Equation}\label{section:derivation}

The multi-scale equations that describe the interaction between turbulent fluctuations
 at ion scales and turbulent fluctuations at electron scales are obtained
 from the $\dlf$ gyrokinetic equation via an asymptotic expansion in the
 electron-ion mass ratio. This is a subsidiary expansion within the
 gyrokinetic ordering \cite{cattoPP78,friemanPoF82,sugamaPoP97,brizardRMP07,parraPPCF08,abelRPP13}.
 As such, we begin by briefly presenting the steps in
 the derivation of the $\dlf$ gyrokinetic equation, which describes the
 evolution of turbulent \enquote{fluctuations}
 around a slowly changing \enquote{equilibrium} plasma profile in a plasma
 which is strongly magnetised. In this paper we only consider electrostatic fluctuations, 
 where we can express the electric field $\evec$ purely in terms of the
 electrostatic potential $\fullptl{}$, i.e. 
 \beqn \evec = -\nabla \fullptl{}. \label{eq:efield} \eeqn 
 We specialise
 to a toroidal geometry. Hence the magnetic field $\bvec$ can always be written as
 \beqn \bvec = \bmagform,\label{eq:bmagform1}\eeqn where the flux label $\flxl$
 and field line label $\fldl$ will act as our coordinates perpendicular to the field line.
 We will take the poloidal angle $\lpar$ as the coordinate which determines position along the field line. 

The $\delta f$ gyrokinetic equation is derived by assuming a separation of space and
 time scales between the equilibrium profile and fluctuating
 parts of the particle distribution function. Explicitly, we take
 \beqn \gyrds / \lscal \rightarrow 0 \label{eq:rhostar}\eeqn
 and 
 \beqn \cycfs \gg \wfreq \gg \teql^{-1},\label{eq:timesorders} \eeqn 
where $\lscal$ is a characteristic equilibrium scale length, 
$\wfreq$ is a typical fluctuation frequency and $\teql$ is a characteristic
 timescale of the equilibrium profiles.
It is conventional to order the collision frequency $\cfreq \sim \wfreq$
as a maximal ordering to allow for the possibility of both
 collisionless and collisional plasmas.
 As the electron-ion mass ratio is treated as an order unity parameter
 in the gyrokinetic ordering the gyrokinetic equation for each species
 has the same form. We thus simplify notation by suppressing the species
 index $\spe$ where it does not introduce ambiguity.
 We assume that the distribution function $\fulldist$ and
 the electrostatic potential $\fullptl{}$
 have negligible amplitude at scales intermediate to $\gyrds$ and $\lscal$. Moreover, we
 assume there is no direct cascade of energy between the scales.
  It is then possible to split the distribution function $\fulldist$,
  \beqn \fulldist = \eqlbf + \dlf, \label{eq:distfnsplit}\eeqn with $\eqlbf$ and $\dlf $
the equilibrium and fluctuating parts of the distribution function respectively.
We introduce a turbulent average $\turbav{\cdot}$, which averages over
 spatial scales and time scales which are intermediate to the fluctuation and equilibrium scales,
 such that \beqn \eqlbf = \turbav{\fulldist},\label{eq:distfnav} \eeqn with the 
assumption of statistical periodicity \beqn \turbav{\dlf} = 0. \label{eq:distfnsolve}\eeqn 
 Note that \refeq{eq:distfnsolve} can be satisfied by imposing
 periodicity on the turbulence in the plane perpendicular the magnetic field.
 This is justified by noting that the turbulent fluctuations
 should have the same statistics everywhere in a domain that is asymptotically
 small compared to $\lscal$. This can be ensured by deriving local evolution equations for the
 turbulence, choosing a simulation domain larger than the correlation length of the turbulence,
 and imposing periodicity as the boundary condition in the perpendicular domain.
 We are able to make the assumption of statistical periodicity because of our assumptions
 of scale separation. This is in contrast to the study of neutral fluids, where the turbulence
 is universal and the inertial range spans the outer scale to the dissipation scale. As there
 is no scale-separation in neutral fluids, sub-grid models for neutral fluid turbulence
 are derived using large eddy simulation (LES) approaches
 (see e.g. \cite{germano_1992,BerselliLES,GarnierLES}).
Let time be $\ttime$, $\nbl$ be the gradient operator, $\bu$ be the unit vector
 in the direction of the magnetic field, and $\eye$ be the identity matrix.  
Then $\nblpar = \bu \cdot \nbl$ is the derivative parallel to the magnetic field,
 and $\nblper = (\eye - \bu \bu ) \cdot \nbl$ 
 is the derivative perpendicular to the magnetic field. 
 With these definitions the gyrokinetic orderings for 
$\eqlbf$, $\dlf$ and the fluctuating electrostatic potential $\ptl{}$ are
\beqn \nblpar \eqlbf \sim \nblper \eqlbf  \sim \frac{\eqlbf}{\lscal}, \quad \drv{\eqlbf}{\ttime} \sim \frac{\eqlbf}{\teql},  \nonumber \eeqn
\beqn \nblpar {\dlf} \sim \frac{\dlf}{\lscal}, \quad \nblper {\dlf} \sim \frac{\dlf}{\gyrd},\quad \drv{\dlf}{\ttime} \sim \wfreq \dlf,  \label{eq:gkordering} \eeqn
\beqn \frac{\dlf}{\eqlbf} \sim  \frac{e \ptl{}}{T} \sim \frac{\gyrd}{\lscal}.\nonumber\eeqn
The ordering \refeq{eq:gkordering} indicates that the fluctuations are highly anisotropic
 with respect to the magnetic field and evolve rapidly in time compared to the equilibrium.
 
 The gyrokinetic ordering \refeq{eq:gkordering} is motivated by
 particle motion in magnetic confinement fusion devices, which consists of rapid helical motion
following magnetic field lines. Particles stream along the field at thermal speed time
 scales $\lscal/\vther \sim \wfreq^{-1}$, which are much longer
 than the gyration time scale $\cycf^{-1}$, i.e.
 $\cycf^{-1} \ll \lscal/\vther$. In order to separate the rapid gyration from the
 particle streaming it is convenient to use gyrokinetic variables \cite{cattoPP78,parraPPCF08}
 rather than the particle
position $\rr$ and particle velocity $\pvel$. We will use the guiding centre
 $\gc = \rr - \gyrdvecs$, where $\gyrdvecs = \fract{\bu \xp \pvel}{\cycfs}$
 is the vector gyroradius; the particle energy $\energy =  \ma \vmag^2/2$,
 where $\vmag = |\pvel|$; the sign of the parallel velocity $\sign = \fract{\vpar}{|\vpar|} $, where 
 $\vpar = \bu \cdot \pvel$;  the pitch angle $\pitch = \fract{\vperp^2}{\vmag^2 \bmag}$,
 where $\vperp = |\pvel - \vpar \bu|$; and the gyrophase $\gyrophase$,
 which identifies the angular position of a particle in its gyromotion, and which is defined by
\beqn \tan{\gyrophase}  = \frac{\etwo \cdot \pvel}{\eone \cdot \pvel}, \eeqn
where $\eone$ and $\etwo$ are unit vectors which form an orthonormal basis with $\bu$.
We define $\eone$ and $\etwo$ in terms of $\flxl$ and $\fldl$ in \refsec{sec:velocitygyroaverage}.
In these variables, variation on $\cycf^{-1}$ time scales occurs only through $\gyrophase$,
 over which one can conveniently average. We define the gyroaverage $\gav{\cdot }{}$ as
 \beqn \gav{\cdot }{}= \frac{1}{2\pi} \int^{2\pi}_0 \cdot \; d \gyrophase  . \label{equation:gav}\eeqn
Note that the gyroaverage is taken at fixed
$\energy$, $\sign$, and $\pitch$. In addition,
 either $\rr$ or $\gc$ is held fixed during the gyroaveraging; 
 we will state explicitly whether $\rr$ or $\gc$ is held fixed in each gyroaverage.

To find equations which determine $\dlf$ and $\eqlbf$ we use the turbulent average \refeq{eq:distfnav},
 the assumption of statistical periodicity \refeq{eq:distfnsolve},
 the orderings \refeq{eq:gkordering}, the gyroaverage \refeq{equation:gav},
 and the Fokker-Planck equation for the plasma. We find
 that the equilibrium piece of the distribution function $\eqlbf = \eqlb$, where  
 $\eqlb$ is a maxwellian distribution of velocities.  
 The fluctuating piece of the distribution function $\dlf$ is determined by 
 the gyrokinetic equation, which
 is written in terms of the non-adiabatic response 
\beqn \hh{}(\gc,\energy,\pitch,\sign)=\dlf(\rr,\pvel)+\frac{\zed\charge\ptl{}(\rr)}{\temp}\eqlb. \label{eq:hh} \eeqn
 Note that $\hh{}$ is independent of $\gyrophase$ when $\hh{}$ is regarded as a function of $\gc$.
 The dependence on $\gyrophase$ in $\dlf$ arises due to $\gyrdvec$
 in the transformation between $\rr$ and $\gc$.
 In terms of $\hh{}$, the gyrokinetic equation is
\beqn \drv{\hh{}}{\ttime} + \vpar \kpar \drv{\hh{}}{\lpar} + (\vm + \ve) \cdot \nbl \hh{} + \ve \cdot \nbl \eqlb = \frac{\zed \charge \eqlb}{\temp}\drv{\gptl{}}{\ttime} + \cop, \label{eq:gke}\eeqn
 where $\gptl{}=\gav{\ptl{}}{\gc}$ is the gyroaverage of the fluctuating potential
  at fixed $\gc$,
$\cop$ is the gyroaveraged,
 linearised Fokker-Planck collision operator for the species,
 $\vm=(\vperp^2/2\cycf) \bu \xp \nbl \bmag + (\vpar^2 / \cycf) \bu \xp \bu \cdot \nbl \bu$
 is the magnetic drift velocity,  and
 \beqn \ve=\frac{\ltsp}{\bmag}\bu\xp\nbl\gptl{}\eeqn 
is the gyroaveraged $E\xp B$ velocity. 
To close the gyrokinetic equations for each species one requires an equation
 for the field; this is the quasineutrality relation,
\beqn \sum_{\spe} \zeds \charge \intv{\big|_\rr}  \hh{\spe}(\gc,\energy,\pitch,\sign)
 = \sum_{\spe} \frac{\zeds^2 \charge^2 \denss}{\temps}\ptl{}(\rr) \label{eq:qnlty},\eeqn
where $\denss = \intv \eqlba$ is the equilibrium plasma density. 
Where it does not introduce ambiguity, we will suppress $(\gc,\energy,\pitch,\sign) $ dependences and species indices.
We will take the collisionless limit $\cfreq \ll \wfreq$
in this paper for simplicity; we demonstrate how the effects of collisions
may be included within the scale-separated framework in \refsec{sec:collisions}.
It is important to note that in gyrokinetics we implicitly assume that the typical
 velocity scale of the distribution function for a given species $\dvels\sim\vthers$.
In a truly collisionless system $\dlf$ and $\eqlbf$ could develop
arbitrarily small velocity space structures through phase mixing
 \cite{KampenPhysica1955,CaseAnnPhys1959,krommesPoP94,krommesPoP99,schekPPCF08,schekApJ09,tatsunoPRL09,barnesPoP10a}, i.e. $\dvel \rightarrow 0$ 
as $\cfreq \rightarrow 0 $. 
The presence of the collision operator $\copNL$, a diffusive operator in velocity space,
 provides the necessary, and physical,
 regularisation of $\dlf$ and $\eqlbf$ \cite{abelPoP08,barnesPoP09}.
In the $\dlf$ gyrokinetic approach, equations \refeq{eq:gke} and \refeq{eq:qnlty} are solved
 in a field-following domain,
termed a flux tube.
We use the field-aligned coordinate system $(\flxl,\fldl,\lpar)$, which is theoretically convenient and
 allows for an efficient
 simulation domain which captures the structure of anisotropic, magnetised plasma turbulence with minimal
 resolution. 
 The flux tube has a narrow extent around a central field line 
 but extends for (typically) a $2 \pi$ poloidal circuit along the field line.
 The assumptions of scale separation in the plane perpendicular to $\bvec$,
 and statistical periodicity \refeq{eq:distfnsolve}, permit the use of
 periodic boundary conditions in the $(\flxl,\fldl)$ plane. However, we must also specify a
 boundary condition in the poloidal angle $\lpar$ that gives the location along the magnetic field.
 The relevant physical boundary condition is periodicity in $\lpar$, but this must be applied at fixed toroidal angle $\toragl$ 
 -- not at fixed field line label $\fldl$.  The presence of shear in the pitch of the magnetic field makes enforcement
 of this boundary condition non-trivial. The standard flux tube parallel boundary condition is the
 so-called \enquote{twist-and-shift} boundary condition~\cite{beerPoP95}. 
 
The gyrokinetic system \refeq{eq:gke} and \refeq{eq:qnlty} is the starting point for our model.
Note that the gyrokinetic ordering allows for the possibility that the
 fluctuations have multiple distinct space and time scales arising
 from differences in the thermal speed  $\vthers$ and thermal gyroradius 
 $\gyrds$ of each species.
 Indeed, equilibrium gradients in plasma density and temperature drive instabilities with
 distinct spatial scales of the ion and electron gyroradii,
 with corresponding time scales of $\lscal/\vtheri$ and $\lscal/\vthere$,
 respectively. We suppose the existence
 of an ion scale (\iscaleNS) and a separated electron scale (\escaleNS) within the turbulence. 
The separation between the scales is governed by the mass ratio
 $\massrt \sim   \gyrde / \gyrdi \sim  \vtheri / \vthere $,
 which we treat as an asymptotic expansion parameter; i.e., $\massrt \rightarrow 0$.
 We will assume that fluctuations in the scales
 intermediate to the \iscaleS and \escaleS have vanishing amplitude.  
Note that our expansion in $\massrt$ is a subsidiary expansion within gyrokinetics,
 so it satisfies $\gyrds/\lscal \ll \massrt$.

 We proceed in analogy to the derivation of the coupled equilibrium-fluctuation
 equations to derive scale-separated, coupled \iscaleNS-\escaleS equations. 
These equations contain the physics of nonlocal (in wave number) cross-scale
 interaction. We propose an extension of the \enquote{twist-and-shift} flux tube parallel
 boundary condition to allow for efficient simulation of the \iscaleNS-\escaleS equations
 in a system of coupled flux tubes. We find the \iscaleNS-\escaleS equations
 are parallelisable in the sense that the \escaleS equations may
be integrated at multiple radial locations within the \iscaleS domain without reference to one another.

 \section{Separation of Scales within the Turbulence}\label{sec:separation}
We assume a separation of scales between \iscaleS fluctuations and \escaleS
 fluctuations in the turbulence, 
i.e., $\massrt\rightarrow 0$. We make the assumption that the turbulent
 fluctuations have negligible amplitude at scales intermediate to the \iscaleS and \escaleNS. 
This allows us to decompose the fluctuating distribution function $\dlf$ into
\beqn \dlf = \idlf + \edlf, \eeqn
where $\idlf$ and $\edlf$ are the fluctuating \iscaleS and \escaleS pieces
 of the particle distribution function, respectively.
 In order to find evolution equations for $\idlf$ and $\edlf$
 we introduce an \escaleS average $\esav{\cdot}$,
 which averages over spatial scales and time scales which are
 intermediate to the \iscaleS and \escaleS,
 such that
\beqn \esav{\dlf} = \idlf, \label{eq:esavg} \eeqn 
with the assumption of statistical periodicity
\beqn\esav{\edlf} = 0.  \label{eq:solvability}\eeqn 
This assumption allows us to asymptotically expand the gyrokinetic equation
 and find unique asymptotic series for $\idlf$ and $\edlf$ in the limit $\massrt \rightarrow 0 $,
 and is analogous to the assumption of statistical periodicity \refeq{eq:distfnsolve}.

 One could derive the coupled multi-scale equations with an implicit assumption of statistical periodicity, as e.g., in \cite{abelRPP13}.
 However, here we explicitly show how \refeq{eq:solvability}
 is satisfied by introducing fast and slow variables for space and time
 after the method of multiple scales (cf.~\cite{BenderOrszag}) and then homogenising the system.
 This approach clarifies how to deal with the nonlocal nature of gyrokinetics introduced by the gyroaverage.
  
    In analogy to $\delta f$ gyrokinetics, we adopt the following orderings for space and time scales,
    \beqn \nabla_{\parallel}\idlf \sim \frac{\idlf}{a}, \quad \nblper \idlf \sim \frac{\idlf}{\gyrdi}, \quad \drv{\idlf}{\ttime} \sim \frac{\vtheri}{\lscal} \idlf, \nonumber \eeqn
    \beqn \nabla_{\parallel}\edlf \sim \frac{\edlf}{a}, \quad \nblper \edlf \sim \frac{\edlf}{\gyrde}, \quad \drv{\edlf}{\ttime} \sim \frac{\vthere}{\lscal} \edlf. \label{order:sptm} \eeqn
    Note that we assume that the parallel scale is always set by the machine size,
and hence we do not assume separation of scales in the parallel direction. 
These assumptions will be justified a posteriori
using a critical balance argument.
Our ordering \refeq{order:sptm} assumes spatial isotropy in the turbulence in the plane
 perpendicular to the magnetic field line; this assumption excludes structures which
 have scales of size $\gyrdi$ in one direction perpendicular to the magnetic field,
 but scales of size $\gyrde$ in another. In particular, this assumption excludes electron
 temperature gradient (ETG) streamers \cite{dorland2000electron,jenko2000electron,jenko2002prediction}
 if their radial extent scales like $\gyrdi$ rather than $\gyrde$. 
Using the orderings \refeq{order:sptm} we proceed to find coupled \iscaleNS -\escaleS equations
including all terms which might be relevant to leading order.
We then use dominant balance arguments to find orderings
 for the size of the fluctuations consistent with \refeq{order:sptm},
 and neglect the terms in the coupled  \iscaleNS -\escaleS equations which are small.

    We introduce a fast spatial variable $\rf$ and a slow spatial variable $\rs$ in the 2-D plane 
    perpendicular to the field line, such that all \escaleS variation in the solution appears
    through dependence on $\rf$, and all \iscaleS variation appears 
    through dependence on $\rs$. 
    Functions of $\rr$ will become functions of $(\lpar,\rs,\rf)$, where the perpendicular
    and parallel spatial dependence is explicitly written out. We will also introduce
    fast and slow guiding centre coordinates, $\gcf=\rf - \gyrdvecs$ and
    $\gcs =\rs - \gyrdvecs$, respectively.
    We introduce the fast and slow times $\tf $
    and $\ts$, respectively. Functions of $\ttime$ will become functions of $(\ts,\tf)$.
    Note that the slow variables do not 
    contain equilibrium variation, as this is ordered out in deriving the gyrokinetic equation.
    We will assume periodicity of the fluctuations in $\rs$ to be consistent with the separation
    of scales between the fluctuations and the equilibrium assumed in $\delta f$ gyrokinetics via
    condition \refeq{eq:distfnsolve}. The gyrokinetic equation must be modified so that, 
    \beqn \dlf ( \ttime , \rr ) \rightarrow \dlf ( \ts, \tf, \lpar,\rs, \rf), \eeqn 
    \beqn \nblper \rightarrow \nbls + \nblf, \quad \nbls \sim \gyrdi^{-1}, \quad \nblf \sim \gyrde^{-1}, \eeqn
    \beqn \drv{}{\ttime} \rightarrow \drv{}{\ts} + \drv{}{\tf}, \quad \drv{}{\ts} \sim  \frac{\vtheri}{\lscal} , \quad \drv{}{\tf} \sim  \frac{\vthere}{\lscal}. \eeqn 
    We then perform an asymptotic expansion on the resulting equations in the mass ratio $\massrt $
    to find the leading order, asymptotically valid equations. With the introduction of the new variables
    we may explicitly define the \escaleS average,
    
    \beqn \fl \idlf (\ts, \lpar, \rs ) = \esav{\dlf( \ts, \tf, \lpar, \rs, \rf) } = \nonumber \eeqn 
    \beqn     \frac{1}{\tcut \esarea} \int^{\ts + \tcut/2}_{\ts - \tcut/2} d \tf \int^{\flxls + \xcut/2}_{\flxls - \xcut/2} d \flxlf \int^{\fldls + \ycut/2}_{\fldls - \ycut/2} d \fldlf \; \jacob \; \dlf( \ts, \tf, \lpar, \rs, \rf),\label{eq:defesav} \eeqn 
   where $\tcut$ is a time intermediate to the \escaleS and \iscaleS correlation times; $\xcut$ and $\ycut$ are intermediate to the
    \escaleS and \iscaleS correlation lengths and angles in the $\flxl$ and $\fldl$ directions, respectively; $\flxlf$ and  $\flxls$
    are the fast and slow flux labels, respectively; $\fldlf$ and  $\fldls$ are the fast and slow field line labels,
    respectively; $\jacob $ is the Jacobian of the transformation from $\rf \rightarrow (\flxlf, \fldlf)$, and
    \beqn    \esarea =  \int^{\flxls + \xcut/2}_{\flxls - \xcut/2} d \flxlf \int^{\fldls + \ycut/2}_{\fldls - \ycut/2} d \fldlf \; \jacob. \label{eq:defesav2} \eeqn
    Note that we regard $\rf=\rf(\flxlf,\fldlf)$ and $\rs = \rs(\flxls,\fldls)$ in the integration in \refeq{eq:defesav} and \refeq{eq:defesav2}.
    To satisfy \refeq{eq:solvability}
    for every $(\ts,\lpar,\rs)$, we impose that the fluctuations are periodic in the fast variable $\rf$. 
    In particular we assume  that
    \beqn \dlf( \ts, \tf, \lpar,\rs, \rf) = \dlf( \ts, \tf,\lpar, \rs, \rf + j\xcut \evflxl + l\ycut \evfldl),\eeqn
     where $j$, $l$ are any integers,  $\evflxl = (\bu \xp \nbl \fldl)/|\nbl \flxl \cdot \bu \xp \nbl \fldl|$,
     and $ \evfldl = (\nbl \flxl \xp \bu )/|\nbl \flxl \cdot \bu \xp \nbl \fldl|$.

    The \escaleS average \refeq{eq:defesav} is an average over areas perpendicular
    to the magnetic field line,
    and times intermediate to $\lscal/\vtheri$ and  $\lscal/\vthere$, with the average taken
    at fixed $\lpar$, $\energy $, $\pitch $, and $\sign $. To find the scale-separated, leading order
    equation for electrons at \iscaleS it is also necessary to average the \iscaleS electron gyrokinetic
    equation over electron orbits parallel to the magnetic field line, cf.\cite{abelEMOD}.
    This necessity arises because electron parallel streaming,
    which introduces frequencies of order $\vthere/\lscal$, is faster
than any other \iscaleS dynamics, which by assumption have 
frequencies of order $\vtheri/\lscal \ll \vthere/\lscal $.
Naively this would appear to break our ordering,
 as it would seem we would need to include \escaleS timescales in the \iscaleS equation.
 Furthermore, it is known that rapid electron streaming can lead to very long tails
 in ballooning modes due to the passing electron response \cite{HallatschekgiantelPRL2005}.
 The long ballooning tails result in very fine radial structure in the turbulence
 near low order mode rational surfaces \cite{DominskinonadPOP015}. 
 This problem is resolved by treating electron parallel streaming
 at \iscaleS as being asymptotically faster than \iscaleS frequencies, consistent with our expansion.
 To find the leading order equation for the electron distribution function at \iscaleS
 we introduce the orbital average $\orbav{\cdot}$.
 After \cite{HazeltineMeiss} we define the orbital average $\orbav{\cdot}$
in the passing and trapped parts of velocity space separately. 
In the passing region of velocity space
\beqn \orbav{\cdot}= \frac{ \int d \fldls d \lpar \; \fract{\cdot}{\vpar \kpar}  }{\int d \fldls d \lpar  \fract{}{\vpar \kpar} }  , \label{eq:orbpassing}\eeqn
where the integration $\int d \fldls d \lpar$ is taken over the whole flux surface.
In the trapped region of velocity space
\beqn \orbav{\cdot}= \frac{ \sum_\sign \int^{\lparp}_{\lparm}   d \lpar \; \fract{\cdot}{|\vpar| \kpar} }{2 \int^{\lparp}_{\lparm} d \lpar  \fract{}{|\vpar| \kpar} }  , \label{eq:orbtrapped}\eeqn
where the limits in the integration $\lparp$ and $\lparm$ 
are the upper and lower bounce points of the trapped particle, respectively.
Note that the orbital average commutes with
fast and slow spatial derivatives, 
\beqn \nbls \orbav{\cdot} = \orbav{\nbls \cdot}  \label{eq:orbcom1}\eeqn 
and
\beqn \nblf \orbav{\cdot} = \orbav{\nblf \cdot}\label{eq:orbcom2}\eeqn
as $\vpar \kpar$ contains only equilibrium spatial variation.
With the definitions \refeq{eq:orbpassing} and
 \refeq{eq:orbtrapped} we show in \refsec{sec:annihilate} that
 the orbital average $\orbav{\cdot}$ has the very useful property
\beqn \orbav{ \vpar \kpar \drv{\func}{\lpar}}=0 \label{eq:orbprop}\eeqn
for any $\func(\lpar,\gcs,\energy,\pitch,\sign)$.
 The property \refeq{eq:orbprop} will allow us to use the orbital average
$\orbav{\cdot}$ to eliminate
 parallel streaming from the electron equation at \iscaleNS.

\section{Coupled Gyrokinetic Equations}\label{sec:coupledgke}
In this Section we will use the ideas of Section \refsec{sec:separation} to find the scale-separated
 \iscaleS and \escaleS gyrokinetic equations, where the largest possible cross-scale
 terms are retained. We take this approach, as it will allow us to consider whether
 or not the presence of cross-scale interaction can lead to novel mass ratio orderings for the
 sizes of the fluctuations.
 
 We now apply the \escaleS average \refeq{eq:defesav} to the gyrokinetic
 equation \refeq{eq:gke} to find the \iscaleS gyrokinetic equation.
 The resulting gyrokinetic equation is a function of $\gcs$. 
 We suppress species indices, but note that what is done here applies
 to both ion and electron species. Upon averaging, the gyrokinetic equation is,
\beqn \fl \drv{\ihh{}}{\ts} + \vpar \kpar \drv{\ihhi{}}{\lpar} + \vm \cdot \nbls \ihh{} + \esav{\ve \cdot \nbl \hh{} } + \ive \cdot \nbl \eqlb = \frac{\zed \charge \eqlb}{\temp}\drv{\igptl{}}{\ts}, \label{eq:gyroeqav}\eeqn
where \beqn \ive = \frac{\ltsp}{\bmag}\bu\xp\nbls \igptl{}, \label{eq:igavexbdrift}\eeqn
and \beqn \igptl{}=\gav{\iptl{}}{\gcs}=\gav{\esav{\ptl{}}}{\gcs}. \label{eq:ispotential}\eeqn 
Note that whilst the gyroaveraged potential $\gptl{} $ is a function of
 $\gcf$, the potential $\ptl{}$ is a function of $\rf$.
To obtain \refeq{eq:gyroeqav}-\refeq{eq:ispotential},
 we used the fact that the the \escaleS average can be taken over
 either the real space variable $\rf$ or the guiding centre $\gcf$,
 and that the \escaleS average commutes with the gyroaverage,
 \beqn\gav{\esav{\cdot}}{\gcf,\gcs} = \esav{\gav{\cdot}{\gcf,\gcs}}, \label{eq:esavtextprop} \eeqn
where the notation $\gav{\cdot}{\gcf,\gcs}$ indicates a gyroaverage
 with both $\gcf$ and $\gcs$ held fixed.
  Both of these properties are proven in \refsec{sec:gavproperties}.
The linear terms in \refeq{eq:gyroeqav} are simply filtered versions
 of the terms in the $\dlf$ gyrokinetic equation~\refeq{eq:gke}.
 However, the nonlinear term yields a new, cross-scale coupling term.
  As shown in \refsec{section:gyroalgebraion}, the form of this new term is
\beqn \esav{\ve \cdot \nbl \hh{} } = \ive \cdot \nbls \ihh{} + \nbls \cdot \esav{ \ehh{}\frac{\ltsp}{\bmag}\bu\xp(\nbls + \nblf) \egptl{}  }, \nonumber \eeqn
where  $\ive \cdot \nbls \ihh{}$ is the usual $E\xp B$ nonlinearity
 appearing in the \iscaleS gyrokinetic equation in the absence of cross-scale coupling.   
Dropping the $\nbls \egptl{}$ term since it is a factor $\massrt$ 
smaller than the $\nblf \egptl{}$ term, we find 
\beqn \esav{\ve \cdot \nbl \hh{} } = \ive \cdot \nbls \ihh{} + \nbls \cdot \esav{ \eve \ehh{} }, \eeqn
 where
\beqn \eve = \frac{\ltsp}{\bmag}\bu\xp\nblf \egptl{}. \label{eq:egavexbdrift} \eeqn
After filtering with the \escaleS average, the \iscaleS equations for ions and electrons
 in the presence of cross-scale coupling are therefore
\beqn \fl \drv{\ihhi{}}{\ts} + \vpar \kpar \drv{\ihhi{}}{\lpar} + \vmi \cdot \nbls \ihh{} +  \ivei \cdot \nbls \ihhi{} \nonumber \eeqn 
 \beqn + \nbls \cdot \esav{ \evei \ehhi{} }
 + \ivei \cdot \nbl \eqlbi = \frac{\zedi \charge \eqlbi}{\temp}\drv{\igptli{}}{\ts}, \label{eq:ionscleq1ion} \eeqn
and
 \beqn \fl \drv{\ihhe{}}{\ts} + \vpar \kpar \drv{\ihhe{}}{\lpar} + \vme \cdot \nbls \ihhe{} +  \ivee \cdot \nbls \ihhe{} \nonumber\eeqn 
 \beqn + \nbls \cdot \esav{ \evee \ehhe{} }
 + \ivee \cdot \nbl \eqlbe = -\frac{ \charge \eqlbe}{\temp}\drv{\igptle{}}{\ts}. \label{eq:ionscleq1el} \eeqn
The only modifications to the usual \iscaleS gyrokinetic equations
 are the inclusion of $\nbls \cdot \esav{ \evei \ehhi{} }$ in \refeq{eq:ionscleq1ion},
 and $\nbls \cdot \esav{ \evee \ehhe{} }$ in \refeq{eq:ionscleq1el}.
 The terms $\nbls \cdot \esav{ \evei \ehhi{} }$ and $\nbls \cdot \esav{ \evee \ehhe{} }$
 physically represent the divergence of the spatial fluxes of $\ihhi{}$ and $\ihhe{}$
 due to \escaleS fluctuations. 
 The reader will notice that \refeq{eq:ionscleq1el} contains electron parallel streaming
 $\vpar \kpar \drvt{\ihhe{}}{\lpar}$. As a consequence, equation \refeq{eq:ionscleq1el} is not properly
 scale-separated, as the equation still contains $\lscal/\vthere$ time scales and $\gyrde$ spatial scales.
In our asymptotic expansion the leading order equation for electrons at \iscaleS is
 \beqn \vpar \kpar \drv{\ihhe{}}{\lpar} =0. \label{eq:evpar} \eeqn To solve this equation we decompose
 \beqn \ihhe {} =  \ihhe{}^{(0)}(\gcs,\energy,\pitch,\sigma) + \ihhe{}^{(1)}(\lpar,\gcs,\energy,\pitch,\sigma), \label{eq:electrondecomp2}\eeqn where 
 $\ihhe{}^{(0)} \sim \fract{ e\iptl{}}{\temp}$ and  $\ihhe{}^{(1)} \sim \massrt \ihhe{}^{(0)}$.
 Note $\ihhe{}^{(0)}$ has no dependence on $\lpar$. 
 We see that if we explicitly make the decomposition \refeq{eq:electrondecomp2}
 when solving \refeq{eq:ionscleq1el} then we can formally order
 \beqn  \vpar \kpar \drv{\ihhe{}}{\lpar} =\vpar \kpar \drv{\ihhe{}^{(1)}}{\lpar}
 \sim \frac{\vtheri}{\lscal} \ihhe{}^{(0)} \sim \vme \cdot \nbls \ihhe{}. \eeqn
  We take only the leading order piece of the electron
distribution function $\ihhe{} = \ihhe{}^{(0)}$. 
 Defining the non-zonal piece of the distribution function
 to be the piece which contains variation in $\fldls$,
we show in \refsec{h0proof} that $\ihhe{}^{(0)}= 0$ for
 the non-zonal passing piece of the electron
 distribution function. 
 This is a result of the usual 
 \enquote{twist-and-shift} boundary condition \cite{beerPoP95} which
 connects modes of different radial wavenumber,
 with the observation that modes which have very
 high radial wavenumber should have $\hh{} \rightarrow 0$.
 To obtain the equation for the non-zero piece of $\ihhe{}^{(0)}$  
 we need to eliminate the parallel streaming term $\vpar \kpar \drvt{\ihhe{}^{(1)}}{\lpar}$ from
\refeq{eq:ionscleq1el}.
We achieve this by averaging over the rapid parallel orbits
of the electrons in \refeq{eq:ionscleq1el}. 
We use the orbital average, defined in \refeq{eq:orbpassing} and \refeq{eq:orbtrapped},
 and the properties \refeq{eq:orbprop},
\beqn \orbav{\ihhe{}^{(0)}} = \ihhe{}^{(0)} \label{eq:orbprop2} \eeqn 
 and \cite{hintonRMP76,helander} \beqn \orbav{\vm \cdot \nbl \flxl} = 0, \label{eq:orbprop3}\eeqn
 shown in \refsec{sec:orbpropmore}.
 The resulting, scale-separated, equation
 for the leading order piece of the electron distribution function
  at \iscaleNS, $\ihhe{} =\ihhe{}^{(0)}$, is 
 \beqn \fl \drv{\ihhe{}}{\ts} + \orbav{\vme \cdot \nbl \fldl} \drv{\ihhe{}}{\fldls} 
+ \orbav{\ivee \cdot \nbls \ihhe{}} + \orbav{\ivee \cdot \nbl \eqlbe } \nonumber \eeqn  \beqn 
  +  \nbls \cdot \orbav{\esav{\evee \ehhe{} }} 
  = -\frac{ \charge \eqlbe}{\tempe}\drv{\orbav{\igptle{}}}{\ts},\label{eq:ionscleq1el2}\eeqn
where it is understood that $\ihhe{} =0$
 for the non-zonal passing piece of the electron distribution function.
 
To find the \escaleS equation we subtract the
 \iscaleS equation~\refeq{eq:gyroeqav} from the full equation \refeq{eq:gke}.
 Again, the linear terms follow easily and  the nonlinear term provides new cross-scale coupling terms:
\beqn \fl \drv{\ehh{}}{\ts} + \drv{\ehh{}}{\tf}+ \vpar \kpar \drv{\ehh{}}{\lpar} + \vm \cdot (\nbls + \nblf) \ehh{} +\left[ \ve \cdot \nbl \hh{} - \esav{\ve \cdot \nbl \hh{} } \right] \nonumber \eeqn 
\beqn  +\left(\frac{\ltsp}{\bmag}\bu\xp \left(\nbls + \nblf \right) \egptl{} \right) \cdot \nbl \eqlb = \frac{\zed \charge \eqlb}{\temp}\left(\drv{\egptl{}}{\ts} + \drv{\egptl{}}{\tf}\right). \label{eq:esproto} \eeqn
We can further simplify~\refeq{eq:esproto} by neglecting sub-dominant terms in $\massrt$; i.e., 
$\nbls+\nblf \approx \nblf$,
 $\partial/\partial t_s + \partial/\partial t_f \approx \partial /\partial t_f$,
 and $\vpar \kpar \drvt{\ehhi{}}{\lpar} \ll \vm \cdot \nblf \ehhi{}$ in the equation for ions.
  As shown in \refsec{section:gyroalgebraelectron}, the nonlinear term reduces to 
\beqn \ve \cdot \nbl \hh{}- \esav{\ve \cdot \nbl \hh{} } = \eve \cdot \nblf \ehh{} + \ive \cdot \nblf \ehh{} + \eve \cdot \nbls \ihh{}, \eeqn 
where $\eve \cdot \nblf \ehh{}$ is the usual $E\xp B$ nonlinearity appearing in the \escaleS gyrokinetic equation in the absence of cross-scale coupling. 
Combining these results, the \escaleS equation \refeq{eq:esproto}, for electrons, becomes  
\beqn \fl \drv{\ehhe{}}{\tf} + \vpar \kpar \drv{\ehhe{}}{\lpar} + (\vme + \ivee ) \cdot \nblf \ehhe{} \nonumber
 \eeqn \beqn + \evee \cdot \nblf \ehhe{} + \evee \cdot (\nbl \eqlbe + \nbls \ihhe{}) =- \frac{\charge \eqlbe}{\temp}\drv{\egptle{}}{\tf}. \label{eq:esel} \eeqn
For ions, equation \refeq{eq:esproto} becomes
\beqn \fl \drv{\ehhi{}}{\tf} + (\vmi + \ivei ) \cdot \nblf \ehhi{} + \evei \cdot \nblf \ehhi{} + \evei \cdot (\nbl \eqlbi + \nbls \ihhi{}) = \frac{\zedi \charge \eqlbi}{\temp}\drv{\egptli{}}{\tf}. \label{eq:esion} \eeqn
Equations \refeq{eq:esel} and \refeq{eq:esion} contain 
the effect of \iscaleS gradients on \escaleS fluctuations through two terms: 
 $\ive \cdot \nblf \ehh{}$,
 which represents advection of \escaleS fluctuations by \iscaleS eddies;
 and $\eve \cdot  \nbls \ihh{}$, which represents the modification of the equilibrium gradient drive
due to gradients in the \iscaleS fluctuations.
 Note that the term $\ive \cdot \nblf \ehh{}$ cannot be eliminated by a 
change of reference frame because the advection velocity $\ive$
 is a function of the parallel-to-the-field coordinate $\lpar $,
 as $\iptl{}$ varies along the magnetic field line.
 The \escaleS equations \refeq{eq:esel} and \refeq{eq:esion} are scale-separated
 in the sense that $\gcs$
 appears only as a label of the \iscaleS gradients
 $\nbls \igptl{} (\gcs)$ and $\nbls \ihh{} (\gcs)$, and 
as a label on the fields $\egptl{}(\gcs, \gcf)$ and $\ehh{}(\gcs, \gcf)$. 

\section{Quasineutrality}\label{sec:qn}

Using~\refeq{eq:esavg} and \refeq{eq:esavtextprop},
 and noting that the integration variable for the \escaleS average can
 be either $\rf$, or $\gcf$ (see \refsec{sec:gavproperties}),
 one can directly find the \iscaleS quasineutrality relation,
\beqn \sum_{\spe} \zeds \charge \intv{|_\rs} \ihh{\spe}(\lpar,\gcs,\energy,\pitch,\sign)
  = \sum_{\spe} \frac{\zeds^2 \charge^2 \denss}{\temps}\iptl{}(\lpar,\rs), \label{eq:iscalqn}\eeqn
which is supplemented by the relation, \beqn \igptl{} = \gav{\iptl{}}{\gcs}. \label{eq:isptlrel}\eeqn
We find the quasineutrality relation for the \escaleS by subtracting \refeq{eq:iscalqn} 
from the full quasineutrality relation \refeq{eq:qnlty} to obtain,
\beqn \sum_{\spe} \zeds \charge \intv{|_{\rs,\rf}} \ehh{\spe}(\lpar,\gcs,\gcf,\energy,\pitch,\sign) 
 = \sum_{\spe} \frac{\zeds^2 \charge^2 \denss}{\temps}\eptl{}(\lpar,\rs,\rf), \label{eq:escalqn}\eeqn
 which is supplemented by the relation, \beqn  \egptl{} = \gav{\eptl{}}{\gcf,\gcs} .\label{eq:esptlrel} \eeqn 
 The system of equations \refeq{eq:ionscleq1ion},
 \refeq{eq:ionscleq1el2}, and \refeq{eq:esel}-\refeq{eq:esptlrel} constitutes a formally
 closed system of equations for the \iscaleS and \escaleS quantities.
 To calculate the ion contribution to the
 \escaleS quasineutrality relation \refeq{eq:escalqn} requires that we take
 a gyroaverage over a scale of the ion gyroradius,
 as does computing the ion gyroaveraged potential $\egptli{}$ in equation \refeq{eq:esptlrel}.
 Naively one might think that the ion species would thus
 introduce ion gyroradius scale correlations in the \escaleS
turbulence. However, by considering the possible dominant balances
 in the \iscaleNS -\escaleS system we will find
 the contributions from ions at \escaleS can always be ignored,
 and so we will arrive at a fully  scale-separated system. 

\section{Sizes of the \iscaleS and \escaleS fluctuations and scale-separation}\label{sec:size}

We now proceed to find all possible self-consistent orderings for $\ihhi{}$, $\ihhe{}$, $\ehhi{}$,
 $\ehhe{}$, $\iptl{}$, and $\eptl{}$ that lead to steady-state solutions to the
 system of equations \refeq{eq:ionscleq1ion},
 \refeq{eq:ionscleq1el2}, and \refeq{eq:esel}-\refeq{eq:esptlrel}. 
  For a statistically steady-state turbulence to be possible
there must be a competition between the growth of linear instabilities
 and nonlinear interactions, which can be cross-scale in nature.
 We look for dominant balances consistent with this observation.

 Note that the presence of the gyroaverages
can introduce mass ratio factors.
Recalling that we have periodicity perpendicular to the field line in both the slow and fast variables,  we write a fluctuation $\fluc{}(\rs,\rf)$ as,
\beqn \fl \fluc{}( \rs, \rf) = \sum_{\kf} \fluc{\kf}(\rs) \expo{\imag \kf \cdot \rf}= \sum_{\kf,\ks} \fluc{\kf,\ks} \expo{\imag \kf \cdot \rf} \expo{\imag \ks \cdot \rs}, \label{eq:fouriernote}\eeqn
where $\kf\sim\gyrde^{-1}$, $\ks\sim\gyrdi^{-1}$, and we have used periodicity in $\rs$ to write $\fluc{\kf}(\rs)=\sum_{\ks}\fluc{\kf,\ks}\expo{\imag\ks\cdot\rs}$. 
If we gyroaverage \refeq{eq:fouriernote},
 recalling $\gcf=\rf -\gyrdvec$ and $\gcs=\rs -\gyrdvec$, we find
\beqn \fl \gav{\fluc{}( \rs, \rf) }{\gcf,\gcs} 
 = \sum_{\kf,\ks} \fluc{\kf,\ks} \expo{\imag \kf \cdot \gcf} \expo{\imag \ks \cdot \gcs} \gav{\expo{\imag (\kf+ \ks) \cdot \gyrdvec} }{} \nonumber \eeqn
\beqn =\sum_{\kf,\ks} \fluc{\kf,\ks} \expo{\imag \kf \cdot \gcf} \expo{\imag \ks \cdot \gcs}\bes{|\kf+\ks||\gyrdvec|},\label{eq:gyroavfourier}\eeqn
where $\bes{z}$ is the 0th Bessel function of the 1st kind,
and we used the result
\beqn \gav{\expo{\imag \kk \cdot \gyrdvec} }{} = \bes{|\kk||\gyrdvec|}, \label{eq:gavresult}\eeqn
shown in \refsec{sec:velocitygyroaverage}. Note that $\bes{z}\sim 1$ for $z\lesssim 1$ 
and $\bes{z}\sim z^{-1/2}$ for $z\gg1$.  Because $|\kf|\gyrde \sim |\ks|\gyrdi \sim 1$ and $|\ks|\gyrde \sim \massrt \ll 1$,
 the gyroaverage introduces no additional mass ratio factors for \iscaleS quantities 
 or for electrons at electron scales.  However, because $|\kf|\gyrdi \sim \massrut \gg 1$,
 the gyroaveraging operation does reduce the ion fluctuation amplitudes at electron scales; i.e.,
\beqn\bes{|\kf+\ks||\gyrdveci|} \sim \bes{|\kf| \gyrdi} \sim \massrs. \label{eq:jiesfac} \eeqn

With these scalings for the Bessel functions, we show in \refsec{sec:eqscalings} that
 the only possibility for achieving a saturated dominant balance 
 in the equations \refeq{eq:ionscleq1ion},
 \refeq{eq:ionscleq1el2}, and \refeq{eq:esel}-\refeq{eq:esptlrel} is for the fluctuations
 to obey the gyro-Bohm ordering
\beqn \frac{e\iptl{}}{\temp} \sim \frac{\gyrdi}{\lscal}, \quad \frac{e\eptl{}}{\temp} \sim \frac{\gyrde}{\lscal}, \nonumber \eeqn 
    \beqn  \frac{\ihhi{}}{\eqlbi} \sim \frac{\ihhe{}}{\eqlbe} \sim \frac{e\igptli{}}{\temp}  \sim \frac{e\igptle{}}{\temp} \sim \frac{e\iptl{}}{\temp}, \label{order:gbohm} \eeqn 
    \beqn \frac{\ehhe{}}{\eqlbe} \sim \frac{e\egptle{}}{\temp} \sim \frac{e\eptl{}}{\temp}, \nonumber \eeqn
    \beqn \frac{\ehhi{}}{\eqlbi} \sim  \frac{e\egptli{}}{\temp} \sim \massrs\frac{e\eptl{}}{\temp}.  \nonumber \eeqn  
 Note that \refeq{order:gbohm} excludes orderings 
 in which either the \iscaleS or \escaleS fluctuations
 have amplitudes which are greater or smaller than  gyro-Bohm levels
 by a factor which scales with mass ratio.
 In \refsec{sec:eqscalings} we consider the possible ways in which cross-scale interaction
 might have allowed for novel mass ratio orderings
 which deviate from the gyro-Bohm scaling \refeq{order:gbohm}.
We looked for balances where the \escaleS turbulence is enhanced,
 with $\fract{e\eptl{}}{\temp} \gg \fract{\gyrde}{\lscal}$ (see \refsec{sec:electronscaleenhanced});
 the \iscaleS turbulence is enhanced,
 with $\fract{e\iptl{}}{\temp} \gg \fract{\gyrdi}{\lscal}$ (see \refsec{sec:ionscaleenhanced});
 the \iscaleS turbulence suppresses the \escaleS turbulence,
 i.e. $\fract{e\eptl{}}{\temp} \ll \fract{\gyrde}{\lscal}$
 and $ \fract{e\iptl{}}{\temp} \sim \fract{\gyrdi}{\lscal}$ (see \refsec{sec:electronscalesuppressed});
and the \escaleS turbulence suppresses the \iscaleS turbulence,
 i.e. $\fract{e\eptl{}}{\temp} \sim \fract{\gyrde}{\lscal}$
 and $ \fract{e\iptl{}}{\temp} \ll \fract{\gyrdi}{\lscal}$ (see \refsec{sec:ionscalesuppressed}).
 Of these possible impacts of cross-scale interactions,
 only the suppression of \escaleS turbulence by \iscaleS turbulence
 is self consistently allowed by dominant balance.
 Hence, the ordering \refeq{order:gbohm}
 is the only ordering which gives a nonlinearly saturated steady-state turbulence.
 We discuss the physical meaning of ordering \refeq{order:gbohm}
in Section \refsec{sec:sclsepcoupledeq}.

 We now consider the relative sizes of the cross-scale terms appearing in \refeq{eq:ionscleq1ion},
 \refeq{eq:ionscleq1el2}, \refeq{eq:esel}, and \refeq{eq:esion} in the ordering
 \refeq{order:gbohm}, and demonstrate that ions at \escaleS can be neglected.
 Finally, we consider critical balance arguments
to show that the orderings \refeq{order:gbohm} for the sizes of fluctuations are consistent
with our orderings for the parallel length scales \refeq{order:sptm}.
We use these considerations in Section \refsec{sec:sclsepcoupledeq},
 where we give the fully scale-separated
\iscaleS and \escaleS equations, with quasineutrality evaluated consistently with scale-separation,
 and all small terms neglected. 
 
 \subsection{Influence of \escaleS fluctuations on ions at \iscaleNS}\label{sec:iIS}
    Equation \refeq{eq:ionscleq1ion} contains 
    a term which is a divergence of a turbulent flux driven at the \escaleNS.
    This cross-scale term is of size
    \beqn \nbls \cdot \esav{ \evei \ehhi{} } \sim  \frac{\ltsp}{\bmag}\frac{\ehhi{}}{\gyrdi \gyrde} \frac{\charge\egptli{} }{\temp}
    \sim  \frac{\ltsp}{\bmag}\frac{\eptl{}}{\gyrdi^2} \frac{\charge\eptl{} }{\temp} \eqlbi,  \eeqn
    and so in the gyro-Bohm ordering~\refeq{order:gbohm}, 
    \beqn \nbls \cdot \esav{ \evei \ehhi{} } 
    \sim \massrl \frac{\ltsp}{\bmag}\frac{\iptl{}}{\gyrdi^2} \frac{\charge\iptl{} }{\temp} \eqlbi
    \sim \massrl \ivei \cdot \nbls \ihhi{}.\eeqn
    Therefore, the cross-scale term is small compared to the single scale nonlinear term
    $\ivei \cdot \nbls \ihhi{}$, which provides the \iscaleS saturation mechanism in the ordering
    \refeq{order:gbohm}.
   We conclude that we can neglect 
    $\nbls \cdot \esav{ \evei \ehhi{} }$ in the equation for ions at \iscaleS \refeq{eq:ionscleq1ion}.

 \subsection{Influence of \escaleS fluctuations on electrons at \iscaleNS}\label{sec:eISterm}
    The cross-scale term in \refeq{eq:ionscleq1el} is of size
 \beqn \nbls \cdot \orbav{\esav{ \evee \ehhe{} }} \sim  \massru\frac{\ltsp}{\bmag}\frac{\eptl{}}{\gyrdi^2} \frac{\charge\eptl{} }{\temp} \eqlbe, \eeqn 
    and so in the gyro-Bohm ordering~\refeq{order:gbohm}, 
    \beqn \nbls \cdot \orbav{\esav{ \evee \ehhe{} }}
    \sim \massr \frac{\ltsp}{\bmag}\frac{\iptl{}}{\gyrdi^2} \frac{\charge\iptl{} }{\temp} \eqlbe
    \sim \massr \ivee \cdot \nbls \ihhe{}.\eeqn
    Hence in the gyro-Bohm ordering~\refeq{order:gbohm}, we can neglect the electron cross-scale term
    $\nbls \cdot \orbav{\esav{ \evee \ehhe{} }} $ to leading order in
    the \iscaleS electron equation \refeq{eq:ionscleq1el}. 
    
  \subsection{Influence of \iscaleS gradients on \escaleS fluctuations}\label{sec:ISgradES}
    The cross-scale terms appearing in the \escaleS gyrokinetic equation have sizes
    \beqn  \ive \cdot \nblf \ehh{} \sim \frac{\ltsp}{\bmag}\frac{\charge \iptl{}}{\temp} \frac{\ehh{}}{\gyrdi \gyrde}, \eeqn
    and 
    \beqn \eve \cdot \nbls \ihh{} \sim \frac{\ltsp}{\bmag}\frac{\charge \eptl{}}{\temp} \frac{\ihh{}}{\gyrdi \gyrde}. \eeqn 
    Hence, in the gyro-Bohm ordering~\refeq{order:gbohm},
    when $\fract{e\iptl{}}{\temp} \sim \fract{\gyrdi}{\lscal}$
    and $\fract{e\eptl{}}{\temp} \sim \fract{\gyrde}{\lscal}$, we find that
     \beqn  \ive \cdot \nblf \ehh{} \sim   \eve \cdot \nbls \ihh{} \sim \eve \cdot \nbl \eqlb. \eeqn 
     The cross-scale terms modify the \escaleS dynamics at leading order. Therefore, these terms can provide the mechanism for enhancement or suppression
     of the \escaleS turbulence in the presence of \iscaleS fluctuations. 
    The \iscaleS gradients can linearly stabilise the \escaleS instability, but the enhancement of
    the \escaleS fluctuation amplitude cannot scale with any power of the mass ratio. 
\subsection{Ions at \escaleNS}\label{sec:iES}  
The contribution of ions to the \escaleS electrostatic potential is small in $\massrt$.
This is seen by comparing the relative sizes of ion and electron
contributions to the \escaleS quasineutrality relation. Observe that
 \beqn \fl \zedi \charge \intv{|_{\rs,\rf}} \ehhi{}(\lpar,\gcs,\gcf,\energy,\pitch,\sign) \sim \massrs \zedi\ehhi{} \sim  \massr \frac{\charge\eptl{} }{\temp}\zedi\densi
  \nonumber \eeqn \beqn  \ll -\charge \intv{|_{\rs,\rf}} \ehhe{}(\lpar,\gcs,\gcf,\energy,\pitch,\sign) \sim \frac{\charge\eptl{} }{\temp}\dense, \label{order:escalqn}\eeqn
 where one factor of $\massrst$ appears because of the velocity integration for ions,
 which introduces a gyroaverage, and the second appears because of the scaling~\refeq{order:gbohm} of
 $\ehhi{}$ with ${\charge\eptl{} /\temp}$. We conclude that the non-adiabatic response of 
 ions at \escaleS
 does not contribute to the \escaleS potential to leading order.
Physically we are able to neglect the ions at \escaleS because the ion
 gyroradius is much larger than the \escaleS domain. 
 The ions rapidly gyrate at the ion cyclotron frequency $\cycfi$,
 which is much larger than the turbulent frequencies $\wfreq$, i.e. $\cycfi \gg \wfreq$,
 and so they rapidly sample many uncorrelated instances of \escaleS
 turbulence because of their larger gyroradius.
 The ions effectively respond to a spatial average of \escaleS
 turbulence at fixed time. In the asymptotic limit  
 $\massrt \rightarrow 0 $, ions can only weakly respond to \escaleS fluctuations because,
 as a consequence of statistical periodicity, spatial averages over \escaleS turbulence should vanish.
 Taken with the conclusion of Section \refsec{sec:iIS}, we see
 that ions at \escaleS can be neglected entirely. 
 
 \subsection{Critical Balance}\label{sec:critical}
 Observe that the gyro-Bohm ordering \refeq{order:gbohm},
 where $\fract{\charge \eptl{}}{\temp} \sim \fract{\gyrde}{\lscal}$,
 is consistent with the critical balance argument \cite{barnesPRL11b}
 at the \iscaleS and \escaleS separately. At both scales the $E \xp B $ drift has the same magnitude 
 \beqn\ve \sim \ive \sim \eve \sim \frac{\vtheri \gyrdi}{\lscal} \sim \frac{\vthere \gyrde}{\lscal}. \eeqn
 By assumption, the \iscaleS perpendicular correlation length $\icorrper\sim\gyrdi$ and the \escaleS perpendicular correlation length $\ecorrper \sim \gyrde$. 
The \iscaleS nonlinear turnover time $\inonlintime$ obeys
\beqn \inonlintime \sim \frac{\icorrper}{\ive} \sim \frac{\lscal}{\vtheri}. \eeqn 
The \escaleS nonlinear turnover time $\enonlintime$ obeys 
\beqn \enonlintime \sim \frac{\ecorrper}{\eve} \sim \frac{\lscal}{\vthere}. \eeqn
In critical balance, the parallel extent of the eddies
is set by how far a particle can stream in one nonlinear
turnover time. This implies that the \iscaleS parallel correlation length $\icorrpar$ is 
\beqn \icorrpar \sim \vtheri\inonlintime \sim \lscal, \eeqn  
where we have used that the ions are the dominant species for communicating
 information in the direction parallel to the field line.
Similarly the \escaleS  parallel correlation length $\ecorrpar$ is 
\beqn \ecorrpar \sim \vthere\enonlintime \sim \lscal, \label{order:lpara}\eeqn
where we have used that the electrons are the dominant species.
This result is consistent with our ordering \refeq{order:sptm}.
 
\section{Scale-separated, Coupled Equations}\label{sec:sclsepcoupledeq}
Following the discussion in the previous Section, we can now resolve how to take the gyroaverages
 in the quasineutrality relation \refeq{eq:escalqn} in a scale-separated way.
 We neglect the non-adiabatic response of ions at \escaleS because the ion contribution to 
 \escaleS quasineutrality \refeq{eq:escalqn} is small by $\massrt$ (see Section \refsec{sec:iES}). 
At leading order, equation \refeq{eq:escalqn} becomes
\beqn -\charge \intv{|_{\rs,\rf}} \ehhe{}(\lpar,\gcs,\gcf,\energy,\pitch,\sign) = \sum_{\spe} \frac{\zeds^2 \charge^2 \denss}{\temps}\eptl{}(\lpar,\rs,\rf). \eeqn

When solving the \escaleS gyrokinetic equation \refeq{eq:esel},
 the quantity that we need to close the equation is $\egptle{}(\gcs,\gcf)$.
Noting that $|\kf||\gyrdvece| \sim |\kf|\gyrde \sim 1$ and
 $|\ks||\gyrdvece| \sim |\ks|\gyrde \sim \massrt$, in \refsec{sec:egavqn}
we expand the Bessel function  $\bes{|\kf+\ks||\gyrdvece|}$, due to electron gyroaverages,
in the expression for $\egptle{}(\gcs,\gcf)$ to find 
\beqn \fl \egptle{}(\gcs,\gcf) =  
 -\Big(\sum_{\spe} \frac{\zeds^2 \charge \denss}{\temps}\Big )^{-1} 
   \sum_{\kf} \expo{\imag \kf \cdot \gcf} \bes{|\kf||\gyrdvece|} \times \nonumber \eeqn \beqn
    \intv{|_\gcs} \ehhe{\kf} (\lpar,\gcs,\energy,\pitch,\sign) \bes{|\kf||\gyrdvece|}\left(1+ \order{\massr} \right).\label{eq:quasiapprox}\eeqn
 Here  $\gcs$ only appears as a label. Therefore, we have found a scale-separated scheme where
 \escaleS fluctuations at different $\gcs$ can be determined independently, as long as there is no
 coupling introduced by the parallel boundary condition (considered in Section \refsec{sec:pbc}).

We now present the full system of scale-separated equations keeping only those terms that 
appear at leading order in the gyro-Bohm ordering \refeq{order:gbohm}. 
At \iscaleS we have
\beqn \fl \drv{\ihhi{}}{\ts} + \vpar \kpar \drv{\ihhi{}}{\lpar} +
 \vmi \cdot \nbls \ihhi{}+ \ivei \cdot \nbls \ihhi{} +\ivei \cdot \nbl \eqlbi
 = \frac{\zedi \charge \eqlbi}{\tempi}\drv{\igptli{}}{\ts}, \label{equation:ionionreal}\eeqn
and
\beqn \fl \drv{\ihhe{}}{\ts} + \orbav{\vme \cdot \nbl \fldl} \drv{\ihhe{}}{\fldls} 
+ \orbav{\ivee \cdot \nbls \ihhe{}} + \orbav{\ivee \cdot \nbl \eqlbe } 
  = -\frac{ \charge \eqlbe}{\tempe}\drv{\orbav{\igptle{}}}{\ts},\label{equation:electronionreal}\eeqn
where $\igptli{} = \gav{\iptl{}}{\gcs,i}$, $\igptle{} = \gav{\iptl{}}{\gcs,e}$,
 $\orbav{\cdot}$ is the orbital average defined in \refeq{eq:orbpassing} and \refeq{eq:orbtrapped},
 with the properties \refeq{eq:orbcom1} and \refeq{eq:orbprop},
 and it is understood that $\ihhe{}=0$
 for non-zonal passing electrons. These equations are closed by the quasineutrality relation
\beqn \sum_{\spe} \zeds \charge \intv{|_\rs} \ihh{\spe}(\lpar,\gcs,\energy,\pitch,\sign) = \sum_{\spe} \frac{\zeds^2 \charge^2 \denss}{\temps}\iptl{}(\lpar,\rs). \label{equation:qnionreal}\eeqn
At \escaleS we have,
\beqn \fl \drv{\ehhe{}}{\tf} + \vpar \kpar \drv{\ehhe{}}{\lpar} + (\vme + \ivee ) \cdot \nblf \ehhe{} +
  \evee \cdot \nblf \ehhe{} \nonumber \eeqn \beqn + \evee \cdot (\nbl \eqlbe + \nbls \ihhe{}) = -\frac{\charge \eqlbe}{\tempe}\drv{\egptle{}}{\tf}, \label{equation:electronelectronreal}\eeqn
which is closed by the quasineutrality relation,
\beqn \fl \egptle{}(\lpar,\gcs,\gcf) =  -\Big(\sum_{\spe} \frac{\zeds^2 \charge \denss}{\temps}\Big )^{-1}
  \gav{ \intv{|_\gcs} \gav{ \ehhe{} (\lpar,\gcs,\gcf,\energy,\pitch,\sign) }{\gcs,\rf,e}}{\gcs,\gcf,e}.\label{equation:qnelectronreal}\eeqn
 Our notation in~\refeq{equation:qnelectronreal} indicates that the gyroaverage does not average over the slow variable $\gcs$, which is left fixed during the integrations.

Inspecting equations \refeq{equation:ionionreal}-\refeq{equation:qnelectronreal}, the reader can see
that the gyro-Bohm ordering \refeq{order:gbohm} is an ordering in which the \iscaleS
is dominant: the \escaleS turbulence
is modified at leading order by the cross scale terms
 $\ivee  \cdot \nblf \ehhe{}$ and $\evee \cdot\nbls \ihhe{}$ in \refeq{equation:electronelectronreal} 
(see Section \refsec{sec:ISgradES}); 
the \iscaleS turbulence evolves independently of the \escaleS fluctuations,
 as the largest possible cross-scale term in the \iscaleS equations, $\nbls \cdot \orbav{\esav{ \evee \ehhe{} }}$,
 is small in our orderings (see Sections \refsec{sec:iIS} and \refsec{sec:eISterm});
and the \iscaleS heat flux
 for ions $\ihfluxi$ and electrons $\ihfluxe$
 dominates the contribution to the heat flux from electrons at \escaleS $\ehfluxe$, i.e.  
 \beqn \ihfluxi \sim  \ihfluxe \sim \dens \temp \vtheri \left(\frac{\gyrdi}{\lscal}\right)^2 
 \gg \ehfluxe \sim \dens \temp \vthere \left(\frac{\gyrde}{\lscal}\right)^2. \label{eq:hfluxi} \eeqn
 We use our model to 
 recover the usual gyro-Bohm estimate for the heat fluxes \refeq{eq:hfluxi} in \refsec{sec:heatflux}.
The reader will notice that the heat flux contribution from non-zonal passing electrons
at \iscaleNS, which we self-consistently neglect, is formally comparable to
$\ehfluxe$. This is not an inconsistency, but the result of
the dominance of \iscaleS  transport in the gyro-Bohm
ordering. Our estimates \refeq{eq:hfluxi} do not capture the behaviour of the larger
 than electron gyro-Bohm heat flux observed in
 \cite{dorland2000electron,jenko2000electron,jenko2002prediction}.
 In \cite{dorland2000electron,jenko2000electron,jenko2002prediction} $\ehfluxe$ 
 was large due to the presence of spatially anisotropic ETG streamers.
 Recall that we have assumed spatial isotropy in the turbulence, and
 regarded all physical parameters besides $\gyrds/\lscal$ and $\massrt$ as of order unity.
 Deviation from the gyro-Bohm ordering \refeq{order:gbohm}, and the consequent dominance of
 \iscaleS transport \refeq{eq:hfluxi}, may be possible in a model which breaks these assumptions.
 We do not consider such a model here. 

\subsection{Fourier Representation}\label{sec:fouriersclsepcoupledeq}
 For equations \refeq{equation:ionionreal}-\refeq{equation:qnelectronreal} to be implemented in a 
 code it is convenient for them to be expressed spectrally. For clarity we give the full system of 
 equations in Fourier space. At \iscaleS we have
\beqn \fl \drv{\ihhi{\ks}}{\ts} + \vpar \kpar \drv{\ihhi{\ks}}{\lpar} + \Big[\imag\vmi \cdot \ks\Big] \ihhi{\ks}-\sum_{\ksp}  \frac{\ltsp}{\bmag}\bu\xp \ksp \cdot \ks \igptli{\ksp} \ihhi{\ks - \ksp }
 \nonumber \eeqn \beqn +\Big[\imag\frac{\ltsp}{\bmag}\bu\xp \ks \cdot  \nbl \eqlbi\Big] \igptli{\ks} = \frac{\zedi \charge \eqlbi}{\tempi}\drv{\igptli{\ks}}{\ts}, \label{equation:ionionfourier}\eeqn
and
 \beqn \fl \drv{\ihhe{\ks}}{\ts} + \Big[\imag \orbav{\vme \cdot \nbl \fldl} \evfldl \cdot \ks\Big] \ihhe{\ks}  
 - \orbav{\sum_{\ksp}  \frac{\ltsp}{\bmag}\bu\xp \ksp \cdot \ks \igptle{\ksp} \ihhe{\ks - \ksp } } \nonumber \eeqn \beqn 
 + \orbav{ \Big[\imag\frac{\ltsp}{\bmag}\bu\xp \ks \cdot  \nbl \eqlbe\Big] \igptle{\ks}}  = -\frac{ \charge \eqlbe}{\tempe}\drv{\orbav{\igptle{\ks}}}{\ts}, \label{equation:electronionfourier}\eeqn
 where $\ihhe{\ks} = 0 $ for the non-zonal ($\ks \cdot \evfldl \neq 0$) passing piece of phase space, 
$\igptli{\ks} = \bes{|\ks||\gyrdveci|}\iptl{\ks}$ and $\igptle{\ks} = \bes{|\ks||\gyrdvece|}\iptl{\ks}$.
 Equations  \refeq{equation:ionionfourier} and \refeq{equation:electronionfourier}
 are closed by the quasineutrality relation
\beqn \sum_{\spe} \zeds \charge \intv{} \bes{|\ks||\gyrdvecs|}\ihh{\spe \ks}(\lpar,\energy,\pitch,\sign) = \sum_{\spe} \frac{\zeds^2 \charge^2 \denss}{\temps}\iptl{\ks}(\lpar). \label{equation:qnionfourier}\eeqn
At \escaleS we have
\beqn \fl \drv{\ehhe{\kf}}{\tf} + \vpar \kpar \drv{\ehhe{\kf}}{\lpar} + \Big[\imag (\vme + \ivee ) \cdot \kf\Big] \ehhe{\kf} - \sum_{\kfp}  \frac{\ltsp}{\bmag}\bu\xp \kfp \cdot \kf \egptle{\kfp} \ehhe{\kf - \kfp } \nonumber \eeqn 
\beqn+ \Big[\imag\frac{\ltsp}{\bmag}\bu\xp \kf \cdot  (\nbl \eqlbe + \nbls \ihhe{})\Big] \egptle{\kf}   = -\frac{\charge \eqlbe}{\tempe}\drv{\egptle{\kf}}{\tf}. \label{equation:electronelectronfourier}\eeqn
Equation \refeq{equation:electronelectronfourier} is closed by the quasineutrality relation
\beqn \fl \egptle{\kf}(\gcs) =  -\charge\Big(\sum_{\spe} \frac{\zeds^2 \charge^2 \denss}{\temps}\Big )^{-1} \bes{|\kf||\gyrdvece|} \intv{|_\gcs} \ehhe{\kf} (\lpar,\gcs,\energy,\pitch,\sign) \bes{|\kf||\gyrdvece|},\eeqn
and the relations for $\ivee$ and $\nbls \ihhe{}$, 
\beqn \ivee = \ivee(\gcs) = \frac{\ltsp}{\bmag}\bu\xp \Big[\sum_{\ks} \expo{i\ks\cdot\gcs} i \ks \igptle{\ks}\Big], \label{eq:exbdef} \eeqn 
\beqn \nbls\ihhe{} = \nbls\ihhe{}(\gcs) = \sum_{\ks} \expo{\imag\ks\cdot\gcs} \imag \ks \ihhe{\ks}. \label{eq:graddef} \eeqn 

Note that the dependence of $\ehhe{\kf}$ and $\egptl{\kf}$ on $\gcs$
 is parametric in equation \refeq{equation:electronelectronfourier},
 as the dependence on $\gcs$ only appears through the quantities $\ivee$ and $\nbls\ihhe{}$.
 The evolution equations for $\ehhe{\kf}(\gcs)$ and $\egptl{\kf}(\gcs)$ may be solved in a
 system of flux tubes, with a single \escaleS flux tube for each of the considered $\gcs$
 locations within the \iscaleS flux tube. The \escaleS turbulence
 in flux tubes at different radial locations $\flxls$ may be evolved independently. 
 As we discuss in the next section,
on any $\flxls$ surface 
 where the safety factor $\saffac(\flxls)$ is not
  an \iscaleS rational,
 \escaleS flux tubes must be coupled in the binormal 
 direction by the parallel boundary condition.
 
\section{Parallel boundary condition}\label{sec:pbc}
 In the previous Section we found the scale-separated,
 coupled \iscaleNS-\escaleS equations \refeq{equation:ionionreal}-\refeq{equation:qnelectronreal}.
 Due to the assumptions of statistical periodicity, \refeq{eq:distfnsolve} and \refeq{eq:solvability},
 these equations are solved
 with periodic boundary conditions in the plane perpendicular to the magnetic field line.
 In this section we propose parallel boundary conditions for the \iscaleNS-\escaleS equations
which allow the \escaleS turbulence to be simulated in a system of flux tubes nested
 within a single \iscaleS flux tube.
 The approach that we will use as the starting point for our treatment is to
 use the so-called \enquote{twist-and-shift} boundary condition~\cite{beerPoP95},
 which we now briefly summarise.

We first use toroidal symmetry of the confining magnetic field to argue that the turbulence on a $\flxl,\toragl$
 plane at a fixed $\lpar$ is statistically identical. Assuming that the correlation length of the turbulence
 is shorter than one poloidal $2\pi$  turn, then a boundary condition which recovers the statistical properties is 
\beqn \hh{}(\lpar, \gcp(\flxl,\fldl(\flxl,\lpar,\toragl) ) )  = \hh{}(\lpar + 2\pi, \gcp(\flxl,\fldl(\flxl,\lpar+2\pi,\toragl)) ),\label{eq:twsh1} \eeqn
where we have decomposed the dependence of $\hh{}$ on $\gc$ into dependence on $(\lpar, \gcp)$, 
with $\gcp$ the guiding centre coordinate in the direction perpendicular to the local magnetic field line.
We suppress the velocity space dependences in $\hh{}$, and regard $\gcp$ as a function of $(\flxl,\lpar,\toragl)$. 
We use the representation for the guiding center $\gcp = (\flxl -\flxl_0)\evflxl + (\fldl -\fldl_0)\evfldl $, where
 $\fldl(\flxl,\lpar,\toragl) -\fldl_0 =\toragl - \saffac(\flxl) \lpar= \toragl - \saffac_0 \lpar - \saffacprim \lpar(\flxl -\flxl_0)  $,
 with $\saffac_0 = \saffac(\flxl_0)$, $\saffacprim = d \saffac / d \flxl |_{\flxl_0}$, and $(\flxl_0,\fldl_0)$ the coordinates of the central field line in the flux tube. Recalling that
$\evflxl = (\bu \xp \nbl \fldl)/|\nbl \flxl \cdot \bu \xp \nbl \fldl|$ and $\evfldl = (\nbl \flxl \xp \bu )/|\nbl \flxl \cdot \bu \xp \nbl \fldl|$, 
and noting that $\kk = \kkflxl \nbl \flxl + \kkfldl \nbl \fldl$, then we have,  
\beqn \fl \hh{}(\lpar, \gcp(\flxl,\fldl) ) = \sum_{\kk}\expo{\imag \kk \cdot \gcp} \hh{\kk}(\lpar) = \nonumber \eeqn \beqn 
 \sum_{\kk}\expo{\imag \kkflxl (\flxl -\flxl_0) + \imag\kkfldl (\fldl -\fldl_0) } \hh{(\kkflxl,\kkfldl)}(\lpar). \label{eq:bcfourier} \eeqn 
Using this, the boundary condition \refeq{eq:twsh1} in Fourier space is
\beqn \hh{(\kkflxl,\kkfldl)}(\lpar) = \underbrace{ \expo{- \imag 2 \pi \saffac_0 \kkfldl } }_{ = 1 }
   \hh{(\kkflxl + 2\pi \saffacprim \kkfldl ,\kkfldl)}(\lpar + 2\pi).  \label{eq:beer} \eeqn
The phase $\expo{- \imag 2 \pi \saffac_0 \kkfldl }$ is set to $1$ because as $\gyrd / \lscal \rightarrow 0 $
 the integer $\kkfldl^{\mbox{\scriptsize min}}\rightarrow\infty$  and hence $\saffac_0 \kkfldl$ can be made
 arbitrarily close to a very large integer for all $\kkfldl$. These arguments and the boundary condition
 \refeq{eq:beer} were first proposed in \cite{beerPoP95}. To allow the reader to familiarise
 themselves with our notation, we reproduce the calculation from \cite{beerPoP95} in~\refsec{sec:isbc}. 
 Note that the expression for the change in $\fldl$ in a poloidal turn is 
\beqn \fldl(\flxl,\lpar+2\pi,\toragl) - \fldl(\flxl,\lpar,\toragl) =- 2\pi \saffac_0 - 2 \pi \saffacprim(\flxl -\flxl_0). \eeqn
As the phase factor $\expo{- \imag 2 \pi \saffac_0 \kkfldl}$ is set to $1$ in \refeq{eq:beer}, we may write
the real space boundary condition corresponding to \refeq{eq:beer} as
\beqn \fl \hh{}(\lpar, \gcp(\flxl,\fldl(\flxl,\lpar,\toragl) ) )  = \hh{}(\lpar + 2\pi, \gcp(\flxl,\fldl(\flxl,\lpar,\toragl)- 2 \pi \saffacprim(\flxl -\flxl_0)) ). \label{eq:isrealbc} \eeqn
We will take the boundary condition \refeq{eq:isrealbc} as the boundary condition for the turbulence
in the \iscaleS flux tube.

We now consider how the boundary condition \refeq{eq:isrealbc} must be modified when
 considering \escaleS turbulence embedded within \iscaleS turbulence.
 To understand why we must provide a new flux tube parallel boundary
 condition for the \escaleS turbulence consider
that on an \iscaleS rational $\saffac$ surface, e.g. $ \flxl=\flxl_0$,
the \iscaleS boundary condition \refeq{eq:isrealbc} ensures that
 the fluctuations are periodic in $\lpar$ at fixed $\fldl$.
 For radial locations in the flux tube 
 with $\flxl \neq \flxl_0$ and $\saffac(\flxl)$ not an \iscaleS rational
 \iscaleS fluctuations are not periodic in $\lpar$ at fixed $\fldl$;
 the \iscaleS boundary condition \refeq{eq:isrealbc}
 encapsulates the effect of local magnetic shear by coupling field lines at different
 $\fldl$ at the boundaries in  $\lpar$. 
 Therefore, at radial locations 
 where $\flxl \neq \flxl_0$ and $\saffac(\flxl)$ is not an \iscaleS
 rational the \escaleS turbulence 
 experiences \iscaleS gradients, $\ivee$ and $\nbls\ihhe{} $
 in equation \refeq{equation:electronelectronreal}, that are
 not periodic in $\lpar$ at fixed $\fldl$.  
 On these surfaces is not possible to use the usual parallel boundary condition
 \refeq{eq:isrealbc}.
 Instead, away from 
 the \iscaleS rational $\saffac $ surfaces,
 the \escaleS flux tube at
 $\fldl$ should couple to the \escaleS flux tube at $\fldl - 2 \pi \saffacprim(\flxl -\flxl_0)$ 
 after one poloidal turn, where $\lpar \rightarrow \lpar + 2\pi$. 
By connecting multiple \escaleS flux tubes in this way we arrive at a chain of \escaleS flux tubes
 with a self-consistent parallel boundary condition. Our chain of \escaleS flux tubes
 here is reminiscent of the 
 \enquote{flux tube train} of \cite{2015Watanabefluxtubetrain}. 
 Our proposed extension to the standard \enquote{twist-and-shift} boundary condition
 for the \escaleS flux tubes is 
\beqn \fl \ehh{}(\lpar, \gcf(\flxlf,\fldl(\flxlf,\lpar,\toragl)), \gcs(\flxls,\fldl(\flxls,\lpar,\toragl) ) ) \nonumber \eeqn
\beqn  = \ehh{}(\lpar + 2\pi, \gcf(\flxlf,\fldl(\flxlf,\lpar+2\pi,\toragl)),\gcs(\flxls,\fldl(\flxls,\lpar+2\pi,\toragl))  ), \label{eq:esrealbc}\eeqn
where $ \gcf = (\flxlf -\flxl_0)\evflxl + (\fldlf -\fldl_0)\evfldl $, and $ \gcs = (\flxls -\flxl_0)\evflxl + (\fldls -\fldl_0)\evfldl $,
with $\flxls$ and $\flxlf$ the slow and fast flux labels, and $\fldls = \fldl(\flxls,\lpar,\toragl)$ and $\fldlf = \fldl(\flxlf,\lpar,\toragl)$. 

\begin{figure}[htb]
\begin{center}

\includegraphics[width=0.7\textwidth]{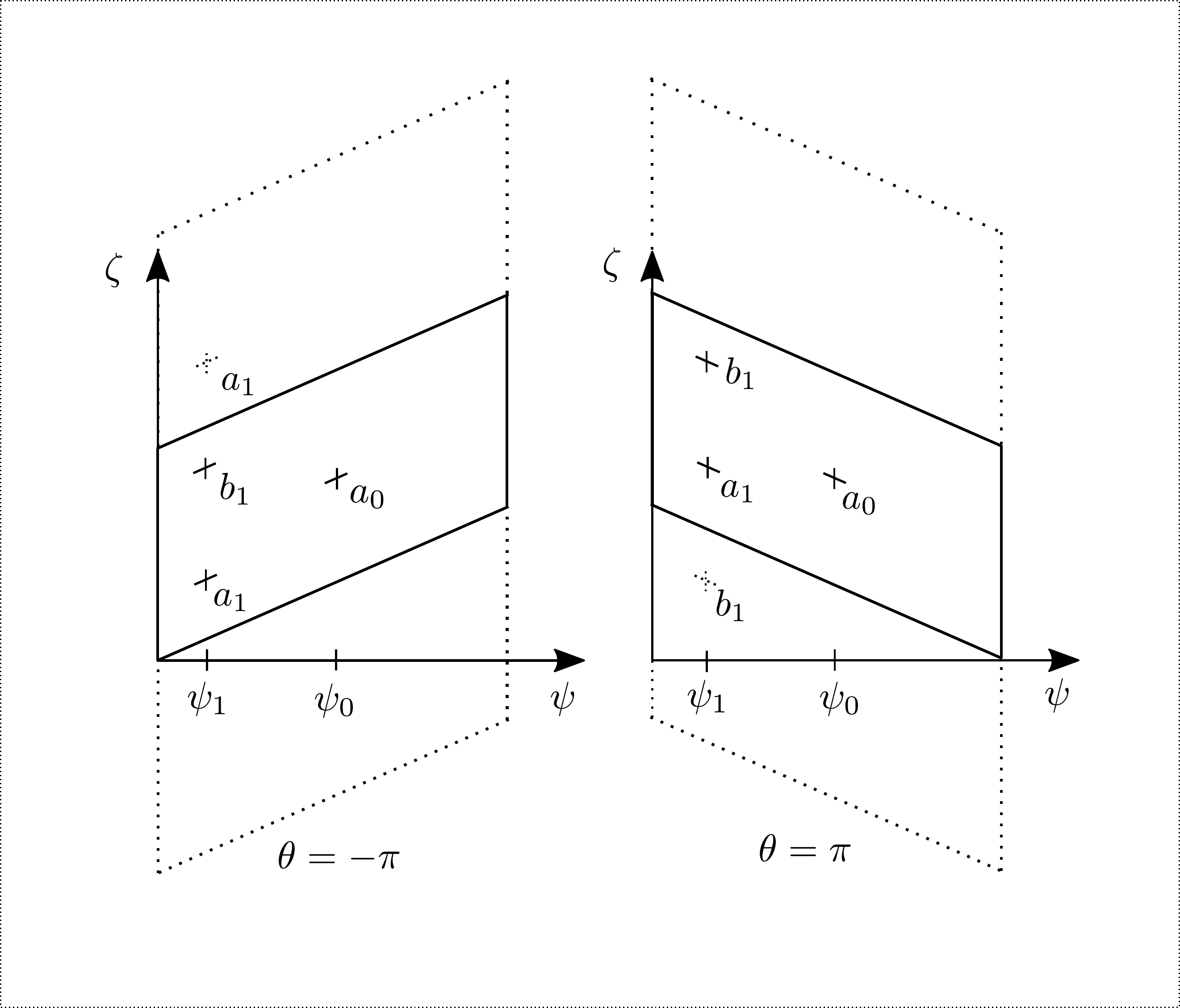}
\caption { The diagram shows the $\lpar = - \pi$  (left) 
 and $\lpar = \pi$ (right) ends of the \iscaleS flux tube,
 represented by solid-lined parallelograms on the inboard mid-plane, which we parameterise
 with $(\toragl, \flxl)$. The $\lpar \pm \pi$ ends of the flux tube are at 
 $\toragl$ points different by $2\pi \saffac_0$. We do not show the 
 $2\pi \saffac_0$ shift in $\toragl$.
 The dotted-lined parallelograms are periodic continuations
 of the \iscaleS flux tube. The boundary condition \refeq{eq:isrealbc}
 enforces that the turbulence on the $(\toragl,\flxl)$ plane mapped out by the parallelograms at
 $\lpar = \pm\pi$ must be the same. 
At $\flxl=\flxl_0$,  
 the field line labelled
 by $a_0$
 begins and ends at points in the $(\toragl,\flxl)$ plane
 where the fluctuations have the same phase in $\toragl$;
 the field line  $a_0$ is coupled to a periodic copy of
 itself after a poloidal turn.
 However, at $\flxl = \flxl_1$, a field line begins and
 ends at points in the $(\toragl,\flxl)$ plane
 where the fluctuations do not have the same phase in $\toragl$.
 Instead of coupling to a periodic copy of itself, 
 the field line labelled by $a_1$ at $\lpar=\pi$
 couples to a different field line,
 $b_1$ at $\lpar=-\pi$. 
 The field line labelled by $b_1$ at $\lpar=\pi$ likewise
 couples to the periodic copy of $a_1$ at $\lpar=-\pi$.
 When the \escaleS turbulence uses the boundary condition \refeq{eq:esrealbc}, the \escaleS flux tubes
within the \iscaleS flux tube (see Figure \ref{figure:cartoonfluxtubes}) 
 couple in the same manner as the field lines that they follow.}
\label{figure:cartoonfluxtubes2}
\end{center}
\end{figure}

In~\refsec{sec:esbc}, we show that boundary condition \refeq{eq:esrealbc} leads to the following spectral boundary condition, 
\beqn \fl  \ehh{(\ekkflxl,\ekkfldl)}(\lpar , \gcs(\flxls,\fldl(\flxls,\lpar,\toragl) ) ) \nonumber \eeqn
 \beqn \fl \qquad = \underbrace{\expo{- \imag 2 \pi \saffac_0 \ekkfldl}}_{ = 1 } \ehh{(\ekkflxl + 2\pi \saffacprim \ekkfldl ,\ekkfldl)}(\lpar + 2 \pi, \gcs(\flxls,\fldl(\flxls,\lpar,\toragl)-2 \pi \saffacprim(\flxls -\flxl_0)) ), \eeqn
where $\ekkflxl$ and $\ekkfldl$ are the \escaleS wave numbers corresponding to $\flxlf$ and $\fldlf$,
 and $\kf = \ekkflxl \nbl \flxl + \ekkfldl \nbl \fldl$. The phase-factor $\expo{-\imag 2 \pi \saffac_0 \ekkfldl}$ is set to $1$, 
because as $\gyrde / \gyrdi \rightarrow 0$ we can again make $\ekkfldl^{\mbox{\scriptsize min}}$ increasingly large
 and hence $\saffac_0 \ekkfldl$ can again be made arbitrarily close to a very large integer for all $\ekkfldl$.
Figure \ref{figure:cartoonfluxtubes2} gives a visualisation of the coupling between \escaleS flux tubes
 introduced by the boundary condition \refeq{eq:esrealbc}. 
  
To demonstrate that it is necessary to satisfy the proposed \escaleS boundary condition 
\refeq{eq:esrealbc}, consider one of the new \escaleS terms appearing in \refeq{equation:electronionreal} in Fourier space,
written in components,

\beqn \fl \eve \cdot \nbls \ihh{} =\Big[\imag\frac{\ltsp}{\bmag}\bu\xp \kf \cdot  \nbls \ihh{}\Big] \egptl{\kf}
 \nonumber \eeqn \beqn =\ltsp\Big[ \ekkflxl \drv{}{\fldls} \ihh{}(\lpar, \gcs(\flxls,\fldls) )-\ekkfldl \drv{ }{\flxls} \ihh{}(\lpar, \gcs(\flxls,\fldls) )   \Big] \egptl{(\ekkflxl,\ekkfldl)}, \eeqn
where the $\flxls$ derivative is taken at fixed $\fldls$, as is our convention unless otherwise stated. 
To ensure continuity of the \escaleS coefficient multiplying $\egptl{(\ekkflxl,\ekkfldl)}$ at the boundaries $(\lpar, \lpar + 2\pi)$ we need to have, 
\beqn  \fl \ekkflxl \drv{}{\fldls} \ihh{}(\lpar, \gcs(\flxl,\fldls ) ) -\ekkfldl \drv{ }{\flxls} \ihh{}(\lpar, \gcs(\flxls,\fldls ) ) = \nonumber \eeqn
\beqn \fl \qquad (\ekkflxl + 2 \pi \saffacprim \ekkfldl ) \drv{}{\fldls} \ihh{}(\lpar + 2\pi, \gcs(\flxl,\fldls + \delta \fldls  ) )
 \nonumber \eeqn 
 \beqn- \ekkfldl \drv{ }{\flxls} \ihh{}(\lpar + 2 \pi, \gcs(\flxls,\fldls + \delta \fldls  ) ),\eeqn
where $\delta \fldls  = - 2 \pi \saffacprim (\flxls - \flxl_0)$ and $\fldls = \fldl(\flxls,\lpar,\toragl)$. The \escaleS wave numbers $\ekkflxl$ and 
$\ekkfldl$ should be independent of each other, and independent of the \iscaleNS, so, 
\beqn \drv{}{\fldls} \ihh{}(\lpar, \gcs(\flxls,\fldls ) )  = \drv{}{\fldls} \ihh{}(\lpar + 2\pi, \gcs(\flxls,\fldls + \delta \fldls  ) ), \label{eq:alphaderivative}\eeqn
and, 
\beqn \fl \drv{ }{\flxls} \ihh{}(\lpar + 2 \pi, \gcs(\flxls,\fldls + \delta \fldls ) )
 -  2 \pi \saffacprim \drv{}{\fldls} \ihh{}(\lpar + 2\pi, \gcs(\flxls,\fldls + \delta \fldls ) ) \nonumber \eeqn 
  \beqn = \drv{ }{\flxls} \ihh{}(\lpar, \gcs(\flxls,\fldls ) ), \label{eq:fluxderivative}\eeqn
should be satisfied independently. As shown in~\refsec{sec:esbcconfirm}, these relations are
 satisfied independently when the \iscaleS turbulence satisfies the boundary condition \refeq{eq:isrealbc}. 

\section{Discussion} \label{sec:physics}

In this paper we have derived a system of coupled gyrokinetic equations,
 closed by quasineutrality, \refeq{equation:ionionreal}-\refeq{equation:qnelectronreal}.
 These equations describe turbulent fluctuations
driven at the scales of the ion and electron gyroradii and the leading order cross-scale
 interactions between the them. The equations \refeq{equation:ionionreal}-\refeq{equation:qnelectronreal} 
 are obtained
via an asymptotic expansion in the smallness of the electron-ion mass ratio $\massrt$, 
subsidiary to the gyrokinetic expansion in $\gyrd / \lscal$.
The derivation relies on the existence of turbulence with a separated ion scale (\iscaleNS),
where $\kperp  \gyrdi\sim 1$ and $\wfreq\sim\fract{\vtheri}{\lscal}$, and electron scale (\escaleNS),
 where $\kperp \gyrde\sim 1$ 
 and $\wfreq\sim\fract{\vthere}{\lscal}$. 
In addition, we assume that the turbulence is spatially isotropic
 in the plane perpendicular to the magnetic field.
Our assumption of scale separation
places limitations on the applicability of the model, but it allows us to 
efficiently capture the dominant cross-scale coupling physics.
In this Section we give a physical description of the cross-scale 
coupling terms, discuss the implications of these terms for
 fluctuation amplitudes, and describe the limitations of the model. \newline

Our model differs from the full gyrokinetic system, 
the $\dlf$ gyrokinetic equation \refeq{eq:gke} closed with quasineutrality \refeq{eq:qnlty},
 in the following key ways: the non-adiabatic ion response 
at \escaleS is neglected; fast electron time scales due to electron parallel streaming
and small radial structures due to passing electrons are ordered out
of the \iscaleS equations; 
 cross-scale interaction terms appear in the equation for the \escaleS turbulence;
 and the \iscaleS decouples from the \escaleNS.
Under our assumptions of isotropic turbulence with
 $\massrt \rightarrow 0$ (subsidiary to $\gyrd / \lscal \rightarrow 0$),
 we have shown that the only possible ordering for the fluctuation
 amplitudes that results in saturated dominant balance is the gyro-Bohm ordering \refeq{order:gbohm}.
 We find that the presence of cross-scale interaction does not allow for steady-state,
 scale-separated \iscaleNS-\escaleS turbulence where the fluctuation amplitudes differ
 from the gyro-Bohm estimate by factors of mass ratio. 
 In contrast to standard flux tube models, in our model the \escaleS
 turbulence is simulated with equations 
 \refeq{equation:electronelectronreal}-\refeq{equation:qnelectronreal}
 and boundary condition \refeq{eq:esrealbc}, using many flux tubes
 embedded within a single \iscaleS flux tube, which is evolved using equations 
 \refeq{equation:ionionreal}-\refeq{equation:qnionreal}
 and boundary condition \refeq{eq:isrealbc} \cite{beerPoP95}.
 The \escaleS flux tubes are coupled 
in the binormal direction $\fldl$ by the parallel boundary condition \refeq{eq:esrealbc}, in a manner which is
consistent with the \iscaleS parallel boundary condition \refeq{eq:isrealbc}.
 Our model allows for the rigorous use
 of single-scale simulations, in a way that allows for efficient parallelisation,
 while still capturing cross-scale interactions. \newline

We are able to neglect the non-adiabatic response of ions at \escaleS
 because of the size of the large ion gyroradius compared to
the scale of the correlation length perpendicular to the field for
 \escaleS fluctuations $\ecorrper$,
\beqn \ecorrper \sim \gyrde \ll \gyrdi. \label{eq:ecorrper}\eeqn
The ions rapidly gyrate perpendicular to the field line at the ion cyclotron frequency $\cycfi$,
which is much larger than any turbulent frequency $\wfreq$, i.e., $ \cycfi \gg \wfreq$.
Consequently the ion orbits rapidly sample many uncorrelated instances
of \escaleS turbulence. 
 In the asymptotic limit, because of our assumption of statistical periodicity
\refeq{eq:solvability},
the averaged \escaleS turbulence which the ions experience
is vanishingly small and so the ions can only respond weakly to the turbulence at \escaleNS.
 Hence, they do not contribute to the \escaleS potential (see Section \refsec{sec:iES}).
Because of the weakness of the ion response at \escaleNS,
 \iscaleS fluctuations are not influenced by
the ion response at \escaleS (see Section \refsec{sec:iIS}).
 It is thus not necessary to evolve
ions at \escaleS in the scale-separated system. \newline

 In order to retain only \iscaleS time and space scales in the \iscaleS equations, it is necessary
to formally remove the fast time scales introduced by electron parallel streaming.
 It is also necessary to formally remove very fine radial structures that are
 introduced by passing electrons near rational surfaces in \iscaleS simulations,
 as a consequence of very long tails in the electron
 response in ballooning modes \cite{HallatschekgiantelPRL2005,DominskinonadPOP015}.
 We achieve this separation by taking parallel streaming for electrons to be asymptotically
 fast compared to \iscaleS frequencies. 
 This allows us to neglect the $\massrt$ small piece of the
 electron distribution function due to non-zonal passing electrons
 and use the orbital average to find the leading
 order equation for electrons at \iscaleNS, cf.\cite{abelEMOD}. \newline

There are two novel cross-scale interaction terms appearing
in the coupled system of equations \refeq{equation:ionionreal}-\refeq{equation:qnelectronreal}.
 These new terms appear in the gyrokinetic
 equation \refeq{equation:electronelectronreal} for electrons at \escaleNS: 
 $\evee \cdot \nbls \ihhe{}$, and $\ivee \cdot \nblf \ehhe{}$.
 The term $\evee \cdot \nbls \ihhe{}$ arises due to gradients 
 in the electron distribution function at \iscaleS
 and is analogous to the typical equilibrium drive term $\evee \cdot \nbl \eqlbe$, 
 with the caveat that the equilibrium distribution is a Maxwellian in velocities 
 and has spatial variation only in $\psi$.  
It may seem surprising that gradients in the \iscaleS distribution function
can drive (or suppress) \escaleS instability on an equal footing with gradients
 of the equilibrium distribution function. However, while the 
 equilibrium distribution function is large, it varies on spatial scales long
 compared to the size of the fluctuations.  The net effect is that
 $\nbls \ihhe{} \sim \nbl \eqlbe$.  
The second novel cross-scale coupling term in~\refeq{equation:electronelectronreal},
 $\ivee  \cdot \nblf \ehhe{}$,
 represents the advection of \escaleS eddies perpendicular
 to the field line by \iscaleS drift wave motion $\ivee$.
The perpendicular correlation length of \iscaleS eddies $\icorrper$
is much larger than the perpendicular correlation length of \escaleS eddies $\ecorrper$; 
\beqn \frac{\ecorrper}{\icorrper}\sim \frac{\gyrde}{\gyrdi}\sim\massr.\eeqn
Furthermore, the nonlinear turnover time of \iscaleS eddies $\inonlintime$
 is much longer than the nonlinear turnover time of \escaleS eddies $\enonlintime$;
 \beqn \frac{\enonlintime}{\inonlintime}\sim \frac{\vtheri}{\vthere}\sim\massr.\eeqn
Therefore, the \escaleS turbulence
sees an \iscaleS drift $\ivee$ and an \iscaleS gradient $\nbls \ihhe{}$
 which are constant in the plane perpendicular to the field line, and constant in time.
 Due to the parallel orbit averaging in \refeq{equation:electronionreal},
 the gradient $\nbls \ihhe{}$ is also constant in
 the parallel-to-the-field coordinate $\lpar$.
Like the magnetic drift $\vme$, the \iscaleS drift $\ivee$ varies
in $\lpar$ along the field line, and so
$\ivee$ causes nontrivial advection of the \escaleS eddies.
 Because \iscaleS and \escaleS eddies have
 the same parallel length scale (see \refeq{order:sptm} and Section \refsec{sec:critical}),
 the \iscaleS drift $\ive$ has variation in $\lpar$ on scales comparable to the \escaleS turbulence.
 The \iscaleS drift is dynamically relevant, 
 and has the effect of shearing \escaleS eddies parallel to the field line.
We emphasise that in our leading order equations the 
\iscaleS drift does not cause a shear perpendicular to the field line,
which is commonly thought of as the relevant cross-scale mechanism for suppressing
turbulence. The shear perpendicular to the field line arising from $\ivee$
 is small by $\massrt$
compared to the drifts that we keep, and must be dropped from the scale-separated 
equations like the
 ion non-adiabatic response at \escaleS and the other terms small by $\massrt$. 
 In \refsec{sec:thetaconstve} we show that the piece
 of $\ivee$ that is constant in $\lpar$ may be removed from the equation for the \escaleS 
 fluctuations \refeq{equation:electronelectronreal}
 by writing equation \refeq{equation:electronelectronreal} in terms of 
the guiding centre distribution functions $\egge{} = \gav{\edlfe}{\gcf,\gcs}$ and 
$\igge{} = \gav{\idlfe}{\gcs}$, and changing coordinates to a rotating frame. \newline
 
There are no cross-scale terms in the leading order \iscaleS equations.
 This means that the \iscaleS turbulence in our model is unaffected by \escaleS eddies,
 and therefore evolves independently. This is in contrast to inferred interactions
 in full multi-scale simulations
 \cite{ howard2016comparison,maeyama2015cross,howard2014synergistic}.
 The largest cross-scale term at the \iscaleNS,
 $\nbls \cdot \orbav{\esav{ \evee \ehhe{} }}$, appeared in the equation \refeq{eq:ionscleq1el2},
 and was small by a factor of $\massrt$.
Our model assumptions result in 
 gyro-Bohm orderings for the fluxes; 
 the \iscaleS heat flux for ions $\ihfluxi$ and electrons $\ihfluxe$
 dominate the contribution to the heat flux from electrons at \escaleS $\ehfluxe$ by a 
 factor of $\massrut$. 
 We note that the heat flux from ions at \escaleS $\ehfluxi$ is small compared to $\ehfluxe$ by
  a factor of $\massrt$ and so $\ehfluxi$ can always be neglected.
 Our orderings and heat flux estimates do not capture the fact that
 turbulent transport is stiff; the numerical values of
$\ihfluxi$ $\ihfluxe$ and $\ehfluxe$ are highly sensitive to order unity
 parameters within our expansion. In addition, due to our assumptions of spatial isotropy,
 our estimates do not allow for spatially
 anisotropic electron temperature gradient (ETG) streamers, which can drive larger
 than expected \escaleS heat flux \cite{dorland2000electron,jenko2000electron,jenko2002prediction}.
 Therefore, whilst strictly outside the
asymptotic orderings used here, it is possible in simulations with non-zero $\massrt$ that
 the formally small \escaleS contribution to the heat flux may not be negligible. 
 We anticipate that a careful asymptotic analysis of the gyrokinetic equation
 might find an ordering and a set of scale-separated coupled equations where the
 orderings for the heat fluxes deviate from the gyro-Bohm estimates, and the
 \iscaleS cross-scale term $\nbls \cdot \orbav{\esav{ \evee \ehhe{} }}$ 
 appears at leading order in equation \refeq{eq:ionscleq1el2}.
 Such a theory could possibly
be constructed by allowing for parameters that we have considered to be of order unity to
be of the same size as some fractional power of the electron-ion mass ratio. Candidate
parameters include the degree of spatial anisotropy of the ES turbulence, the distance
of the \iscaleS and \escaleS turbulence from marginal stability, and the ratio of the zonal to non-zonal
fluctuation amplitudes. \newline
 
We stress that the key assumption in deriving the model equations 
\refeq{equation:ionionreal}-\refeq{equation:qnelectronreal}
is the assumption of scale separation between the ion and the electron space
 and time scales in the turbulence. We assumed that the turbulent wave 
 number and frequency spectrum had vanishingly small amplitude
in the intermediate range between the \iscaleS 
and the \escaleNS. We assumed scale separation
 for both the radial ($\flxl$) and binormal ($\fldl$) directions perpendicular to the
 magnetic field line, and we assumed that the turbulence
was isotropic at both scales. Altogether, this means that our model is
 unable to describe cases where there are significant
 fluctuations between the \iscaleS and the \escaleNS,
 which for example might be driven by trapped electron mode instability (TEM) or 
 ion temperature gradient instability (ITG) and ETG when
 the macroscopic temperature gradients are far above marginal stability.
 The assumption of spatial isotropy of the turbulence in the plane
 perpendicular to the magnetic field means that our model
 cannot describe individual modes with both scales of order the
 electron gyroradius and scales of order the ion gyroradius.
For example, the model
 cannot describe ETG streamers if their spatial scale in the radial direction
is not of order the electron gyroradius in the mass ratio expansion. \newline
 
 We expect there are cases where our model can give quantitatively accurate
predictions for transport, for example, 
where the turbulence between the \iscaleS and the \escaleS is suppressed.
The true purpose of the
model presented in this paper is as a tool to aid understanding of the cross-scale interactions observed in full
multi-scale turbulence. It is intended that the cross-scale interaction terms derived here
can be used in conjunction with single-scale simulations to help assess whether
a full multi-scale simulation is necessary. For example, an \iscaleS simulation
may not need to be extended to \escaleS if gradients of the \iscaleS
fluctuations consistently suppress the \escaleS linear instability. \newline
 
 \section{Acknowledgements}\label{sec:ack}
The authors would like to thank A. A. Schekochihin, P. Dellar, W. Dorland, S. C. Cowley,
 J. Ball, A. Geraldini, N. Christen, A. Mauriya, P. Ivanov and J. Parisi
 for useful discussion. M. R. Hardman would like to thank the Wolfgang Pauli Institute for providing a 
 setting for discussion and funding for travel.
 \par \textit{This work has been carried out within the framework of the EUROfusion Consortium and
 has received funding from the Euratom research and training programme 2014-2018 and 2019-2020 under grant
 agreement No 633053 and from the RCUK Energy Programme [grant number EP/P012450/1].
 The views and opinions expressed herein do not necessarily reflect those of the European Commission.
 The authors acknowledge EUROfusion, the EUROfusion High Performance Computer (Marconi-Fusion),
 and the use of ARCHER through the Plasma HEC Consortium EPSRC grant number EP/L000237/1 under the
 projects e281-gs2.}

\appendix

\section{The velocity space and gyroaverages}\label{sec:velocitygyroaverage}
In this Section we show that in Fourier space taking the gyroaverage
 leads to the appearance of Bessel functions.
Recalling $ \bvec = \bmagform$ is the representation for the magnetic field,
  where $\flxl$ is a flux label and $\fldl$ is the field line label, we can define
an orthonormal field aligned coordinate system
with basis vectors $\bu$, $\eone$, and $\etwo$, 
\beqn \bu = \frac{\bvec}{\bmag}, \quad 
\eone = \frac{\nbl \flxl}{|\nbl \flxl|}, 
\quad \etwo = \bu \xp \frac{\nbl \flxl } { |\nbl \flxl|  }, \label{eq:orthonormal} \eeqn
with the properties,
\beqn \eone \cdot \bu = 0, \quad \etwo \cdot \bu = 0,\quad \eone \cdot \etwo = 0, \eeqn
\beqn \eone \cdot \eone = 1, \quad \etwo \cdot \etwo = 1,\quad \bu \cdot \bu = 1, \eeqn
\beqn \eone \xp \bu = \etwo, \quad \etwo \xp \bu = -\eone. \eeqn
Using this coordinate system, and using the energy $\energy $,
pitch angle $\pitch $, sign $\sign$, and gyrophase $\gyrophase $ as coordinates,
 we can express the particle velocity $\pvel$ in the following manner,
\beqn \pvel = \vpar \bu + \vperp ( \cos{\gyrophase}\; \eone + \sin{\gyrophase}\; \etwo ), \label{eq:pvel}\eeqn
where $\vpar = \sign \left(\fract{2 \energy }{\ma}\right)^{1/2}(1 - \pitch \bmag)^{1/2}$, 
and $\vperp = \fract{2 \energy \pitch \bmag }{\ma}$.
The form \refeq{eq:pvel} is especially convenient for magnetised plasmas,
 where there is rapid gyromotion in the plane perpendicular to the magnetic field.
Using the representation \refeq{eq:pvel} we can write the vector gyroradius $\gyrdvec$ as
\beqn \fl \gyrdvec = \frac{\bu \xp \pvel}{\cycf} 
= -\frac{\vperp}{\cycf}( \cos{\gyrophase}\; \eone \xp \bu+ \sin{\gyrophase} \;\etwo \xp \bu)
= -\frac{\vperp}{\cycf}( \cos{\gyrophase} \;\etwo -  \sin{\gyrophase} \;\eone)\eeqn
Taking the dot product of $\gyrdvec$ with the wave vector $\kk$, a wave vector 
in the plane perpendicular to the direction of the magnetic field, i.e. $\bu \cdot \kk = 0$,
we find
\beqn \kk \cdot \gyrdvec = -\frac{\vperp}{\cycf}( \cos{\gyrophase} \; \kk \cdot \etwo -  \sin{\gyrophase} \; \kk \cdot \eone) \nonumber \eeqn
\beqn = -|\kk|\frac{\vperp}{\cycf}  \sin{(\gyrophase + \phase) } = -|\kk||\gyrdvec|  \sin{(\gyrophase + \phase) }, \eeqn
where
\beqn \tan {\phase} = -\frac{\kk \cdot \etwo}{\kk \cdot \eone}.\eeqn
Using the definition of the gyroaverage operator \refeq{equation:gav}, 
we take the gyroaverage over the phase $\expo{\imag \kk \cdot \gyrdvec}$ appearing in the
 gyrokinetic equations
 when the gyrokinetic equations are expressed in Fourier components. 
We find
\beqn \gav{\expo{\imag \kk \cdot \gyrdvec} }{}= \frac{1}{2\pi} \int^{2\pi}_0 d \gyrophase \; \expo{\imag \kk \cdot \gyrdvec} 
 \nonumber \eeqn \beqn = \frac{1}{2\pi} \int^{2\pi}_0 d \gyrophase \; \expo{ -\imag|\kk||\gyrdvec|  \sin{(\gyrophase + \phase) }}.\eeqn
Noting that $\sin{(\gyrophase + \phase)}$ is periodic in $[0,2\pi]$, we can rewrite this integral as
\beqn \gav{\cdot }{}=\frac{1}{2\pi} \int^{2\pi}_0 d \gyrophase \; \expo{ -\imag|\kk||\gyrdvec|  \sin{\gyrophase  }} = \bes{|\kk||\gyrdvec|}, \eeqn
where in the final equality we have recognised the definition of the 0th Bessel function of the 1st kind.
Therefore, we have shown that
\beqn \gav{\expo{\imag \kk \cdot \gyrdvec} }{}=\bes{|\kk||\gyrdvec|}. \eeqn

\section{Useful properties of the \escaleS average}\label{sec:gavproperties}
In this Section we prove the properties of the \escaleS average necessary
 to derive the coupled \escaleS and \iscaleS gyrokinetic equations. 
    Periodicity perpendicular to the field line in both slow and fast variables allows us to write a fluctuation $\hh{}$ as
    \beqn \fl \hh{}( \rs, \rf) = \sum_{\kf} \hh{\kf}(\rs) \expo{\imag \kf \cdot \rf} \nonumber \eeqn 
    \beqn= \sum_{\kf,\ks} \hh{\kf,\ks} \expo{\imag \kf \cdot \rf} \expo{\imag \ks \cdot \rs}. \eeqn
    This form allows us to show that one may take the \escaleS average using real space $\rf$ or guiding centre $\gcf$ as the integration variable.
    First using $\rf$ as the integration variable,
    \beqn \fl \esav{\hh{}( \rs, \rf)} = \sum_{\kf} \hh{\kf}(\rs) \esav{\expo{\imag \kf \cdot \rf}} \nonumber \eeqn
    \beqn = \sum_{\kf} \hh{\kf}(\rs) \kron{\vzero,\kf} = \hh{\vzero}(\rs). \eeqn 
    Then using $\gcf= \rf - \gyrdvec$ as the integration variable, where $\gyrdvec= (\bu\xp\pvel )/\cycf $ is the vector gyroradius,
    and noting that for fixed $\energy$, $\pitch$, and $\gyrophase$, $\gyrdvec$ is a constant vector shift which does not depend on either $\rf$ or $\rs$,
    \beqn \fl \esav{\hh{}( \rs, \rf)} = \sum_{\kf} \hh{\kf}(\rs) \esav{\expo{\imag \kf \cdot \gcf}} \expo{\imag \kf \cdot \gyrdvec}  \nonumber \eeqn
    \beqn = \sum_{\kf} \hh{\kf}(\rs) \expo{\imag \kf \cdot \gyrdvec} \kron{\vzero,\kf} = \hh{\vzero}(\rs). \eeqn
    Hence the choice of integration variable is unimportant.
    We are also able to show that the \escaleS average commutes with the gyroaverage, 
    \beqn \gav{\esav{\cdot}}{\gcf,\gcs} = \esav{\gav{\cdot}{\gcf,\gcs}} \eeqn
    To show this we apply the operations and find an identical result in the two cases:
    \beqn \fl \gav{\esav{\hh{}( \rs, \rf)}}{\gcf,\gcs} = \gav{\hh{\vzero}( \rs)}{} = \sum_{\ks} \hh{\vzero,\ks} \gav{\expo{\imag \ks \cdot \gyrdvec} }{} \expo{\imag \ks \cdot \gcs}, \eeqn
    and
    \beqn \fl \esav{\gav{\hh{}( \rs, \rf)}{\gcf,\gcs}} = \esav{\sum_{\kf,\ks} \hh{\kf,\ks} \gav{\expo{\imag (\kf+\ks) \cdot \gyrdvec}}{} \expo{\imag \ks \cdot \gcs}\expo{\imag \ks \cdot \gcf}} \nonumber \eeqn
    \beqn = \sum_{\kf,\ks} \hh{\kf,\ks} \gav{\expo{\imag (\kf+\ks) \cdot \gyrdvec}}{} \expo{\imag \ks \cdot \gcs} \esav{\expo{\imag \ks \cdot \gcf}}  \nonumber \eeqn
    \beqn= \sum_{\kf,\ks} \hh{\kf,\ks} \gav{\expo{\imag (\kf+\ks) \cdot \gyrdvec}}{} \expo{\imag \ks \cdot \gcs} \kron{\vzero,\kf} \nonumber\eeqn
    \beqn = \sum_{\ks} \hh{\vzero,\ks} \gav{\expo{\imag \ks \cdot \gyrdvec} }{} \expo{\imag \ks \cdot \gcs}. \eeqn 
    Hence, the \escaleS average commutes with the gyroaverage.

 \section{Electrons at \iscaleS and orbital averaging} \label{sec:he0ISmodel}
In this Section we prove the statements that we use to remove
the $\lscal/\vthere$ time scales and $\gyrde$ spatial scales
from the  \iscaleS equation for electrons.

\subsection{Proving the property \refeq{eq:orbprop} of the orbital average}\label{sec:annihilate}
To prove the property \refeq{eq:orbprop} in the passing region
we first note that because the integration $\int d \fldls d \lpar$
in the definition of the orbital average in the passing region \refeq{eq:orbpassing}
is taken over the whole flux surface we are free to write
the integration in terms of the toroidal angle $\toragl $ in place of
the field line label $\fldls$. Hence, 
an equivalent definition of the orbital average in the passing region is
\beqn  \orbav{\cdot}=
\frac{ \int d \toragl d \lpar \; \fract{\cdot}{\vpar \kpar}  }{\int d \toragl d \lpar  \fract{}{\vpar \kpar} }  . \label{eq:orbpassing2}\eeqn
 Using the definition \refeq{eq:orbpassing2}, we see that
 \beqn \orbav{ \vpar \kpar \drv{\func}{\lpar}\Big|_{\fldl}} = \frac{ \int d \toragl d \lpar \;  \drvt{\func}{\lpar}|_{\fldl}  }{\int d \toragl d \lpar  \fract{}{\vpar \kpar} }  = 0 , \label{eq:orbpassing3}\eeqn
 where we have used the relation
\beqn \drv{\func}{\lpar}\Big|_{\fldl} = \drv{\func}{\lpar}\Big|_{\toragl} + \drv{\toragl}{\lpar}\Big|_{\fldl} \drv{\func}{\toragl}\Big|_{\lpar}; \eeqn
 physical periodicity in the poloidal and toroidal directions,  $\func(\lpar, \toragl)=\func(\lpar+2\pi,\toragl)$,
 and $\func(\lpar, \toragl)=\func(\lpar,\toragl+2\pi)$;
 and that $\drvt{\toragl}{\lpar}|_{\fldl}$ is only a function of $(\flxl,\lpar)$ in axisymmetric devices.
 In the trapped region
 \beqn \orbav{ \vpar \kpar \drv{\func}{\lpar}}=
 \frac{ \sum_\sign \int^{\lparp}_{\lparm}   d \lpar \; \sign  \drvt{\func}{\lpar} }{2 \int^{\lparp}_{\lparm} d \lpar  \fract{}{|\vpar| \kpar} } = 
\frac{ \sum_\sign  \sign  \left[\func(\lpar)\right]^{\lparp}_{\lparm} }{2 \int^{\lparp}_{\lparm} d \lpar  \fract{}{|\vpar| \kpar} } = 0 , \label{eq:orbtrapped2}\eeqn
 where we have used that in the trapped region $\func $
satisfies the bounce condition $\func(\lparpm,\sign = 1)= \func(\lparpm,\sign = -1)$,
 where $\lparp$ and $\lparm$ are the poloidal coordinates for the
 upper and lower bounce points respectively.

 \subsection {Showing that $\ihhe{}^{(0)}= 0$ for non-zonal passing electrons}\label{h0proof}
The leading order equation for electrons at \iscaleS is \refeq{eq:evpar}.
This equation is solved using the decomposition \refeq{eq:electrondecomp2}
so that now \refeq{eq:evpar} reads 
\beqn \vpar \kpar \drv{\ihhe{}^{(0)}}{\lpar} =0. \label{eq:evpar2} \eeqn
Using the Fourier decomposition for \iscaleS fluctuations,
we find that
\beqn \vpar \kpar \drv{\ihhe{\ks}^{(0)}}{\lpar} =0. \label{eq:evpar3} \eeqn
Equation \refeq{eq:evpar3} implies that $\ihhe{\ks}^{(0)}$ is constant in $\lpar$,
for both passing and trapped pieces of the distribution function.
As the radial wave number $|\kkflxl| \rightarrow \infty$ we expect that 
$\hh{\ks}\rightarrow 0$ due to the presence of magnetic shear,
 which leads to dissipation for large $|\kkflxl|$. As such,
it is conventional to take $\hh{\ks} = 0$ 
as the boundary condition for passing particles 
in the parallel direction to the magnetic field.
In the Fourier representation passing particles are free to travel between
the modes labelled by the radial wave number $\kkflxl$
at fixed wave number in the $\fldls$ direction $\kkfldl$,
 due to the \enquote{twist-and-shift} parallel boundary condition \cite{beerPoP95}
 discussed in \refsec{sec:isbc}.
Taken with the boundary conditions \refeq{eq:isfourierbc} 
equation \refeq{eq:evpar3} gives the result that 
the leading piece of the \iscaleS electron distribution function
$\ihhe{\ks}^{(0)} =0$ for non-zonal $(\evfldl\cdot\ks \neq 0)$ passing particles.  

\subsection {Further properties of the orbital average}\label{sec:orbpropmore}
We now show that $\orbav{\ihhe{}^{(0)}} = \ihhe{}^{(0)}$.
In the passing region only the zonal component
of the electron distribution function is non-zero.
This means that in the passing region we can write
\beqn \ihhe{}^{(0)} = \ihhe{}^{(0)}(\flxls,\energy,\lambda, \sign), \eeqn
i.e. $\ihhe{}^{(0)}$ is constant in $\lpar$ and $\fldls$. 
Hence, applying the orbital average in the passing region \refeq{eq:orbpassing} we find
\beqn \orbav{\ihhe{}^{(0)}}= \frac{ \int d \fldls d \lpar \; \fract{\ihhe{}^{(0)}}{\vpar \kpar}  }{\int d \fldls d \lpar  \fract{}{\vpar \kpar} } = \ihhe{}^{(0)} , \label{eq:orbpassinghh}\eeqn
 where the constancy of $\ihhe{}^{(0)}$ in $\lpar$ and $\fldls$ allows us to take $\ihhe{}^{(0)}$
out of the integral in the numerator.
In the trapped region $\ihhe{}^{(0)}$ is constant in $\lpar$,
 and obeys the bounce condition $\ihhe{}^{(0)}(\lparpm,\sign = 1)= \ihhe{}^{(0)}(\lparpm,\sign = -1)$.
 This has the consequence that $\ihhe{}^{(0)}$ is also a constant in $\sign$.
We can therefore write
\beqn \ihhe{}^{(0)} = \ihhe{}^{(0)}(\flxls,\fldls,\energy,\lambda). \eeqn
The orbital average in the trapped region \refeq{eq:orbtrapped} 
only averages over $\lpar$ and $\sign$; hence,
\beqn \orbav{\ihhe{}^{(0)}}= \frac{ \sum_\sign \int^{\lparp}_{\lparm}   d \lpar \; \fract{\ihhe{}^{(0)}}{|\vpar| \kpar} }{2 \int^{\lparp}_{\lparm} d \lpar  \fract{}{|\vpar| \kpar} } = \ihhe{}^{(0)}  , \label{eq:orbtrappedhh}\eeqn
where we are again able to extract $\ihhe{}^{(0)}$ from the integral in the numerator.
Using identical arguments one can show that $\orbav{\drvt{\ihhe{}^{(0)}}{\fldls}}=\drvt{\ihhe{}^{(0)}}{\fldls}$ and 
$\orbav{\drvt{\ihhe{}^{(0)}}{\flxls}}=\drvt{\ihhe{}^{(0)}}{\flxls}$.

Finally, we show that $\orbav{\vm \cdot \nbl \flxl} = 0$ in an axisymmetric magnetic field.
In an axisymmetric device the magnetic field $\bvec$ 
may be expressed in a more restrictive form than \refeq{eq:bmagform1} \cite{hintonRMP76,HazeltineMeiss,helander},
\beqn \bvec = \bmagformaxi, \label{eq:bmagform2} \eeqn
where $\bcur$ is a flux function. 
The magnetic drift in the radial direction $\vm \cdot \nbl \flxl$ may be written as
 \cite{hintonRMP76,helander}
\beqn \vm \cdot \nbl \flxl = \frac{\bu}{\cycf} \xp \frac{\nbl \bmag}{\bmag} \cdot \nbl \flxl
 \left(\vpar^2 +  \frac{\vperp^2}{2}\right).  \eeqn
 Using the form of the magnetic field \refeq{eq:bmagform2},
$\vm \cdot \nbl \flxl$ can be expressed as
\beqn \vm \cdot \nbl \flxl = \vpar \kpar \drv{}{\lpar} 
\left( \frac{\bcur \vpar}{\cycf} \right) \label{equation:conmagdrift}.\eeqn
Using the property of the orbital average \refeq{eq:orbprop} 
 and \refeq{equation:conmagdrift}, we see that
\beqn\orbav{\vm \cdot \nbl \flxl}=0.\eeqn

\section{Obtaining the cross-scale terms}\label{section:gyroalgebra}
In this Section we derive the form of the cross-scale terms
 appearing in the coupled \iscaleS and \escaleS gyrokinetic equations.
\subsection{\iscaleS cross-scale terms}\label{section:gyroalgebraion}

Applying the \escaleS average to the nonlinear term, and using
\beqn \gptl{} = \igptl{} + \egptl{}, \nonumber \eeqn 
\beqn \hh{} = \ihh{} + \ehh{}, \nonumber \eeqn 
\beqn \nbl = \nbls + \nblf, \nonumber \eeqn 
we find that
\beqn \fl \esav{\ve \cdot  \nbl \hh{}} = \frac{\ltsp}{\bmag} \esav{\bu \xp \nbl \gptl{} \cdot \nbl \hh{} } \nonumber \eeqn
\beqn =\frac{\ltsp}{\bmag} \bu \xp \nbls \igptl{} \cdot\nbls \ihh{} + \frac{\ltsp}{\bmag} \esav{\bu \xp (\nbls + \nblf) \egptl{} \cdot (\nbls + \nblf) \ehh{}}. \label{eq:isal1} \eeqn
Note that
\beqn \esav{\bu \xp \nblf \egptl{} \cdot \nblf \ehh{}} = \esav{\nblf\cdot(\bu \xp \nblf \egptl{}  \ehh{})}= 0, \label{eq:ftv} \eeqn
where we have used the fact that the equilibrium does not depend on the fast spatial variable $\rs$.
 Furthermore note that
\beqn \fl \esav{\bu \xp \nbls \egptl{} \cdot \nblf \ehh{}} = -\esav{\ehh{} \nblf\cdot(\bu \xp \nbls \egptl{})  }= \esav{\ehh{}\nbls\cdot(\bu \xp \nblf \egptl{})}, \eeqn
where first we integrated by parts, and then performed an anti-cyclic permutation of the gradients acting on $\egptl{}$ recalling that the equilibrium does not depend on the slow spatial variable $\rs$.  
Finally, note that the slow derivative $\nbls$ can pass through the average:
\beqn \esav{\bu \xp \nbls \egptl{}{} \cdot \nbls \ehh{}{}} =  \nbls\cdot\esav{(\bu \xp \nbls \egptl{}{}) \ehh{}{}}, \eeqn
and
\beqn \fl  \esav{\ehh{}\nbls\cdot(\bu \xp \nblf \egptl{})} + \esav{(\bu \xp \nblf \egptl{}) \cdot \nbls  \ehh{}} = \nbls\cdot\esav{(\bu \xp \nblf \egptl{}{}) \ehh{}{}}. \eeqn
Thus, we find that
\beqn \esav{\bu \xp (\nbls + \nblf) \egptl{} \cdot (\nbls + \nblf) \ehh{}} = \nbls \cdot \esav{\left(\bu \xp (\nbls+\nblf) \egptl{}\right) \ehh{} }. \eeqn
Equation \refeq{eq:isal1} now becomes
\beqn \esav{\ve \cdot  \nbl \hh{}} =\frac{\ltsp}{\bmag} \bu \xp \nbls \igptl{} \cdot \nbls \ihh{} +  \nbls \cdot \esav{\ehh{}\frac{\ltsp}{\bmag}\bu \xp (\nbls+ \nblf) \egptl{}}. \eeqn
Dropping the term $\nbls \cdot \esav{\ehh{}(\fract{\ltsp}{\bmag})\bu \xp \nbls \egptl{}}$, which is small by $\massrt$, we find that
\beqn \esav{\ve \cdot \nbl \hh{}}=\ive \cdot \nbls\ihh{} +  \nbls \cdot \esav{\eve \ehh{}},\label{eq:isalF} \eeqn
where we have used the definitions \refeq{eq:igavexbdrift} and \refeq{eq:egavexbdrift}.

Physically the cross-scale term $\nbls \cdot \esav{\eve \ehh{}}$
  represents a divergence of a flux of particle density. 
 One might have expected that the \iscaleS
cross-scale term would have contained two fast derivatives, noting that 
 \beqn \eve \cdot \nblf \ehh{} \gg \nbls \cdot \esav{ \eve \ehh{} },\eeqn 
  and so the cross-scale term would have been of the form $\esav{\eve \cdot \nblf \ehh{}}$.
 However, because of our assumption of statistical periodicity \refeq{eq:solvability},
 the term $\esav{\eve \cdot \nblf \ehh{}} $ vanishes as shown in \refeq{eq:ftv},
 and the leading order term in the \iscaleS cross-scale term is the one given in \refeq{eq:isalF}.

\subsection{\escaleS cross-scale terms}\label{section:gyroalgebraelectron}
The \escaleS nonlinear term is
\beqn \fl \ve \cdot  \nbl \hh{} - \esav{\ve \cdot  \nbl \hh{}} = \nonumber \eeqn
\beqn \fl \qquad \frac{\ltsp}{\bmag} \bu \xp \nbls \igptl{} \cdot (\nbls + \nblf) \ehh{} + \frac{\ltsp}{\bmag} \bu \xp (\nbls + \nblf)\egptl{} \cdot \nbls  \ihh{} \nonumber \eeqn
\beqn \fl \qquad+ \frac{\ltsp}{\bmag} \bu \xp (\nbls+\nblf) \egptl{} \cdot (\nbls + \nblf) \ehh{}- \nbls \cdot \esav{(\frac{\ltsp}{\bmag}\bu \xp (\nbls+\nblf) \egptl{}) \ehh{} }. \eeqn
Keeping only leading order terms from each group of terms,
 and using definitions \refeq{eq:igavexbdrift} and \refeq{eq:egavexbdrift},
we have that
\beqn \ve \cdot  \nbl \hh{} - \esav{\ve \cdot  \nbl \hh{}} = \ive \cdot \nblf \ehh{} + \eve \cdot \nbls \ihh{} + \eve \cdot \nblf \ehh{}. \label{eq:esalF} \eeqn

\section{The sizes of the \iscaleS and \escaleS turbulent fluctuations} \label{sec:eqscalings}

In this Section we determine the allowed scalings for the quantities
 $\ihhi{}$, $\ihhe{}$, $\ehhi{}$, $\ehhe{}$, $\iptl{}$ and $\eptl{}$
in the coupled system of equations \refeq{eq:ionscleq1ion}, \refeq{eq:ionscleq1el2},
\refeq{eq:esel}, \refeq{eq:esion}, \refeq{eq:iscalqn}, and \refeq{eq:escalqn}. 

Writing the coupled gyrokinetic equations in real space, with the size of each term in terms of
 $\ihhi{}$, $\ihhe{}$, $\ehhi{}$, $\ehhe{}$, $\iptl{}$ and $\eptl{}$,
we find 
the equation for ions at \iscaleNS,
\beqn \fl \myoverbrace{\drv{\ihhi{}}{\ts}}{\frac{\vtheri}{\lscal} \ihhi{} } 
+ \myoverbrace{\vpar \kpar \drv{\ihhi{}}{\lpar}}{\frac{\vtheri}{\lscal}\ihhi{}}
+ \myoverbrace{\vmi \cdot \nbls \ihhi{}}{\frac{\vtheri}{\lscal} \ihhi{}}  
+ \myoverbrace { \ivei \cdot \nbls \ihhi{} }{ \frac{\vtheri}{\gyrdi}\frac{\charge \iptl{}}{\temp}\ihhi{}}
\nonumber \eeqn \beqn+ \myoverbrace { \nbls \cdot \esav{\evei \ehhi{} }   }{\frac{\vthere}{\gyrdi}\frac{\charge \eptl{} }{\temp} \jies \ehhi{} }
+\myoverbrace {\ivei \cdot \nbl \eqlbi}{\frac{\vtheri}{\lscal}\frac{\charge \iptl{}}{ \temp} \eqlbi} 
= \myoverbrace{\frac{\zedi \charge}{\tempi}\eqlbi\drv{\igptli{}}{\ts} }{\frac{\vtheri}{\lscal}\frac{\charge \iptl{}}{ \temp} \eqlbi}  
\label{eq:ionationsclordering}, \eeqn
where $\jies \sim \bes{|\kf| \gyrdi} \sim \massrst$. We remind the reader that an
 ion gyroaverage over an \escaleS quantity introduces the additional mass ratio
 scaling factor $\jies$. See equations \refeq{eq:fouriernote}-\refeq{eq:jiesfac} for a full discussion. 
The equation for electrons at \iscaleS is
\beqn  \fl \myoverbrace{\drv{\ihhe{}}{\ts}}{\frac{\vtheri}{\lscal}\ihhe{}^{(0)}} 
    + \myoverbrace{\orbav{\vme \cdot \nbl \fldl} \drv{\ihhe{}}{\fldls} }{\frac{\vtheri}{\lscal}\ihhe{}^{(0)}} 
  + \myoverbrace{ \orbav{\ivee \cdot \nbls \ihhe{}}}{\frac{\vtheri}{\gyrdi}\frac{\charge \iptl{}}{\temp} \ihhe{}^{(0)}} 
  \nonumber \eeqn \beqn + \myoverbrace{\orbav{\ivee \cdot \nbl \eqlbe}}{\frac{\vtheri}{\lscal}\frac{\charge \iptl{}}{ \temp} \eqlbe} 
  +  \myoverbrace{\nbls \cdot \orbav{\esav{\evee \ehhe{}} }}{\frac{\vthere}{\gyrdi}\frac{\charge \eptl{}}{\temp} \ehhe{}} 
  = \myoverbrace{-\frac{\charge}{ \temp}\eqlbe\drv{\orbav{\igptle{}}}{\ts}}{\frac{\vtheri}{\lscal} \frac{\charge \iptl{}}{ \temp} \eqlbe}   
 \label{eq:electronationsclordering} ,\eeqn
where $\ihhe{}^{(0)}$ is the leading order piece of the electron
distribution function at \iscaleNS, defined in equation \refeq{eq:electrondecomp2}.
The equation for electrons at \escaleS is
\beqn \fl \myoverbrace{ \drv{\ehhe{}}{\tf}}{\frac{\vthere}{\lscal} \ehhe{}}
+ \myoverbrace{\vpar \kpar \drv{\ehhe{}}{\lpar}}{\frac{\vthere}{\lscal} \ehhe{}}
+ \myoverbrace{\vme \cdot \nblf \ehhe{}}{\frac{\vthere}{\lscal} \ehhe{}}
+ \myoverbrace{\ivee  \cdot \nblf \ehhe{}}{\frac{\vtheri}{\gyrde} \frac{\charge \iptl{}}{\temp} \ehhe{}}
+ \myoverbrace{\evee \cdot \nblf \ehhe{}}{\frac{\vthere}{\gyrde}\frac{\charge \eptl{}}{\temp} \ehhe{}} 
\nonumber \eeqn \beqn + \myoverbrace{\evee \cdot \nbl \eqlbe}{\frac{\vthere}{\lscal}  \frac{\charge \eptl{}}{\temp} \eqlbe}
+ \myoverbrace{\evee \cdot \nbls \ihhe{}}{\frac{\vthere}{\gyrdi}\frac{\charge \eptl{}}{\temp} \ihhe{}^{(0)}} 
 = \myoverbrace{ -\frac{ \charge \eqlbe}{\temp}\drv{\egptle{}}{\tf}}{\frac{\vthere}{\lscal} \eqlbe  \frac{\charge \eptl{}}{\temp}} 
\label{eq:electronatelectronsclordering}. \eeqn     
Finally we find the equation for ions at \escaleNS,
\beqn \fl \myoverbrace{\drv{\ehhi{}}{\tf}}{\frac{\vthere}{\lscal} \ehhi{}} 
+ \myoverbrace{\vmi \cdot \nblf \ehhi{}}{\frac{\vthere}{\lscal} \ehhi{}} 
+ \myoverbrace{\ivei  \cdot \nblf \ehhi{}}{\frac{\vtheri}{\gyrde}\frac{\charge \iptl{}}{\temp} \ehhi{}} 
+ \myoverbrace{\evei \cdot \nblf \ehhi{}}{\frac{\vthere}{\gyrde} \jies \frac{\charge \eptl{}}{\temp} \ehhi{}}
\nonumber \eeqn \beqn+ \myoverbrace{\evei \cdot \nbl \eqlbi}{\frac{\vthere}{\lscal} \jies\frac{ \charge \eptl{}}{\temp} \eqlbi } 
+ \myoverbrace{\evei \cdot \nbls \ihhi{}}{\frac{\vthere}{\gyrdi} \jies \frac{ \charge \eptl{}}{\temp} \ihhi{}}
 = \myoverbrace{\frac{\zedi \charge} {\temp}\eqlbi\drv{\egptli{}}{\tf}}{\frac{\vthere}{\lscal} \eqlbi \jies \frac{ \charge \eptl{}}{\temp}} 
 \label{eq:ionatelectronsclordering}, \eeqn
where again the factor $\jies$ appears due to ion gyroaverages over the \escaleS fluctuations. 
We will use equations \refeq{eq:ionationsclordering}-\refeq{eq:ionatelectronsclordering}
 for the remainder of the discussion in this Section. 
  
The gyrokinetic ordering \refeq{eq:gkordering} naturally suggests the gyro-Bohm ordering, where
\beqn \nblper \idlf \sim \nblper \edlf. \label{eq:ambiturb}\eeqn
 Ordering \refeq{eq:ambiturb} allows for a separation of scales between electron and ion spatial scales,
 in analogy to the separation between turbulent and equilibrium scales in ordinary gyrokinetics.
 In the usual picture used to motivate the gyrokinetic orderings
 there are eddies of size $\gyrd$ stirring up a background equilibrium
 gradient of scale $\lscal $ and size $\eqlb / \lscal$,
 where $\eqlb$ is the typical equilibrium amplitude, which
 results in turbulent fluctuations of amplitude $ (\gyrd/\lscal) \eqlb$.
 In the gyro-Bohm ordering the picture is the same with 
 $ \gyrd \rightarrow \gyrde $, $\lscal \rightarrow \gyrdi$,
 $\eqlb \rightarrow \idlf$, i.e. the \escaleS plays the role of the
 fluctuation and the \iscaleS plays the role of the equilibrium.
 With our assumption of length and time scales, ordering \refeq{order:sptm},
 we arrive at the gyro-Bohm ratio for the turbulent amplitudes,
\beqn \edlf \sim \frac{\gyrde}{\gyrdi} \idlf \sim \frac{\vtheri}{\vthere} \idlf
 \sim \massr \idlf \label{eq:gyrobohmapp}. \eeqn

 We use the ordering \refeq{eq:gyrobohmapp} and look for a saturated dominant
 balance for ions at the \iscaleS in equation \refeq{eq:ionationsclordering}, where
 \beqn \ivei \cdot \nbls \ihhi{} \sim  \ivei \cdot \nbl \eqlbi \Rightarrow \frac{\vtheri}{\gyrdi}\frac{\charge \iptl{}}{\temp}\ihhi{} \sim \frac{\vtheri}{\lscal} \frac{\charge \iptl{}}{\temp} \eqlbi ,\eeqn 
 and similarly for electrons at \iscaleS in equation \refeq{eq:electronationsclordering}, where
 \beqn \ivei \cdot \nbls \ihhe{} \sim  \ivee \cdot \nbl \eqlbe \Rightarrow \frac{\vtheri}{\gyrdi}\frac{\charge \iptl{}}{\temp}\ihhe{} \sim \frac{\vtheri}{\lscal} \frac{\charge \iptl{}}{\temp} \eqlbe .\eeqn 
 We do the same for electrons at \escaleS in equation \refeq{eq:electronatelectronsclordering}, where 
 \beqn \evee \cdot \nbls \ehhe{} \sim  \evee \cdot \nbl \eqlbe \Rightarrow \frac{\vthere}{\gyrde}\frac{\charge \eptl{}}{\temp}\ehhe{} \sim \frac{\vthere}{\lscal} \frac{\charge \eptl{}}{\temp} \eqlbe, \eeqn 
 and note that the scale of $\ehhi{}$ in \refeq{eq:ionatelectronsclordering}
 is set by the drive terms, rather than the advection terms, 
 \beqn \vmi \cdot \nblf \ehhi{} \sim \evei \cdot \nbl \eqlbi \Rightarrow  \frac{\vthere}{\lscal} \ehhi{} \sim \frac{\vthere}{\lscal} \jies \frac{\charge \eptl{}}{\temp} \eqlbi.  \eeqn 
 Then with quasineutrality equations \refeq{eq:iscalqn} and \refeq{eq:escalqn}, which indicate
\beqn \frac{e\iptl{}}{\temp}  \sim \mbox{max}\left\{ \frac{\ihhi{}}{\eqlbi},\frac{\ihhe{}}{\eqlbe} \right\}, \quad \mbox{ and } \quad \frac{e\eptl{}}{\temp}  \sim \mbox{max}\left\{ \frac{\ehhi{}}{\eqlbi},\frac{\ehhe{}}{\eqlbe} \right\}, \label{eq:quasiorder}\eeqn
  we arrive at the self-consistent scalings \refeq{order:gbohm}.

Using the scalings \refeq{order:gbohm} we see that the gyro-Bohm ordering captures
 \escaleS turbulence where $\ehhe{} /\eqlbe$ is modified at leading order by \iscaleS gradients,
 where ions at \escaleS can be ignored, and where the largest possible 
 cross-scale terms can be neglected in the \iscaleS equations. 
The ion cross-scale term $\nbls \cdot \esav{\evei \ehhi{} }$ in \refeq{eq:ionationsclordering} 
 is neglected
 because it is small by $\order{\massrlt}$. The electron cross-scale term
 $\nbls \cdot \orbav{\esav{\evee \ehhe{} }}$
 in \refeq{eq:electronationsclordering} is small by $\order{\massrt} $, and is neglected along with 
the small correction $\ihhe{}^{(1)}$ to the electron distribution function at \iscaleNS. 
 We now consider if it is possible to modify the scalings with mass ratio,
 to explore which saturation mechanisms exist.
 We conclude that \refeq{equation:ionionreal}-\refeq{equation:qnelectronreal} 
 are the most general set of leading order, scale-separated equations. 

\subsection{\escaleS turbulence enhanced by cross-scale interaction: 
${\charge \eptl{} } /{\temp} \gg \gyrde / \lscal $}\label{sec:electronscaleenhanced}
First let us consider the possibility that the presence of cross-scale interactions
 causes the \escaleS turbulent amplitude to be enhanced by a factor of mass ratio
 compared to the gyro-Bohm estimate, i.e. ${\charge \eptl{} } /{\temp} \gg \gyrde / \lscal $.
 
 The dominant terms in the \escaleS equation for electrons \refeq{eq:electronatelectronsclordering}
 would be $\evee \cdot \nblf \ehhe{}$, $ \evee \cdot \nbl \eqlbe$, and $ \evee \cdot \nbls \ihhe{}$.
 The dominant terms in the \escaleS equation for ions \refeq{eq:ionatelectronsclordering}
 would be $\evei \cdot \nblf \ehhi{}$,
 $ \evei \cdot \nbl \eqlbi$, and $ \evei \cdot \nbls \ihhi{}$. 
 If the equilibrium drive terms are dominant then we would have
 a balance between $\evee \cdot \nbl \eqlbe$ 
 and the nonlinear term $\evee \cdot \nblf \ehhe{}$
 in \refeq{eq:electronatelectronsclordering} and
 a balance between $\evei \cdot \nbl \eqlbi$ 
 and the nonlinear term $\evei \cdot \nblf \ehhi{}$
 in \refeq{eq:ionatelectronsclordering}.
 This results in 
 \beqn \ehhe{} \sim \frac{\gyrde}{\lscal} \eqlbe, \quad \ehhi{} \sim \frac{\gyrde}{\lscal} \eqlbi, \eeqn
 and so with quasineutrality \refeq{eq:quasiorder}
 we would find \beqn \frac{\charge \eptl{} } {\temp} \sim \frac{\gyrde}{\lscal}, \eeqn
 inconsistent with our assumption.

 This tells us that
 for  ${\charge \eptl{} } /{\temp} \gg \gyrde / \lscal $ to be possible
 we would need to have dominant
 cross-scale interaction terms. For electrons, this implies that 
 \beqn \evee \cdot \nblf \ehhe{} \sim  \evee \cdot \nbls \ihhe{} \Rightarrow \frac{\vthere}{\gyrde} \frac{\charge \eptl{} } {\temp} \ehhe{} \sim  \frac{\vthere}{\gyrdi}  \frac{\charge \eptl{} } {\temp} \ihhe{}, \eeqn
 and for ions that
 \beqn \evei \cdot \nblf \ehhi{} \sim  \evei \cdot \nbls \ihhi{} \Rightarrow \frac{\vthere}{\gyrde}\jies \frac{\charge \eptl{} } {\temp} \ehhi{} \sim  \frac{\vthere}{\gyrdi}  \jies \frac{\charge \eptl{} } {\temp} \ihhi{}. \eeqn
 Therefore we need,
 \beqn \ehhe{} \sim \frac{\gyrde}{\gyrdi} \ihhe{}, \quad \ehhi{} \sim \frac{\gyrde}{\gyrdi} \ihhi{}  \label{eq:escalggcond}. \eeqn 
 In addition, we must have that the cross-scale interaction 
 is much stronger than the equilibrium drive, and so we must have that
 \beqn \frac{\ihhe{}}{\eqlbe} \gg \frac{\gyrdi}{\lscal} \quad \mbox{ and } \quad \frac{\ihhi{}}{\eqlbi} \gg \frac{\gyrdi}{\lscal}. \eeqn
 With \refeq{eq:quasiorder} this requires
 \beqn \frac{\charge \iptl{} }{\temp} \gg \frac{\gyrdi}{\lscal}. \label{eq:iscalebig} \eeqn
 We now determine if \refeq{eq:iscalebig} is possible.
 
 \subsection {\iscaleS turbulence enhanced by cross-scale interaction:
 $ {\charge \iptl{} } /{\temp} \gg \gyrdi / \lscal  $}\label{sec:ionscaleenhanced}
 Let us consider if it is possible for the \iscaleS turbulent amplitude to be enhanced
 by a mass ratio factor compared to the gyro-Bohm estimate, i.e.
$ {\charge \iptl{} } /{\temp} \gg \gyrdi / \lscal  $.
We consider the possibility that the \iscaleS cross-scale term 
$\nbls \cdot \orbav{\esav{\evee \ehhe{} }}$ is larger than 
the equilibrium drive term $\orbav{\ivee \cdot \nbl \eqlbe}$,
 and is sufficiently large
 to balance the \iscaleS nonlinear term $\orbav{\ivee \cdot \nbls \ihhe{}}$
 in the electron equation
 \refeq{eq:electronationsclordering}.
This is the only possible way to obtain
 $ {\charge \iptl{} } /{\temp} \gg \gyrdi / \lscal$, as 
 all other balances lead to gyro-Bohm scaling.
 
The dominant terms in \refeq{eq:electronationsclordering}
 would now be $\orbav{\ivee \cdot \nbls \ihhe{}} $ and $ \orbav{\nbls \cdot \esav{\evee \ehhe{} }}$,
 which would imply 
\beqn \orbav{\ivee \cdot \nbls \ihhe{}} \sim \nbls \cdot \orbav{\esav{\evee \ehhe{}}}\Rightarrow \frac{\vtheri}{\gyrdi} \frac{\charge \iptl{}}{\temp} \ihhe{} \sim \frac{\vthere}{\gyrdi}\frac{\charge \eptl{}}{\temp} \ehhe{}. \eeqn
With \refeq{eq:escalggcond} this leads to  \beqn  \frac{\charge \eptl{}}{\temp} \frac{\temp}{\charge \iptl{}} \sim 1 , \label{eq:iscalpotratio}\eeqn
 which is inconsistent with \beqn  \frac{\charge \eptl{}}{\temp} \frac{\temp}{\charge \iptl{}} \sim \frac{\gyrde}{\gyrdi} ,  \label{eq:escalpotratio}\eeqn
which is implied by \refeq{eq:escalggcond} and therefore neither
 $\fract{\charge \eptl{}}{\temp} \gg \fract{\gyrde}{\lscal} $
 nor $ \fract{\charge \iptl{}}{\temp}\gg \fract{\gyrdi}{\lscal}$ are possible.

 We could repeat this argument by attempting to balance the \iscaleS
 nonlinear term for ions, $ \ivei \cdot \nbls \ihhe{} $,
 with the \iscaleS cross-scale term for ions, 
$ \nbls \cdot \esav{\evei \ehhi{} }$, and we would find that for
 the cross-scale term to compete we would need
 \beqn  \frac{\charge \eptl{}}{\temp} \frac{\temp}{\charge \iptl{}} \sim \jies^{-1} \gg 1 \label{eq:iscalpotratio2}, \eeqn
 which is again inconsistent with \refeq{eq:escalpotratio} and  therefore not an allowed balance.
 
\subsection{\iscaleS turbulence suppresses \escaleS fluctuations:
 ${\charge \iptl{} } /{\temp} \sim \gyrdi / \lscal $
 and $ {\charge \eptl{} } /{\temp} \ll \gyrde / \lscal $}\label{sec:electronscalesuppressed}
Now let us consider the possibility that, as a result of cross-scale interaction,
the \escaleS fluctuation amplitude is
 suppressed by a mass ratio factor compared to the gyro-Bohm estimate,
 i.e. ${\charge \iptl{} } /{\temp} \sim \gyrdi / \lscal $
 and $ {\charge \eptl{} } /{\temp} \ll \gyrde / \lscal $.
 In this case, at the \escaleS the usual nonlinearity $\eve \cdot \nblf \ehh{}$ is negligible.
 The \escaleS equations are still modified at leading order by the \iscaleS gradients, and so
 linear and cross-scale physics is now dominant in the
 \escaleS equation \refeq{eq:electronatelectronsclordering}, which gives us
 \beqn \fl \vpar \kpar \drv{\ehhe{}}{\lpar} \sim \evee \cdot \nbl \eqlbe \sim \evee \cdot \nbls \ihhe{} \nonumber \eeqn
 \beqn \Rightarrow \frac{\vthere}{\lscal} \ehhe{} \sim \frac{\vthere}{\lscal} \frac{\charge \eptl{}}{\temp} \eqlbe. \eeqn
 Ions at \escaleS are still ignorable. Since linear and cross-scale physics are dominant, we still find
 \beqn \ehhi{} \sim \jies \frac{\charge \eptl{}}{\temp} \eqlbi, \label{eq:eisclsmscl}\eeqn
 as in the gyro-Bohm regime \refeq{eq:gyrobohmapp}.
 Since we retain scaling \refeq{eq:eisclsmscl},
 the cross-scale term in the  \iscaleS ion equation \refeq{eq:ionationsclordering} may be ignored.
 We can ignore the \iscaleS cross-scale term in the \iscaleS electron equation
 \refeq{eq:electronationsclordering} which now appears at an order smaller than
 $\order{\massrt} $ in our expansion,
 \beqn \fl \nbls \cdot \esav{\evee \ehhe{} } \sim \frac{\vthere}{\gyrdi}\frac{\charge \eptl{}}{\temp} \ehhe{} \sim \frac{\vthere^2 }{\gyrdi \lscal}\left(\frac{\charge \eptl{}}{\temp}\right)^2 \eqlbe \ll \massr \frac{\vtheri}{\lscal} \ihhe{}. \eeqn
 Hence, the ordering ${\charge \iptl{} } /{\temp} \sim \gyrdi / \lscal $
 and $ {\charge \eptl{} } /{\temp} \ll \gyrde / \lscal $ 
 captures the modification of \escaleS linear physics by \iscaleS profiles,
 where the \iscaleS is unaffected by the \escaleNS, even at sub-dominant order.
 To realise this ordering the \escaleS fluctuations must be linearly stable,
 and have vanishing amplitude.
 Note that the \iscaleS gradients $\nbls \igptle{}$ and $\nbls \ihhe{}$ do not depend on $\tf$,
 the \escaleS time coordinate. Hence, in this ordering the \escaleS turbulence
 cannot be saturated. Either the presence of \iscaleS gradients suppresses the
 \escaleS instability, in which case the \escaleS turbulence vanishes,
 or the \iscaleS gradients enhance the \escaleS instability, 
 in which case the \escaleS fluctuation amplitudes grow until they
 saturate at gyro-Bohm levels through the usual nonlinear term $\evee \cdot \nblf \ehhe{}$.
 
 \subsection{\escaleS turbulence suppresses \iscaleS fluctuations:
 ${\charge \iptl{} } /{\temp} \ll \gyrdi / \lscal $
 and $ {\charge \eptl{} } /{\temp} \sim \gyrde / \lscal $}\label{sec:ionscalesuppressed}
Finally let us consider the possibility that, as a result of cross-scale interaction,
 the \iscaleS fluctuation amplitude is
 suppressed by a mass ratio factor compared to the gyro-Bohm estimate,
 i.e. ${\charge \iptl{} } /{\temp} \ll \gyrdi / \lscal $
 and $ {\charge \eptl{} } /{\temp} \sim \gyrde / \lscal $. 
 Here, the \escaleS turbulence is nonlinearly saturated by the single
 scale nonlinearity $\evee \cdot \nblf \ehhe{}$ in
 \refeq{eq:electronatelectronsclordering}. The \escaleS turbulence
 is unmodified by the \iscaleS fluctuations as the ion gradient terms
 are small. 
 The single scale nonlinear terms of the \iscaleS are negligible; both
 $\ivei \cdot \nbls \ihhi{}$ in equation \refeq{eq:ionationsclordering} and
 $\orbav{\ivee \cdot \nbls \ihhe{}}$ in equation \refeq{eq:electronationsclordering} are small  
 because ${\charge \iptl{} } /{\temp} \ll \gyrdi / \lscal $.
 The only way for the \iscaleS to be saturated
 is through the cross-scale term $\nbls \cdot \orbav{\esav{\evee \ehhe{} }}$
 in \refeq{eq:electronationsclordering}, which is a time varying source.
 This would require
 \beqn \orbav{\vme \cdot \nbl \fldl} \drv{\ihhe{}}{\fldls} \sim \nbls \cdot \orbav{\esav{\evee \ehhe{} }}
 \Rightarrow  \frac{\vtheri}{\lscal} \ihhe{} \sim \frac{\vthere}{\gyrdi}\frac{\charge \eptl{}}{\temp} \ehhe{}, \eeqn
 which here implies \beqn \frac{\ihhe{}}{\eqlbe} \sim \frac{\gyrde}{\lscal}. \label{eq:electronionsclsmlscl} \eeqn
 If we assume that the electrons set the scale of ${\charge \iptl{} } /{\temp}$ then the ion \iscaleS equation \refeq{eq:ionationsclordering}
 has a dominant balance between only linear terms,
 \beqn \vmi \cdot \nbls \ihhi{} \sim \ivei \cdot \nbl \eqlbi \Rightarrow  \frac{\vtheri}{ \lscal} \ihhi{} \sim \frac{\vtheri} {\gyrdi}\frac{\charge \iptl{}}{\temp} \eqlbi, \eeqn
 and so \beqn \frac{\ihhi{}}{\eqlbi} \sim \frac{\gyrde}{\lscal}. \label{eq:ionionsclsmlscl}\eeqn
 Using \refeq{eq:quasiorder}, \refeq{eq:electronionsclsmlscl} and \refeq{eq:ionionsclsmlscl} we see that,
 \beqn \frac{\charge \eptl{}}{\temp} \sim \frac{\gyrde}{ \lscal}, \eeqn is a consistent scaling.
Therefore, naively ${\charge \iptl{} } /{\temp}\sim \gyrde / \lscal \ll \gyrdi / \lscal $
 and $ {\charge \eptl{} } /{\temp} \sim \gyrde / \lscal $
 appears to be a consistent ordering for the \iscaleNS-\escaleS system.
 However, note that in this regime the \escaleS
 cross-scale terms, $\ivee \cdot \nblf \ehhe{}$ and $\evee \cdot \nbls \ihhe{}$ in
 \refeq{eq:electronatelectronsclordering}
 and $\ivei \cdot \nblf \ehhi{}$ and $\evei \cdot \nbls \ihhi{}$ in \refeq{eq:ionatelectronsclordering},
 are small. 
 The \escaleS turbulence only has dependence on the \iscaleS spatial coordinate
 $\gcs$ through the \iscaleS gradients appearing in the \escaleS cross-scale terms,
 and the parallel boundary condition \refeq{eq:esrealbc}. If we assume that the parallel boundary
 condition does not introduce significant spatial inhomogeneity,
 then in the regime that ${\charge \iptl{} } /{\temp} \ll \gyrdi / \lscal $
 and $ {\charge \eptl{} } /{\temp} \sim \gyrde / \lscal $ 
 the fluxes $\esav{\evee\ehhe{}}$ and $\esav{\evei\ehhe{}}$ are not functions of $\gcs$. Hence,
 \beqn \nbls \cdot \esav{\evei \ehhi{} } =\nbls \cdot \orbav{ \esav{\evee \ehhe{} }} = 0 \label{eq:nogbohmisub} \eeqn
 identically. Therefore, the regime where 
 ${\charge \iptl{} } /{\temp} \ll \gyrdi / \lscal $
 and $ {\charge \eptl{} } /{\temp} \sim \gyrde / \lscal $
 can only exist when the \iscaleS fluctuations are linearly stable, 
 and therefore have no amplitude. In this ordering
 there is no saturation mechanism for the \iscaleS turbulence.
 If the \iscaleS fluctuations are linearly unstable they will 
 therefore grow linearly until the gyro-Bohm saturation level is reached.
 We conclude that only the gyro-Bohm scaling \refeq{order:gbohm}
 gives saturated dominant balance in the equations
 \refeq{eq:ionationsclordering}-\refeq{eq:ionatelectronsclordering}. 

 \section{Obtaining the scale-separated \escaleS quasineutrality relation} \label{sec:egavqn}

In this Section we show how to evaluate \escaleS quasineutrality in a local, scale-separated way
for the electron species.
Recalling that for electrons  $\gcf= \rf - \gyrdvece$ and $\gcs= \rs- \gyrdvece$,
 where $\gyrdvece= (\bu\xp\pvel )/\cycfe $ is the vector gyroradius,
and using the Fourier representation \refeq{eq:fouriernote}, then
\beqn \fl \egptle{}(\gcs,\gcf) = \gav{\eptl{}(\rs,\rf)}{\gcs,\gcf} \nonumber \eeqn 
\beqn \fl \qquad  = -\Big(\sum_{\spe} \frac{\zeds^2 \charge \denss}{\temps}\Big )^{-1}
  \sum_{\kf,\ks} \expo{\imag \ks \cdot \gcs}\expo{\imag \kf \cdot \gcf} \bes{|\kf+\ks||\gyrdvece|} \times \nonumber \eeqn \beqn \intv{} \ehhe{\kf,\ks} (\lpar,\energy,\pitch,\sign) \bes{|\kf+\ks||\gyrdvece|},\label{eq:quasinele}\eeqn
where we have used \refeq{eq:gavresult}. By noting that $|\kf||\gyrdvece| \sim |\kf|\gyrde \sim 1$ and $|\ks||\gyrdvece| \sim |\ks|\gyrde \sim \massrt$, we can expand the Bessel function in its argument,
 \beqn \fl \bes{|\kf+\ks||\gyrdvece|} = \bes{|\kf||\gyrdvece|} + \order{ \underbrace{\frac{\kf}{|\kf|}\cdot \ks |\gyrdvece|}_{\sim \massrt} \underbrace{\frac{d \bes{z}}{ d z}\Big|_{z = |\kf||\gyrdvece|} }_{\sim 1} }
 \nonumber \eeqn \beqn  =\bes{|\kf||\gyrdvece|} + \order{  \massr }, \eeqn
 and so we are able to rewrite \refeq{eq:quasinele} in the form
\beqn \fl \egptle{}(\gcs,\gcf) =  -\Big(\sum_{\spe} \frac{\zeds^2 \charge \denss}{\temps}\Big )^{-1}
 \times \nonumber \eeqn \beqn  \sum_{\kf,\ks} \expo{\imag \ks \cdot \gcs}\expo{\imag \kf \cdot \gcf} \bes{|\kf||\gyrdvece|} \times \nonumber \eeqn  
 \beqn \intv{} \ehhe{\kf,\ks} (\lpar,\energy,\pitch,\sign) \bes{|\kf||\gyrdvece|}\left(1+ \order{\massr} \right).\eeqn
This allows us to resum the slow Fourier series, and thus return to a parametric representation where $\gcs$ only appears as a label,
\beqn \fl \egptle{}(\gcs,\gcf) =  -\Big(\sum_{\spe} \frac{\zeds^2 \charge \denss}{\temps}\Big )^{-1} 
\sum_{\kf} \expo{\imag \kf \cdot \gcf} \bes{|\kf||\gyrdvece|} \times \nonumber \eeqn  \beqn \intv{|_\gcs} \ehhe{\kf} (\lpar,\gcs,\energy,\pitch,\sign) \bes{|\kf||\gyrdvece|}\left(1+ \order{\massr} \right). \eeqn 
Therefore, we have found a scale-separated scheme for evaluating quasineutrality; at the \escaleS we can
 parallelise over the label $\gcs$. Consequently, \escaleS flux tubes which are labelled by different
 $\gcs$ may be integrated in isolation, up to coupling introduced by the parallel boundary condition.    

\section{\iscaleS and \escaleS heat fluxes} \label{sec:heatflux}

In this Section we discuss the scaling of the heat flux predicted by
 the coupled system of equations in the ordering \refeq{order:gbohm}.
 The heat flux $\hflux$ is defined as
\beqn \hflux=\fsav{\intv{\big|_\rr} \energy\dlf\frac{\ltsp}{\bmag}\bu \xp \nbl \ptl{} }
=\fsav{\intv{\big|_\rr} \energy\hh{}\frac{\ltsp}{\bmag}\bu \xp \nbl \ptl{} }, \label{eq:qdef}\eeqn
where $\fsav{\cdot}$ is a spatial average over the entire flux tube domain.
The average $\fsav{\cdot}$ is the flux surface average with an additional average in the
$\flxl$ direction, as appropriate to a calculation of turbulent fluxes
 within the scale-separated framework
of $\dlf$ gyrokinetics. 
In equation \refeq{eq:qdef} we used the fact that in
 electrostatic turbulence $\dlf$ is related to $\hh {} $ by \refeq{eq:hh},
 and integration by parts with the periodic flux tube boundary conditions,
 to show that the heat flux is purely due to the non-adiabatic response $\hh{}$. 
In our formalism we can write,
\beqn \fl \hflux=\fsav{\intv{\big|_\rr}\energy (\ihh{}+\ehh{})\left(\frac{\ltsp}{\bmag}\bu \xp (\nbls+\nblf) \ptl{}\right) } \nonumber \eeqn
\beqn \fl \qquad =\frac{1}{\isclvol} \intrpar \intrs \intv{\big|_\rr} \energy\ihh{}\frac{\ltsp}{\bmag}\bu \xp \nbls \iptl{}  \nonumber \eeqn
\beqn + \frac{1}{\nescl\esclvol} \sum_{\rs} \intrpar \intrf \intv{\big|_\rr} \energy\ehh{}(\rs)\frac{\ltsp}{\bmag}\bu \xp \nblf \eptl{}(\rs) ,\eeqn
where the cross terms vanished because of the assumption of statistical periodicity \refeq{eq:solvability},
$\isclvol=\intrpar\intrs$ is the volume of the \iscaleS flux tube,
 $\esclvol=\intrpar\intrf$ is the volume of each \escaleS flux tube, 
and $\nescl = \sum_{\rs} 1$ is the number of  \escaleS flux tubes.
 The sum $\sum_{\rs} $ is over all the \escaleS flux tubes within the \iscaleS flux tube.
 We can identify the \iscaleS heat flux
\beqn \ihflux = \frac{1}{\isclvol} \intrpar \intrs \intv{\big|_\rr} \energy\ihh{}\frac{\ltsp}{\bmag}\bu \xp \nbls \iptl{}  \nonumber \eeqn
\beqn = \frac{1}{\isclvol} \intrpar \isarea  \sum_{\ks} \intv{\big|_\rr} \energy\ihh{-\ks}\frac{\ltsp}{\bmag} \bu \xp i \ks \iptl{\ks} , \eeqn
and the \escaleS heat flux
\beqn \ehflux = \frac{1}{\nescl\esclvol} \sum_{\rs} \intrpar \intrf \intv{\big|_\rr}\energy \ehh{}(\rs)\frac{\ltsp}{\bmag}\bu \xp \nblf \eptl{}(\rs) \nonumber \eeqn
\beqn = \frac{1}{\nescl\esclvol} \sum_{\rs} \intrpar \esarea  \sum_{\kf} \intv{\big|_\rr} \energy\ehh{-\kf}(\rs)\frac{\ltsp}{\bmag} \bu \xp i \kf \eptl{\kf}(\rs) , \eeqn
where $\isarea = \intrs $ and $\esarea = \intrf$ are the areas of the cross sections of the ion and 
\escaleS flux tubes respectively. 
 With these observations, we are able to write down the scaling
 of the fluxes of each species with the potentials $\fract{\charge \iptl{}}{\temp}$
 and $\fract{\charge \eptl{}}{\temp}$, 
 \beqn \ihfluxi \sim  \ihfluxe \sim  \dens \temp \vtheri \left(\frac{\charge \iptl{}}{\temp}\right)^2,   \eeqn
 \beqn \ehfluxi \sim  \dens \temp \vthere \left(\frac{\charge \eptl{}}{\temp}\right)^2 \massr, \quad \ehfluxe \sim  \dens \temp \vthere \left(\frac{\charge \eptl{}}{\temp}\right)^2,  \eeqn
Here the extra factor of $\massrt $ in $\ehfluxi$ appears due to the
 smallness of the ion response at \escaleS and the Bessel functions
 in the ion gyroaverage at \escaleNS.

 In the gyro-Bohm scaling \refeq{order:gbohm}, 
 \beqn \ihfluxi \sim  \ihfluxe \sim  \dens \temp \vtheri \left(\frac{\gyrdi}{\lscal}\right)^2,   \eeqn
 \beqn \ehfluxi \sim \dens \temp \vthere \left(\frac{\gyrde}{\lscal}\right)^2 \massr, \quad \ehfluxe \sim \dens \temp \vthere \left(\frac{\gyrde}{\lscal}\right)^2.  \eeqn
 Hence
  \beqn  \ehfluxe \sim \massr \ihfluxi \sim \massr  \ihfluxe . \eeqn
  This means that the heat flux from the \escaleS
 should be small, but for finite
 mass ratio there can still be a finite contribution due to the stiffness 
 of turbulent transport. 
 We expect that the heat flux from non-zonal passing electrons at \iscaleNS,
 which we neglect, to always be negligible.
 The heat flux from non-zonal passing electrons at \iscaleS is a small correction to $\ihfluxe$.
 In scenarios where the \iscaleS heat flux dominates the \escaleS heat flux
 neglecting a small piece of the \iscaleS heat flux is justified, providing we neglect $\ehfluxe$.
 In cases where the \escaleS heat flux $\ehfluxe$ is comparable to or dominates the heat flux from \iscaleNS,
then again, neglecting a small piece of $\ihfluxe$ is justified.
 The ion \escaleS contribution to the heat flux can always be neglected as $\ehfluxi \sim \massrt \ehfluxe$.
 
\section{The \enquote{twist-and-shift} parallel boundary condition} \label{sec:isbc}
In this Section we reproduce the calculation of the \iscaleS spectral parallel boundary condition first proposed in \cite{beerPoP95},
to allow the reader to familiarise themselves with our notation.
The real space statement of the \iscaleS boundary condition is
\beqn \ihh{}\left(\lpar, \gc \left(\flxl,\fldl \left(\flxl,\lpar,\toragl\right)\right)\right)
   = \ihh{}(\lpar + 2\pi, \gc(\flxl,\fldl(\flxlf,\lpar+2\pi,\toragl)) ).\label {equation:twistshift}\eeqn

Expanding the guiding centre variable in Fourier modes after \refeq{eq:bcfourier}, we find that 
equation \refeq{equation:twistshift} implies
\beqn \fl  \sum_{\kkflxl,\kkfldl} \ihh{(\kkflxl,\kkfldl)}(\lpar) \expo{ \imag \kkflxl (\flxl -\flxl_0) +\imag \kkfldl (\fldl(\flxl,\lpar,\toragl) -\fldl_0)} \nonumber \eeqn 
 \beqn = \sum_{\kkflxl^\prime,\kkfldl} \ihh{(\kkflxl^\prime,\kkfldl)}(\lpar +2\pi ) \expo{\imag \kkflxl^\prime (\flxl -\flxl_0) +\imag \kkfldl (\fldl(\flxl,\lpar+2\pi,\toragl)-\fldl_0) }. \label{eq:bcfourier2}\eeqn
Using the definition of
 $\fldl(\flxl,\lpar,\toragl)-\fldl_0 = \toragl - \saffac_0 \lpar - \saffacprim(\flxl -\flxl_0)$,
 equation \refeq{eq:bcfourier2} can be written as
\beqn \fl \sum_{\kkflxl,\kkfldl} \ihh{(\kkflxl,\kkfldl)}(\lpar) \expo{\imag \kkflxl (\flxlf -\flxl_0) +\imag \kkfldl (\fldl -\fldl_0)} 
\nonumber \eeqn \beqn \fl \qquad= \sum_{\kkflxl^\prime,\kkfldl} \ihh{(\kkflxl^\prime,\kkfldl)}(\lpar +2\pi) \expo{\imag \kkflxl^\prime (\flxl -\flxl_0) +\imag \kkfldl (\fldl-\fldl_0  - 2\pi \saffac_0 - 2\pi \saffacprim (\flxl -\flxl_0)) }
 \nonumber \eeqn \beqn\fl \qquad = \sum_{\kkflxl^\prime,\kkfldl} \ihh{(\kkflxl^\prime,\kkfldl)}(\lpar +2\pi) \expo{\imag (\kkflxl^\prime   - 2\pi \saffacprim) (\flxl -\flxl_0) +\imag \kkfldl (\fldl-\fldl_0)}\expo{- \imag 2\pi \saffac_0\kkfldl} \nonumber \eeqn
 \beqn \eeqn
Equating Fourier coefficients with the same exponent,
 we find that the Fourier space boundary condition is
\beqn \ihh{(\kkflxl,\kkfldl)}(\lpar)= 
 \ihh{(\kkflxl+2\pi \saffacprim \kkfldl,\kkfldl)}(\lpar +2\pi ) \underbrace{\expo{- \imag2\pi \saffac_0 \kkfldl}}_{\sim 1 }, \label{eq:isfourierbc}\eeqn
where we argue that the phase factor $\expo{- \imag 2 \pi \saffac_0 \kkfldl}$ can be taken to be $1$, 
because as $\gyrd / \lscal \rightarrow 0 $ we can make $\kkfldl^{\mbox{\scriptsize min}}$ increasingly large,
 and hence $\saffac_0 \kkfldl$ can be made arbitrarily close to a very large integer for all $\kkfldl$. 

\section{The \escaleS \enquote{twist-and-shift} parallel boundary condition} \label{sec:esbc}
In this Section we calculate the spectral boundary condition for our proposed 
\escaleS parallel boundary condition consistent with equation \refeq{eq:isfourierbc}.
The real space statement of the proposed \escaleS boundary condition is
\beqn \fl \ehh{}(\lpar, \gcf(\flxlf,\fldl(\flxlf,\lpar,\toragl)), \gcs(\flxls,\fldl(\flxls,\lpar,\toragl) ) ) \nonumber \eeqn
\beqn  = \ehh{}(\lpar + 2\pi, \gcf(\flxlf,\fldl(\flxlf,\lpar+2\pi,\toragl)),\gcs(\flxls,\fldl(\flxls,\lpar+2\pi,\toragl))  ). \label {equation:eltwistshift}\eeqn
Expanding the fast guiding centre variable $\gcf$ in Fourier modes as in \refeq{eq:fouriernote},
 we find that equation \refeq{equation:eltwistshift} implies
\beqn  \fl \sum_{\ekkflxl,\ekkfldl} \ehh{(\ekkflxl,\ekkfldl)}(\lpar , \gcs(\flxls,\fldl(\flxls,\lpar,\toragl) ) ) \expo{\imag \ekkflxl (\flxlf -\flxl_0) + \imag \ekkfldl (\fldl(\flxlf,\lpar,\toragl) -\fldl_0)} \nonumber \eeqn
\beqn \fl \qquad = \sum_{\ekkflxl^\prime,\ekkfldl} \ehh{(\ekkflxl^\prime,\ekkfldl)}(\lpar +2\pi , \gcs(\flxls,\fldl(\flxls,\lpar+2\pi,\toragl) ) ) \times \nonumber\eeqn
\beqn  \expo{\imag \ekkflxl^\prime (\flxlf -\flxl_0) + \imag\ekkfldl (\fldl(\flxlf,\lpar+2\pi,\toragl)-\fldl_0) }, \label{eq:bcfourier3}\eeqn 
where $\ekkflxl$ and $\ekkfldl$ are the \escaleS wave numbers corresponding to $\flxlf$ and $\fldlf$.
Using the definitions of $\fldls(\flxls,\lpar,\toragl)$ and $\fldlf(\flxlf,\lpar,\toragl)$,
 equation \refeq{eq:bcfourier3} can be written as
\beqn \fl \sum_{\ekkflxl,\ekkfldl} \ehh{(\ekkflxl,\ekkfldl)}(\lpar , \gcs(\flxls,\fldls )) \expo{\imag \ekkflxl (\flxlf -\flxl_0) +\imag\ekkfldl (\fldlf -\fldl_0)} \nonumber \eeqn
\beqn \fl \qquad = \sum_{\ekkflxl^\prime,\ekkfldl} \ehh{(\ekkflxl^\prime,\ekkfldl)}(\lpar +2\pi , \gcs(\flxls,\fldls - 2\pi \saffac_0 - 2\pi \saffacprim (\flxls -\flxl_0)) ) \times \nonumber \eeqn
 \beqn \expo{\imag \ekkflxl^\prime (\flxlf -\flxl_0) +\imag\ekkfldl (\fldlf-\fldl_0  - 2\pi \saffac_0 - 2\pi \saffacprim (\flxls -\flxl_0)) }. \eeqn
Equating Fourier coefficients with the same exponent, we find that the Fourier space boundary condition is 
\beqn \fl \ehh{(\ekkflxl,\ekkfldl)}(\lpar , \gcs(\flxls,\fldls ) )= \nonumber \eeqn
\beqn \fl \qquad \ehh{(\ekkflxl+2\pi \saffacprim \ekkfldl,\ekkfldl)}(\lpar +2\pi , \gcs(\flxls,\fldls \underbrace{- 2\pi \saffac_0}_{\mbox{neglected}} - 2\pi \saffacprim (\flxls -\flxl_0)) )  \underbrace{\expo{- \imag 2\pi \saffac_0 \ekkfldl}}_{\sim 1 }, \label{eq:esfourierbc}\eeqn
where we neglect $- 2\pi \saffac_0$ in the $\fldls$ coordinate to be consistent with taking
  $\expo{- \imag 2 \pi \saffac_0 \kkfldl}=1$ in equation \refeq{eq:isfourierbc}.

\section{Satisfaction of relations \refeq{eq:alphaderivative} and \refeq{eq:fluxderivative}}\label{sec:esbcconfirm}
In this Section we show that the relations~\refeq{eq:alphaderivative} and \refeq{eq:fluxderivative}
are satisfied when the \iscaleS turbulence satisfies the boundary condition \refeq{eq:isrealbc},
and hence demonstrate that equation \refeq{eq:esrealbc} is a sensible parallel boundary condition 
for the \escaleS turbulence. 
First, write the $\flxls$ derivatives at fixed $\toragl$ instead of fixed $\fldls$,
\beqn \drv{}{\flxls}\Big|_{\fldls,\lpar} = \drv{}{\flxls}\Big|_{\toragl,\lpar} + 
\saffacprim \lpar \drv{}{\toragl}\Big|_{\flxls,\lpar}, \quad \drv{}{\fldls}\Big|_{\flxls,\lpar}
 = \drv{}{\toragl}\Big|_{\flxls,\lpar}. \label{eq:changevar} \eeqn
Then,
\beqn \fl \drv{ }{\flxls}\Big|_{\fldls,\lpar+2\pi} \ihh{}(\lpar + 2 \pi, \gcs(\flxls,\fldl(\flxls,\lpar+2\pi,\toragl)) )  \nonumber \eeqn 
\beqn - 2 \pi \saffacprim \drv{}{\fldls}\Big|_{\flxls,\lpar+2\pi} \ihh{}(\lpar + 2\pi, \gcs(\flxls,\fldl(\flxls,\lpar + 2 \pi,\toragl)  ) ) \nonumber\eeqn
\beqn =\Big[\drv{}{\flxls}\Big|_{\toragl,\lpar+2\pi} + \saffacprim \lpar \drv{}{\toragl}\Big|_{\flxls,\lpar+2\pi} \Big] \ihh{}(\lpar + 2 \pi, \gcs(\flxls,\fldl(\flxls,\lpar + 2 \pi,\toragl)  ) ) \nonumber \eeqn
\beqn =\Big[\drv{}{\flxls}\Big|_{\toragl,\lpar} + \saffacprim \lpar \drv{}{\toragl}\Big|_{\flxls,\lpar} \Big] \ihh{}(\lpar, \gcs(\flxls,\fldl(\flxls,\lpar,\toragl)  ) ) \nonumber \eeqn
\beqn =\drv{}{\flxls}\ihh{}(\lpar, \gcs(\flxls,\fldl(\flxls,\lpar,\toragl)  ) ), \eeqn
where we have used the relations \refeq{eq:changevar} to rewrite the derivatives
 in a way that we can use relation \refeq{eq:esrealbc}, and used that 
\beqn \drv{}{\flxls}\Big|_{\toragl,\lpar+2\pi} = \drv{}{\flxls}\Big|_{\toragl,\lpar}, \mbox { and } \drv{}{\toragl}\Big|_{\flxls,\lpar+2\pi} = \drv{}{\toragl}\Big|_{\flxls,\lpar}. \eeqn
 We can therefore see that relations \refeq{eq:alphaderivative} and \refeq{eq:fluxderivative} are satisfied.

\section{Boosting away the piece of $\ive$ that is constant in $\lpar$} \label{sec:thetaconstve}

In this Section we show that if we write the equation for the \escaleS fluctuations
 \refeq{equation:electronelectronreal} in terms of 
 the guiding centre distribution functions $\egge{} = \gav{\edlfe}{\gcf,\gcs}$ and 
$\igge{} = \gav{\idlfe}{\gcs}$ then we can remove the component of the
 \iscaleS advection $\ivee$ due to 
 the component $\ciptl $ of the potential $ \iptl{} $ which is constant in
 $\lpar$. 
 By writing equation \refeq{equation:electronelectronreal}
 in terms of $\egge{}$ and 
$\igge{} $  we find that
\beqn \fl \drv{\egge{}}{\tf} + \vpar \kpar \drv{\egge{}}{\lpar} + (\vme + \ivee ) \cdot \nblf \egge{} +
  \evee \cdot \nblf \egge{} \nonumber \eeqn \beqn + \evee \cdot (\nbl \eqlbe + \nbls \igge{}) =
   -\frac{\zed \charge \eqlbe}{\temp}\vpar \kpar \drv{\egptl{}}{\lpar}
    -\frac{\zed \charge \eqlbe}{\temp}\vme \cdot \nblf \egptl{}   , \label{equation:electronelectronrealgg}\eeqn
 where we have used that $\ivee \cdot \nblf \egptl{} = - \evee \cdot \nbls \igptl{}$
  and $\evee \cdot \nblf \egptl{} = 0$.
 Note that if $\ciptl$ is constant in $\lpar$ this means that
 $\drvt{\ciptl}{\lpar}|_{\fldl}=0$.
 For irrational values of the field line pitch $\drvt{\toragl}{\lpar}|_{\fldl}$, this implies
 that $\ciptl$ is also a constant in $\fldl$. We will henceforth make this assumption.
  We write \refeq{equation:electronelectronrealgg} as
\beqn \drv{\egg{}}{\tf} + \cive \cdot \nbl \egg{} = \source, \label{eq:toyadv}\eeqn
where 
\beqn \fl \cive = \frac{\ltsp}{\bmag} \bu \xp \nbl \ciptl,
 \quad \nbl \simeq  \nbl \flxl \left( \drv{}{\flxls} + \drv{}{\flxlf}\right)
 + \nbl \fldl \left(\drv{}{\fldls} + \drv{}{\fldlf}\right), \eeqn
 and $\source$ contains the terms in \refeq{equation:electronelectronrealgg} 
 which do not explicitly appear in \refeq{eq:toyadv}.
Expanding the vector expression $\cive \cdot \nbl \egg{}$ in \refeq{eq:toyadv}, we find that
\beqn \cive \cdot \nbl \egg{} = -\frac{\ltsp}{\bmag} \bu \cdot \underbrace{\bmagform}_{\bvec} \drv{\ciptl}{\flxls}\drv{\egg{}}{\fldlf}
 = -\ltsp  \drv{\ciptl}{\flxls}\drv{\egg{}}{\fldlf}. \eeqn
Here $\partial \ciptl / \partial \flxls$ has no dependence on $\lpar$. 
This means that at a given \iscaleS location $(\flxls, \fldls)$ the drift velocities
 are constant within the \escaleS flux tube. The equation \refeq{eq:toyadv} is now
\beqn \drv{\egg{}}{\tf} - \underbrace{\ltsp \drv{\ciptl}{\flxls}}_{\mbox{constant}} \drv{\egg{}}{\fldlf}  =\source. \eeqn
Let
\beqn \tf^\prime = \tf, \quad \flxlf^\prime = \flxlf,\quad \fldlf^\prime = \fldlf - \ltsp\drv{\ciptl}{\flxls} \tf. \eeqn
Changing to the coordinates $(\tf^\prime, \flxlf^\prime, \fldlf^\prime)$, we find
\beqn \drv{\egg{}}{\tf^\prime} = \source , \eeqn
i.e. we have boosted to a frame rotating with the constant drift $\cive$.

\section{\iscaleS and \escaleS collision operators}\label{sec:collisions}
  In this Section we discuss how to include the effect of collisions
in the scale-separated model of coupled \iscaleNS-\escaleS turbulence
\refeq{equation:ionionreal}-\refeq{equation:qnelectronreal}.
We first discuss the orderings for collision frequencies.
 Then in \refsec{sec:colform} we find the forms of the \iscaleS and \escaleS collision operators.
 Using the same techniques as used in \refsec{sec:egavqn} 
to find the scale-separated quasineutrality relation \refeq{equation:qnelectronreal}
 for the \escaleS gyrokinetic equation \refeq{equation:electronelectronreal},
 we find in \refsec{sec:colsep} the scale-separated \escaleS collision operator
 which should appear in \refeq{equation:electronelectronreal}.
 Finally in \refsec{sec:fouriercol} we give the Fourier representation of the collision operators.
 
We have noted that collisions are required to regularise
velocity space. In order to retain the regularising effect of collisions
without obtaining a purely adiabatic electron response at
\iscaleNS, we order
\beqn \cfreqee \sim \cfreqei \sim \frac{\vtheri}{\lscal}, \label{order:electroncollisions}\eeqn
where $\cfreqee $ is the electron-electron collision frequency
and $ \cfreqei $ is the electron-ion collision frequency.
To be consistent, the ion-ion collision frequency 
\beqn \cfreqii\sim \massr \frac{\vtheri}{\lscal} \label{order:ioncollisions}\eeqn
because ions collide with other ions at a rate $\massrt$
times slower than the rate at which electrons collide with other electrons.
The ion-electron collision frequency is negligible,
\beqn \cfreqie \sim \massrl \cfreqei \sim \massr \cfreqii. \label{order:nuie}\eeqn 
 Due to the diffusive nature of the collision operator \cite{helander}, we have that,
 for ions, at both spatial scales,
\beqn \copi\sim\cfreqii\vtheri^2 (\dveli)^{-2} \hhi{},\label{order:ioncollisions2} \eeqn 
 and for electrons, at both spatial scales,
\beqn \cope\sim\cfreqee\vthere^2 (\dvele)^{-2} \hhe{}. \label{order:electroncollisions2} \eeqn
In the orderings \refeq{order:electroncollisions} and \refeq{order:ioncollisions}
the bulk of the ion distribution functions, where $\dveli\sim\vtheri$,  
and the bulk of the electron distribution function at \escaleNS, 
 where $\dvele\sim\vthere$, are unaffected by collisions at leading order.
 This can be observed by inspecting \refeq{order:ioncollisions2} and \refeq{order:electroncollisions2}
 and comparing the size of the collision terms with
the linear timescale at each scale, i.e.
\beqn \copi \sim \massr\frac{\vtheri}{\lscal} \hhi{} \ll \frac{\vtheri}{\lscal} \hhi{}, \eeqn
and 
\beqn \cope \sim \massr\frac{\vthere}{\lscal} \hhe{} \sim \frac{\vtheri}{\lscal} \hhe{}
 \ll \frac{\vthere}{\lscal} \hhe{}. \eeqn
For electrons at \iscaleS in the bulk of the distribution, 
where $\dvele\sim\vthere$, the frequency of collisions is fast enough
to modify the trapped piece of velocity space at leading order,
but not fast enough to detrap the electrons entirely,
 which would require $\cope \gg (\fract{\vtheri}{\lscal}) \hhe{}$.
 This allows us to keep a fully kinetic description for the electrons.
Nonetheless, pieces of the distribution functions
inevitably develop small velocity space structures
due to the parallel streaming terms, which introduce phase mixing.
In the presence of collisions these small velocity space structures 
are damped, because the collision terms, which introduce dissipation,
 become as large as the parallel streaming terms for each species. 
 If $\dveli\lesssim \massrst \vtheri$ for a piece of the ion distribution function
then by \refeq{order:ioncollisions2} the effect of collisions will be significant: 
for this piece of the ion distribution function, 
$\copi \gtrsim (\fract{\vtheri}{\lscal}) \hhi{}\sim  \vpar \kpar \drvt{\hhi{}}{\lpar} $.
Similarly if $\dvele\lesssim \massrst \vthere$ for a piece of the electron distribution function,
then by \refeq{order:electroncollisions2} the effect of collisions will be significant:
 for this piece of the electron distribution function,
 $\cope \gtrsim (\fract{\vthere}{\lscal}) \hhe{} \sim \vpar \kpar \drvt{\hhe{}}{\lpar}$.
Altogether, this allows us to assume that $\dvele \sim \vthere$ and $\dveli \sim \vtheri$ 
 at both spatial scales.

\subsection{The forms of the collision operators}\label{sec:colform}
 To order $\order{\massrt(\fract{\vtheri}{\lscal})\hhi{}}$,
  the ion collision operator has the form \cite{helander}
\beqn \copi = \gav{\copii{\hhi{}}}{\gc}, \label{eq:fullioncoll}\eeqn
where $\copnaii$ is the linearised ion-ion self collision operator, 
 and the velocity space derivatives acting on $\hhi{}$
are taken at fixed $\lpar$ and $\rr$. Note that because of
ordering \refeq{order:nuie}, the contribution of ion-electron collisions, of order
$(\massrlt)(\fract{\vtheri}{\lscal})\hhi{}$, can be neglected
because it is $\order{\massrlt}$ small compared to the linear terms.
To find the \iscaleS collision operator which should appear in \refeq{equation:ionionreal},
 we apply the \escaleS average to \refeq{eq:fullioncoll}. We find that
     \beqn \icopi = \gav{\copii{\ihhi{}(\lpar,\rs - \gyrdveci,\energy,\pitch,\sign)}}{\gcs},\label{eq:ISioncoll}\eeqn
     where our notation indicates velocity derivatives are held at fixed $\lpar$ and $\rs$.
     To derive \refeq{eq:ISioncoll} we have used the properties proved in \refsec{sec:gavproperties}:
the \escaleS average 
 commutes with the gyroaverage \refeq{equation:gav}; and either $\gcf$ or $\rf$ can be
used as the \escaleS average integration variable.
     The ions at \escaleS are adiabatic and so we will not require $\ecopi$.
     Keeping terms in the electron collision operator relevant in our ordering,
     $\order{\massrt(\fract{\vtheri}{\lscal})\hhe{}}$,
     we have \cite{helander}
     \beqn \cope = \gav{\copee{\hhe{}}}{\gc} +
     \gav{\copei{\hhe{}-\frac{\me \uveli\cdot \pvel}{\tempe}\eqlbe}}{\gc},\label{eq:ecfull} \eeqn
     where $\copnaee$ is the electron-electron self collision operator,
$\copnaei$ is the Lorentz pitch angle scattering operator,
which appears due to electron-ion collisions,
and 
\beqn \uveli=\frac{1}{\dens}\intv{\big|_{\rr}} \; \pvel \hhi{}, \label{eq:uidef} \eeqn
is the the mean ion velocity.
Again using the \escaleS average \refeq{eq:defesav} and the properties proved in
 \refsec{sec:gavproperties}, we find 
 \beqn \icope = \gav{\copee{\ihhe{}}}{\gcs} +
     \gav{\copei{\ihhe{}-\frac{\me \iuveli\cdot \pvel}{\tempe}\eqlbe}}{\gcs}, \label{eq:ecIS} \eeqn
where 
\beqn \iuveli=\frac{1}{\dens}\intv{\big|_{\rr}} \; \pvel \ihhi{}, \label{eq:uidef2} \eeqn
 and the velocity derivatives in $\copnaee$ and $\copnaei$ are held at fixed $\lpar$ and $\rs$.
 In \refeq{eq:ecIS} we regard $\ihhe{}=\ihhe{}(\lpar,\rs - \gyrdvece,\energy,\pitch,\sign)$.
 The collision operator which should appear in \refeq{equation:electronionreal} is
 \beqn\orbav{\icope}= \orbav{\gav{\copee{\ihhe{}^{(0)}}}{\gcs}} +
\orbav{\gav{\copei{\ihhe{}^{(0)}}}{\gcs}}, \label{eq:ecIS2} \eeqn
where $\ihhe{}^{(0)}$ is the leading order piece of the \iscaleS electron distribution function,
 and we have neglected 
 \beqn \frac{\me \iuveli\cdot \pvel}{\tempe}\eqlbe \sim \massr \frac{\gyrdi}{\lscal} \eqlbe \ll \ihhe{}^{(0)}. \eeqn
 By taking the difference between \refeq{eq:ecfull} and \refeq{eq:ecIS}, we find the
  \escaleS electron collision operator
 \beqn \ecope = \gav{\copee{\ehhe{}}}{\gcs,\gcf} +
     \gav{\copei{\ehhe{}-\frac{\me \euveli\cdot \pvel}{\tempe}\eqlbe}}{\gcs,\gcf}, \label{eq:ecES} \eeqn
 where \beqn \euveli=\frac{1}{\dens}\intv{\big|_{\rr}} \; \pvel \ehhi{}. \label{eq:uidef3} \eeqn
 In \refeq{eq:ecES} the velocity derivatives in $\copnaee$ and
 $\copnaei$ are held at fixed $\lpar$, $\rs$ and $\rf$,
 and we regard
 $\ehhe{}=\ehhe{}(\lpar,\rs - \gyrdvece,\rf - \gyrdvece,\energy,\pitch,\sign)$.
 Note that in the electron collision operator at \escaleNS,
we only need to retain terms of
$\order{  \cfreqee \ehhe{}  } \sim \order{(\fract{\vtheri}{\lscal}) \ehhe{}}$. 
As discussed in Section \refsec{sec:separation},
 in our ordering for the electron collision frequency \refeq{order:electroncollisions},
we imposed that collisions at \escaleS appear as a sub-dominant term
designed to regularise sharp gradients in velocity space.
Note that $\euveli/\vtheri$ is small,
\beqn \frac{\euveli}{\vtheri} \sim \massr \frac{\charge \eptl{}}{\temp}, \label{eq:euvelismall}\eeqn
where in \refeq{eq:euvelismall} one factor of $\massrst$ is due to the velocity integration 
 in the definition of $\euveli$ \refeq{eq:uidef3},
 which introduces an ion gyroaverage, and the second appears because 
 $\ehhi{} \sim \massrst\fract{\charge\eptl{}}{\temp}$ by the scaling~\refeq{order:gbohm}.
Hence, we can neglect the contribution to the electron-ion collision operator
from the ion response at \escaleS $\ehhi{}$,
\beqn \frac{\me \euveli \cdot \pvel}{\tempe} \eqlbe \sim \frac{\euveli}{\vthere}\eqlbe
   \ll \ehhe{}. \label{eq:csmall} \eeqn
 The \escaleS electron collision operator neglecting ions at \escaleS is therefore
 \beqn \fl \ecope = \gav{\copee{\ehhe{}(\lpar,\rs - \gyrdvece,\rf - \gyrdvece,\energy,\pitch,\sign)}}{\gcs,\gcf} \nonumber \eeqn
    \beqn + \gav{\copei{\ehhe{}(\lpar,\rs - \gyrdvece,\rf - \gyrdvece,\energy,\pitch,\sign)}}{\gcs,\gcf}, \label{eq:ecES2} \eeqn
where the reader should note that \refeq{eq:ecES2} does not yet
 represent a scale-separated collision operator due to
the gyroaverages and velocity derivatives held at fixed $\rs$.

\subsection{Scale-separation in the presence of collisions}\label{sec:colsep}
    To obtain a scale-separated \escaleS gyrokinetic equation
the slow spatial coordinate $\gcs$ must appear only as a label.
Gyroaverages and velocity derivatives held at fixed $\rs$
introduce coupling between perpendicular locations
in the \iscaleS flux tube which naively appear to break scale separation.

To deal with the gyroaverages and velocity derivatives held at fixed $\rs$ in \refeq{eq:ecES},
 we perform the same operation of approximating the Bessel functions
 as in the \escaleS quasineutrality relation \refeq{eq:quasiapprox}
 to write the gyroaverage and velocity derivatives in $\ecope$ at fixed $\lpar$, $\rf$
and $\gcs$ with only $\order{\massrt} $ error (as with quasineutrality)
  \beqn \fl \ecope = \gav{\copee{\ehhe{}(\lpar,\gcs,\rf - \gyrdvece,\energy,\pitch,\sign)}}{\gcs,\gcf}\left(1+ \order{\massr}  \right) \nonumber \eeqn 
   \beqn+     \gav{\copei{\ehhe{}(\lpar,\gcs,\rf - \gyrdvece,\energy,\pitch,\sign)}}{\gcs,\gcf}\left(1+ \order{\massr}  \right), \label{eq:ecES3} \eeqn
and so obtain a scale-separated collision operator for \refeq{equation:electronelectronreal}.
 
 \subsection{Fourier representation for collisions}\label{sec:fouriercol} 
 We now give the Fourier representation of the collision operators
 which appear in gyrokinetic equations 
 \refeq{equation:ionionfourier}, \refeq{equation:electronionfourier}
 and \refeq{equation:electronelectronfourier} 
 when the effect of collisions is included.
 In the equation for ions at \iscaleS  \refeq{equation:ionionfourier}
 the collision operator is
 \beqn \icopik = 
 \gav{\expo{\imag \ks \cdot \gyrdveci} \copii{\expo{-\imag \ks \cdot \gyrdveci}\ihhi{\ks}}}{}.
 \label{equation:colionionfourier}\eeqn
 In the equation for electrons at \iscaleS \refeq{equation:electronionfourier} 
the collision operator is
 \beqn \fl \icopek=  
  \orbav{\gav{\expo{\imag \ks \cdot \gyrdvece} \copee{\expo{-\imag \ks \cdot \gyrdvece}\ihhe{\ks}}}{}}
\nonumber \eeqn
 \beqn \fl \; \; 
 + \orbav{\gav{\expo{\imag \ks \cdot \gyrdvece} \copei{\expo{-\imag \ks \cdot \gyrdvece}\ihhe{\ks}
 }}{}},
 \label{equation:colelectronionfourier}\eeqn
 where $\ihhe{\ks} = \ihhe{\ks}^{(0)}$
and the reader should note the presence of the orbital average $\orbav{\cdot}$.
At \escaleS in \refeq{equation:electronelectronfourier}
  we also only need the electron collision operator
 \beqn \fl \ecopek = 
 \gav{\expo{\imag \kf \cdot \gyrdvece} \copee{\expo{-\imag \kf \cdot \gyrdvece}\ihhe{\kf}}}{\gcs} \nonumber \eeqn \beqn
 \gav{\expo{\imag \kf \cdot \gyrdvece} \copei{\expo{-\imag \kf \cdot \gyrdvece}\ihhe{\kf}}}{\gcs},
 \label{equation:colelectronelectronfourier}\eeqn
 where the velocity derivatives appearing in
$\copnaee$ and $\copnaei$ are held at fixed $\gcs$.

\bibliography{hardman_MS_theory_IOP11}

\providecommand{\newblock}{}
\begin{thebibliography}{10}
\expandafter\ifx\csname url\endcsname\relax
  \def\url#1{{\tt #1}}\fi
\expandafter\ifx\csname urlprefix\endcsname\relax\def\urlprefix{URL }\fi
\providecommand{\eprint}[2][]{\url{#2}}

\bibitem{sugamaPoP97}
Sugama H and Horton W 1997 {\em Phys. Plasmas\/} {\bf 4} 405

\bibitem{SugamaPoP98}
Sugama H and Horton W 1998 {\em Physics of Plasmas\/} {\bf 5} 2560--2573

\bibitem{trantimecatto08PPCF}
Catto P~J, Simakov A~N, Parra F~I and Kagan G 2008 {\em Plasma Physics and
  Controlled Fusion\/} {\bf 50} 115006

\bibitem{abelRPP13}
Abel I~G, Plunk G~G, Wang E, Barnes M, Cowley S~C, Dorland W and Schekochihin
  A~A 2013 {\em Reports on Progress in Physics\/}  116201

\bibitem{candyPoP09}
Candy J, Holland C, Waltz R~E, Fahey M~R and Belli E 2009 {\em Phys. Plasmas\/}
  {\bf 16} 060704

\bibitem{barnesPoP10b}
Barnes M, Abel I~G, Dorland W, Goerler T, Hammett G~W and Jenko F 2010 {\em
  Phys. Plasmas\/} {\bf 17} 056109 arxiv:0912.1974

\bibitem{longwavecalvo12PPCF}
Calvo I and Parra F~I 2012 {\em Plasma Physics and Controlled Fusion\/} {\bf
  54} 115007

\bibitem{surkoS83}
Surko C~M and Slusher R~E 1983 {\em Science\/} {\bf 221} 817

\bibitem{cowleyPoFB91}
Cowley S~C, Kulsrud R~M and Sudan R 1991 {\em Phys. Fluids B\/} {\bf 3} 2767

\bibitem{hortonRMP99}
Horton W 1999 {\em Rev. Mod. Phys.\/} {\bf 71} 735--778

\bibitem{dorland2000electron}
Dorland W, Jenko F, Kotschenreuther M and Rogers B~N 2000 {\em Phys. Rev.
  Lett.\/} {\bf 85} 5579--5582

\bibitem{jenko2000electron}
Jenko F, Dorland W, Kotschenreuther M and Rogers B~N 2000 {\em Physics of
  Plasmas\/} {\bf 7} 1904--1910

\bibitem{jenko2002prediction}
Jenko F and Dorland W 2002 {\em Phys. Rev. Lett.\/} {\bf 89} 225001

\bibitem{STroach2009PPCF}
Roach C~M, Abel I~G, Akers R~J, Arter W, Barnes M, Camenen Y, Casson F~J,
  Colyer G, Connor J~W, Cowley S~C, Dickinson D, Dorland W, Field A~R,
  Guttenfelder W, Hammett G~W, Hastie R~J, Highcock E, Loureiro N~F, Peeters
  A~G, Reshko M, Saarelma S, Schekochihin A~A, Valovic M and Wilson H~R 2009
  {\em Plasma Physics and Controlled Fusion\/} {\bf 51} 124020

\bibitem{ResolvingWGuttenfelder2011POP}
Guttenfelder W and Candy J 2011 {\em Physics of Plasmas\/} {\bf 18} 022506

\bibitem{ProgressWGuttenfelder2013NF}
Guttenfelder W, Peterson J, Candy J, Kaye S, Ren Y, Bell R, Hammett G, LeBlanc
  B, Mikkelsen D, Nevins W and Yuh H 2013 {\em Nuclear Fusion\/} {\bf 53}
  093022

\bibitem{2016colyer}
Colyer G~J, Schekochihin A~A, Parra F~I, Roach C~M, Barnes M~A, c~Ghim Y and
  Dorland W 2017 {\em Plasma Physics and Controlled Fusion\/} {\bf 59} 055002

\bibitem{ren2017recent}
Ren Y, Belova E, Gorelenkov N, Guttenfelder W, Kaye S, Mazzucato E, Peterson J,
  Smith D, Stutman D, Tritz K, Wang W, Yuh H, Bell R, Domier C and LeBlanc B
  2017 {\em Nuclear Fusion\/} {\bf 57} 072002

\bibitem{waltz2007coupled}
Waltz R, Candy J and Fahey M 2007 {\em Physics of plasmas\/} {\bf 14} 056116

\bibitem{candy2007effect}
Candy J, Waltz R~E, Fahey M~R and Holland C 2007 {\em Plasma Physics and
  Controlled Fusion\/} {\bf 49} 1209

\bibitem{gorler2008scale}
G{\"o}rler T and Jenko F 2008 {\em Physical review letters\/} {\bf 100} 185002

\bibitem{howard2014synergistic}
Howard N~T, Holland C, White A~E, Greenwald M and Candy J 2014 {\em Physics of
  Plasmas\/} {\bf 21} 112510

\bibitem{howard2015fidelity}
Howard N~T, Holland C, White A~E, Greenwald M and Candy J 2015 {\em Plasma
  Physics and Controlled Fusion\/} {\bf 57} 065009

\bibitem{howard2016enhanced}
Howard N, Holland C, White A, Greenwald M and Candy J 2016 {\em Nuclear
  Fusion\/} {\bf 56} 014004

\bibitem{howard2016comparison}
Howard N, Holland C, White A, Greenwald M, Candy J and Creely A 2016 {\em
  Physics of Plasmas\/} {\bf 23} 056109

\bibitem{maeyama2015cross}
Maeyama S, Idomura Y, Watanabe T~H, Nakata M, Yagi M, Miyato N, Ishizawa A and
  Nunami M 2015 {\em Physical review letters\/} {\bf 114} 255002

\bibitem{maeyama2017supression}
Maeyama S, Watanabe T~H and Ishizawa A 2017 {\em Phys. Rev. Lett.\/} {\bf 119}
  195002

\bibitem{howard2014multi}
Howard N, White A, Greenwald M, Holland C and Candy J 2014 {\em Physics of
  Plasmas\/} {\bf 21} 032308

\bibitem{beerPoP95}
Beer M~A, Cowley S~C and Hammett G~W 1995 {\em Phys. Plasmas\/} {\bf 2} 7

\bibitem{cattoPP78}
Catto P~J 1978 {\em Plasma Phys.\/} {\bf 20} 719

\bibitem{friemanPoF82}
Frieman E~A and Chen L 1982 {\em Phys. Fluids\/} {\bf 25} 502

\bibitem{brizardRMP07}
Brizard A~J and Hahm T~S 2007 {\em Rev. Mod. Phys.\/} {\bf 79} 421

\bibitem{parraPPCF08}
Parra F~I and Catto P~J 2008 {\em Plasma Phys. Control. Fusion\/} {\bf 50}
  065014

\bibitem{germano_1992}
Germano M 1992 {\em Journal of Fluid Mechanics\/} {\bf 238} 325–336

\bibitem{BerselliLES}
Berselli L, Iliescu T and Layton W 2005 {\em Mathematics of Large Eddy
  Simulation of Turbulent Flows, First Edition\/} (Springer)

\bibitem{GarnierLES}
Garnier E, Adams N and Sagaut P 2009 {\em Large Eddy Simulation for
  Compressible Flows\/} (Springer)

\bibitem{KampenPhysica1955}
Van~Kampen N~G 1955 {\em Physica (Amsterdam)\/} {\bf 21} 949

\bibitem{CaseAnnPhys1959}
Case K~M 1959 {\em Ann. Phys.\/} {\bf 7} 349

\bibitem{krommesPoP94}
Krommes J~A and Hu G 1994 {\em Phys. Plasmas\/} {\bf 1} 3211

\bibitem{krommesPoP99}
Krommes J~A 1999 {\em Phys. Plasmas\/} {\bf 6} 1477

\bibitem{schekPPCF08}
Schekochihin A~A, Cowley S~C, Dorland W, Hammett G~W, Howes G~G, Plunk G~G,
  Quataert E and Tatsuno T 2008 {\em Plasma Phys. Control. Fusion\/} {\bf 50}
  124024 arXiv: 0806.1069

\bibitem{schekApJ09}
Schekochihin A~A, Cowley S~C, Dorland W, Hammett G~W, Howes G~G, Quataert E and
  Tatsuno T 2009 {\em Astrophys. J. Suppl\/} {\bf 182} 310 arXiv: 0704.0044

\bibitem{tatsunoPRL09}
Tatsuno T, Dorland W, Schekochihin A~A, Plunk G~G, Barnes M, Cowley S~C and
  Howes G~G 2009 {\em Phys. Rev. Lett.\/} {\bf 103} 015003

\bibitem{barnesPoP10a}
Barnes M, Dorland W and Tatsuno T 2010 {\em Phys. Plasmas\/} {\bf 17} 032106

\bibitem{abelPoP08}
Abel I~G, Barnes M, Cowley S~C, Dorland W, Hammett G~W and Schekochihin A~A
  2008 {\em Phys. Plasmas\/} {\bf 15} 122509 arXiv:0806.1069

\bibitem{barnesPoP09}
Barnes M, Abel I~G, Dorland W, Ernst D~R, Hammett G~W, Ricci P, Rogers B~N,
  Schekochihin A~A and Tatsuno T 2009 {\em Phys. Plasmas\/} {\bf 16} 072107

\bibitem{BenderOrszag}
Bender C~M and Orszag S~A 1999 {\em Advanced Mathematical Methods for
  Scientists and Engineers I: Asymptotic Methods and Perturbation Theory.\/}
  (New York: Springer)

\bibitem{abelEMOD}
Abel I~G and Cowley S~C 2013 {\em New Journal of Physics\/} {\bf 15} 023041

\bibitem{HallatschekgiantelPRL2005}
Hallatschek K and Dorland W 2005 {\em Phys. Rev. Lett.\/} {\bf 95} 055002

\bibitem{DominskinonadPOP015}
Dominski J, Brunner S, Görler T, Jenko F, Told D and Villard L 2015 {\em
  Physics of Plasmas\/} {\bf 22} 062303

\bibitem{HazeltineMeiss}
Hazeltine R~D and Meiss J~D 2003 {\em Plasma Confinement\/} (New York: Dover)

\bibitem{hintonRMP76}
Hinton F~L and Hazeltine R~D 1976 {\em Rev. Mod. Phys.\/} {\bf 25} 239--308

\bibitem{helander}
Helander P and Sigmar D~J 2002 {\em Collisional transport in magnetized
  plasmas\/} (Cambridge: Cambridge University Press)

\bibitem{barnesPRL11b}
Barnes M, Parra F~I and Schekochihin A~A 2011 {\em Phys. Rev. Lett.\/} {\bf
  107} 115003 arxiv:1104.4514

\bibitem{2015Watanabefluxtubetrain}
Watanabe T~H, Sugama H, Ishizawa A and Nunami M 2015 {\em Physics of Plasmas\/}
  {\bf 22} 022507

\end{thebibliography}

\end{document}